\documentclass[twocolumn]{aastex63}

\graphicspath{{./}{figures/}}
\usepackage{pifont}
\usepackage{multirow}
\usepackage{amsmath}
\usepackage{color}
\usepackage{graphicx}
\usepackage{subfigure}
\usepackage{boxedminipage}
\usepackage{threeparttable}
\usepackage{cases}
\usepackage{CJKutf8}
\usepackage{lineno}
\usepackage{booktabs}

\newcommand{\vpecp}{v_\mathrm{pec, p}}
\newcommand{\masyr}{\mathrm{mas~yr^{-1}}}
\newcommand{\gaia}{{\it Gaia}}

\newcommand{\pkv}[1][]{v_\mathrm{pec #1}^{z=0}}

\shorttitle{}
\shortauthors{Wang et al.}

\begin{document}
\begin{CJK*}{UTF8}{gbsn}

\title{On Neutron Star Natal Kicks in High-Mass X-Ray Binaries: Insights from Population Synthesis}
\correspondingauthor{Xiang-Dong Li}
\email{lixd@nju.edu.cn}

\author[0000-0002-9738-1238]{Xiangyu Ivy Wang (王翔煜)}
\affiliation{School of Astronomy and Space Science, Nanjing University, Nanjing 210093, China}
\author{Xiang-Dong Li}
\affiliation{School of Astronomy and Space Science, Nanjing University, Nanjing 210093, China}
\affiliation{Key Laboratory of Modern Astronomy and Astrophysics (Nanjing University), Ministry of Education, China}

\begin{abstract}
The motion of neutron stars (NSs) in the Galaxy is largely dependent on natal kicks received by the NSs during supernova explosions. Thus, the measured peculiar velocities of NS high-mass X-ray binaries (HMXBs) provide valuable clues to natal kicks, which also play an important role in the evolution of HMXBs. In this work, we collect proper motions, radial velocities and parallaxes for 36 NS HMXBs to derive their peculiar velocities at the birth of the NSs. We then use binary population synthesis to simulate the velocities of NS HMXBs with various choices of the kick velocity distribution for both core-collapse and electron-capture supernovae. Comparing the simulated and measured velocities, orbital periods, and eccentricities, we show that the natal kick distribution that can best match the observations is characterized by a bimodal Maxwellian distribution with $\sigma_1$ = 320 km\,s$^{-1}$ (for core-collapse supernovae) and $\sigma_2$ = 80 km\,s$^{-1}$ (for electron-capture supernovae) and the He core mass for the latter in the range of $(1.83-2.25)$ $M_{\odot}$.  Our findings provide useful insights for further population synthesis and binary evolution studies of NS binaries.
\end{abstract}

\keywords{Neutron stars (1108); High mass x-ray binary stars (733); Supernovae (1668)}

\section{Introduction} \label{sec:intro}
It is generally accepted that a massive star in the final state will explode as a supernova (SN), forming a neutron star (NS) or black hole (BH). During the SN explosion, the remnant NS or BH receives a momentum kick, referred to as natal kick    \citep{1970SvA....13..562S,Lyne1982MNRAS,Brandt1995MNRAS,Jonker2004MNRAS}, which is thought to be related to the asymmetries in the SN ejecta imprinted by the nonradial hydrodynamic instabilities within the stellar core \citep{Janka2012ARNPS, Wongwathanarat2013A&A}. Understanding natal kicks is crucial for elucidating the explosion mechanisms of SNe \citep[ejecta driven or neutrino driven;][]{2001LNP...578..424L,Fryer2006ApJS}. Furthermore, it can provide insights into the birth channels of NSs and BHs, for instance, the core-collapse supernovae (CCSNe) or the electron-capture supernovae \citep[ECSNe;][]{Miyaji1979ICRC, Nomoto1984ApJ, Nomoto1987ApJ} for NSs, and the direct collapse or fallback-assisted formation for BHs \citep{Gourgoulhon1991A&A,Fryer2012ApJ,Reynolds2015MNRAS}.  

Kick velocities can be inferred from the observations of both isolated and binary NSs/BHs.
Proper motions and parallaxes of radio pulsars provide useful probes to their birth velocities. Since it is challenging to measure the realistic motion of a pulsar, 3D velocity distribution of pulsars is usually derived from the observed 1D and 2D velocities assuming an isotropic velocity vector \citep{Lyne1994Natur,Hansen1997, Cordes1998ApJ}. 
However, this method is subject to the acceleration of pulsars in the Galactic potential that varies between their birthplaces and their current locations, as well as other selection effects. By limiting the study to relatively young pulsars, one may reduce the effects of these complications.
\cite{Hobbs2005MNRAS} analyzed the velocity distribution of 233 single pulsars and derived a Maxwellian distribution with dispersion $\sigma = 265$ km\,s$^{-1}$. \cite{Verbunt2017A&A} focused on pulsars with accurate very long baseline interferometry measurements and showed that a bimodal Maxwellian distribution with $\sigma_1 = 316$ km\,s$^{-1}$ and $\sigma_2 = 75$ km\,s$^{-1}$ provided a better fit. A bimodal kick velocity distribution for young pulsars was also reached by other studies \citep{2002ApJ...568..289A,2003AJ....126.3090B,2020MNRAS.494.3663I}, which assumed that the observed velocities of young pulsars could be regarded as their kick velocities. Since most massive stars are likely in binary systems, the orbital motion before the SN explosion may have an imprint on the pulsar's motion. Thus, inferring natal kick information from the observed velocities of radio pulsars may be subject to considerable uncertainties.

Another way to constrain NS's natal kick is measuring the velocities of X-ray binaries (XRBs) with luminous optical companion stars \citep{1999MNRAS.310.1165T,Mirabel2001Natur,2017MNRAS.467..298R, Atri2019MNRAS,Fortin2022A&A, 2023MNRAS.521.2504O,Zhaoyue2023MNRAS}.  In particular, high-mass X-ray binaries (HMXBs) present a potentially ideal population for studying natal kicks. Unlike low-mass X-ray binaries (LMXBs), which are older and more concentrated toward the Galactic center \citep{van_Paradijs1995ApJ}, HMXBs are a younger stellar population, distributed along the Galactic plane and following the spiral arm structures \citep{van_Paradijs1998ASIC, Grimm2002A&A, Coleiro2013ApJ}. The NSs/BHs in HMXBs are typically less than 20 Myr old \citep{Tauris2023book}. The relatively young ages of HMXBs make them a preferable choice for constraining natal kicks, as their evolution and the Galactic gravitational potential have less impact on peculiar velocities compared to LMXBs. In addition, LMXBs are subject to the selection bias that they are not sensitive to large kicks, which would have unbound the binaries \citep{2023MNRAS.521.2504O}. Another advantage of HMXBs over other types of NS binary systems is that the mass transfer between the progenitor of the NS and the companion star is usually dynamically stable, meaning that NS HMXBs generally have not experienced common envelope (CE) evolution before the SN explosions, which can dramatically change the binary orbit but is poorly understood \citep{2013A&ARv..21...59I}. For the study of HMXBs, \cite{2021MNRAS.508.3345I} compiled a sample of 45 Be XRBs with proper motions and parallaxes from $Gaia$ Early Data Releases 3 ($Gaia$ EDR3). Combining the analyses of Be XRBs and young isolated radio pulsars, they found that NS natal kicks can be best described by a bimodal Maxwellian distribution, with low ($\sigma_1=45^{+25}_{-15}$ kms$^{-1}$)- and high ($\sigma_2=336$ km\,s$^{-1}$)-velocity components. 
\cite{Fortin2022A&A} used the  data of 35 NS HMXBs  to derive their peculiar velocities using their positions (3D) and proper motions (2D) from the {\em Gaia} data. By retrieving their systemic radial velocity (1D) from the literature (when available), these authors found that the distribution of the inferred kick velocity is better represented by gamma distribution than by Maxwellian distribution. Notably, radial velocities for about half of the sources in their study were not available. 
\cite{Zhaoyue2023MNRAS} compiled a catalog of HMXBs and LMXBs with astrometry and radial velocity measurements and calculated their present-day peculiar velocities. By extrapolating peculiar velocities at the Galactic disk, which were treated as their potential peculiar velocities at birth, they found a bimodal Maxwellian distribution of kick velocity (with $\sigma_1\simeq 21$ km\,s$^{-1}$ and $\sigma_2\simeq 107$ km\,s$^{-1}$) for the whole sample.

While the abovementioned investigations provided phenomenological functional forms of natal kicks, it is necessary to reconstruct the velocity distributions based on the physical mechanisms of SN explosions. Theoretically, the high- and low-kick components are often associated with CCSNe and ECSNe, respectively, motivated in part by hydrodynamical simulations \citep{Gessner2018ApJ,2020LRCA....6....3M}. However, there are few population studies that systematically compare modeled velocity distributions with observationally derived velocity samples. In this work, we use the binary population synthesis (BPS) method to investigate the kick velocity distribution for NSs in HMXBs. Our goal is to examine how different assumptions on NS formation manifest themselves in the distribution of peculiar velocities. Our method is similar to \cite{2017MNRAS.467..298R}, but our study is limited to HMXBs, because the nearby kicked objects that are older than tens of Myr will obtain the Galactocentric speeds that are not representative of their kicks, due to their motion within the Galactic gravitational potential \citep{2024A&A...687A.272D}. Besides the 25 HMXBs sample from \cite{Zhaoyue2023MNRAS}, we collect the data for 11 additional sources  from the latest HMXBs catalog \citep{Neumann2023A&A, Fortin2024A&A}. We also adopt updated recipes of binary evolution in the BPS code.

The rest of the paper is organized as follows. In Section \ref{sec:sample}, we describe the sample we collect and the method to calculate current peculiar velocity and peculiar velocity at the NS's birth. In Section \ref{sec:method}, we introduce the BPS method and the adopted assumptions for binary evolution. In Section \ref{sec:result}, we present the simulated data and the statistical method employed for comparing the simulated results with the observed data. Our summary and discussion are in Section \ref{sec:conclusion}.

\begin{table*}[]
\begin{center}
\renewcommand{\arraystretch}{1.5}
\begin{threeparttable}
\caption{The Parameters of (Possible) NS HMXBs with Parallaxes and Proper Motions.}
\label{tab:table1}
\begin{tabular*}{\hsize}{@{}@{\extracolsep{\fill}}llcccccc@{}}
\toprule
Name & Class    & \textit{Gaia} DR3 & $\varpi$ & $\mu_\alpha\cos\delta$ & $\mu_\delta$ \\
&          &          & (mas) & ($\masyr$) & ($\masyr$) \\
\midrule
gam Cas \tnote{a} & ??-HMXB & 426558460884582016 & $5.94\pm 0.12$ & $25.17\pm 0.08$ & $-3.92\pm 0.08$\\
IGR J11305-6256 & ??-HMXB & 5333660129603575808 & $0.57\pm 0.05$ & $-6.10\pm 0.05$ & $1.37\pm 0.05$ \\
HD 119682 & ??-HMXB & 5864664975280537728 & $0.61\pm 0.03$ & $-4.75\pm 0.02$ & $-2.26\pm 0.02$ \\
IGR J18406-0539 & ??-HMXB &  4253473896214500096 & $0.23\pm 0.01$ & $-1.03\pm 0.02$ & $-2.94\pm 0.02$ \\
IGR J20155+3827 & ??-HMXB & 2060941573135965824 & 	$0.08\pm 0.02$ & $-2.18\pm 0.02$ & $-3.77\pm 0.02$ \\
SGR 0755-2933 & NS-HMXB & 5597252305589385984 & $0.29\pm 0.01$ & $-2.65\pm 0.01$ & $2.89\pm 0.01$ \\
2MASS J08504008-4211514 & NS-HMXB & 5524532148408036096 & 	$0.08\pm 0.01$ & $-3.53\pm 0.02$ & $4.22\pm 0.02$ \\
IGR J16207-5129 & NS-HMXB & 5934776158877214848 & $-0.01\pm 0.05$ \tnote{b} & $-4.88\pm 0.05$ & $-6.02\pm 0.04$ \\
4U 1954+31 & NS-HMXB & 2034031438383765760 & 	$0.26\pm 0.02$ & $-2.16\pm 0.02$ & $-6.07\pm 0.03$ \\
Swift J0243.6+6124 & NS-HMXB & 465628193526364416 & $0.18\pm 0.01$ & $-0.73\pm 0.01$ & $0.13\pm 0.01$ \\
IGR J06074+2205 & NS-HMXB & 3423526544838563328 & $0.14\pm 0.02$ & $0.57\pm 0.02$ & $-0.61\pm 0.01$ \\
IGR J11215-5952 & NS-HMXB & 5339047221168787712 & $0.12\pm 0.01$ & $-5.15\pm 0.01$ & 	$2.73\pm 0.01$ \\
IGR J18027-2016 & NS-HMXB & 4070968778561141760 & $-0.15\pm 0.13$ \tnote{b} & $-3.41\pm 0.12$ & 	$-5.49\pm 0.09$ \\
IGR J21343+4738 & NS-HMXB & 1978365123143522176 & 	$0.08\pm 0.01$ & $-2.21\pm 0.01$ & $-2.56\pm 0.01$ \\
EXMS B1210-645 & NS-HMXB & 6053076566300433920 & $0.26\pm 0.02$ & $-5.95\pm 0.02$ & $0.45\pm 0.02$ \\
Cir X-1 & NS-HMXB & 5883218164517055488 & $-0.26\pm 0.14$ \tnote{b} & $-5.55\pm 0.13$ & $-4.07\pm 0.16$ \\

\toprule	
\end{tabular*}
\begin{tablenotes}
\item[a] The parallax and proper motion of this source were not recorded by \gaia, so we took these values \citep{1953GCRV..C......0W, 2007A&A...474..653V} from the \href{https://simbad.cds.unistra.fr/simbad/sim-basic?Ident=gam+Cas&submit=SIMBAD+search}{SIMBAD} database.
\item[b] The parallax was poorly constrained by $Gaia$, indicated by a negative value \citep{Luri2018A&A}. 
\end{tablenotes}
\end{threeparttable}
\end{center}
\end{table*}

\begin{table*}[]
\begin{center}
\renewcommand{\arraystretch}{1.5}
\begin{threeparttable}
\caption{Measured and Derived Parameters for 16 (Possible) NS HMXBs.}
\label{tab:table2}
\begin{tabular*}{\hsize}{@{}@{\extracolsep{\fill}}lclllcc@{}}
\hline
\toprule
Name & $\gamma$ & $d$ \tnote{a}   & $\vpecp$ \tnote{a} &  $\pkv$ \tnote{a} & $P_{\rm orb}$ & $e$ \\
& (km\,s$^{-1}$) & (kpc) & (km\,s$^{-1}$) & (km\,s$^{-1}$) & (day) &\\
\midrule
gam Cas & $-$0.018$\pm$0.075 [1] & $0.17_{-0.00}^{+0.00}$ &  $13.64_{-2.23}^{+3.33}$ & $12.72_{-2.56}^{+4.99}$ & 203.59 [2] & 0.26 [2]\\
IGR J11305-6256 & $-$17.94$\pm$4.48 [3] & $1.80_{-0.15}^{+0.18}$ & $16.60_{-5.60}^{+6.25}$ & $18.82_{-7.79}^{+9.31}$ & 120.83 [4] & ... \\
HD 119682 & $-$18.73$\pm$0.09 [5] & $1.65_{-0.08}^{+0.09}$ & $12.50_{-3.26}^{+4.23}$ & $12.16_{-3.83}^{+5.69}$ & 90.0 [6] & ... \\ 
IGR J18406-0539 & 2.28$\pm$28.67 [3] & $4.36_{-0.18}^{+0.20}$ & $63.33_{-6.51}^{+6.74}$ & $67.31_{-30.26}^{+30.12}$ & ... & ... \\
IGR J20155+3827 & $-$105.98$\pm$2.94 [3] & $11.21_{-1.81}^{+2.45}$ & $31.05_{-13.87}^{+24.31}$ & $39.13_{-18.48}^{+35.02}$ & ... & ...\\
SGR 0755-2933 & 58.62$\pm$0.11 [7] & $3.46_{-0.11}^{+0.12}$ & $8.35_{-2.81}^{+4.91}$ & $8.61_{-3.93}^{+6.96}$ & 59.69 [7] & 0.06 [7]\\
2MASS J08504008-4211514 & 40.85$\pm$1.43 [3] & $12.09_{-1.24}^{+1.53}$ & $175.66_{-32.12}^{+33.04}$ & $194.67_{-26.54}^{+20.41}$ & ... & ...\\
IGR J16207-5129 &  38.79$\pm$4.94 [3] & $6.58_{-2.79}^{+4.02}$ \tnote{b} & $152.29_{-52.95}^{+22.29}$ & $176.93_{-69.81}^{+32.09}$ & 9.726 [8] & ... \\
4U 1954+31 & 	4.74$\pm$0.51 [3] & $3.89_{-0.28}^{+0.33}$ & $19.24_{-3.15}^{+3.72}$ & $18.91_{-3.58}^{+5.13}$ & 1296.64 [9] & ...\\
Swift J0243.6+6124 & 325.71 [10] & $5.57_{-0.29}^{+0.33}$ & $393.76_{-5.96}^{+5.99}$
 & $312.48_{-13.16}^{+12.20}$ & 28.3 [11] & 0.092 [11]\\
IGR J06074+2205 &  18.9$\pm$4.1 [12] & $7.19_{-0.87}^{+1.13}$ & $20.84_{-2.36}^{+2.63}$ & $17.13_{-3.59}^{+4.50}$ & ... & ... \\
IGR J11215-5952 & $-$0.20$\pm$3.10 [3] & $8.32_{-0.63}^{+0.73}$ & $57.51_{-7.96}^{+8.94}$ & $52.87_{-11.38}^{+12.14}$ & 164.6 [13] & $>$0.8 \tnote{c} [13]\\
IGR J18027-2016 & 51.7$\pm$2.4 [14] & $5.68_{-4.71}^{+6.97}$ \tnote{b} & $89.92_{-23.91}^{+31.25}$ & $96.47_{-35.50}^{+35.05}$ & 4.57 [14]  & $\lesssim$0.2 \tnote{c} [14] \\ 
IGR J21343+4738 & $-$127$\pm$30 [15] & $12.09_{-1.24}^{+1.53}$ & $30.34_{-13.05}^{+18.53}$ & $48.36_{-19.63}^{+26.28}$ & ... & ... \\
EXMS B1210-645 & $-$42.0$\pm$11.0 [16] & $3.89_{-0.28}^{+0.33}$ & $26.54_{-6.88}^{+6.99}$ & $30.91_{-10.43}^{+13.83}$ & 6.7 [17] & 0.0 [16] \\
Cir X-1 & $-$26.0$\pm$3.0 [18] & $9.4_{-1.0}^{+0.8}$ \tnote{b} & $49.39_{-15.64}^{+13.52}$ & $57.62_{-20.56}^{+17.45}$ & 16.68 [18] & 0.45 [18]\\

\toprule
\end{tabular*}
\begin{tablenotes}
\item[] \textbf{References.} [1]\,\cite{2012A&A...537A..59N}; [2]\,\cite{2000A&A...364L..85H}; [3]\,\cite{2023A&A...674A...1G}; [4]\,\cite{2013arXiv1305.3916L}; [5]\,\cite{2021A&A...647A..19T};  [6]\,\cite{2022MNRAS.510.2286N}; [7]\,\cite{2023Natur.614...45R}; [8]\,\cite{2011ATel.3785....1J};  [9]\,\cite{2020ApJ...904..143H}; [10]\,\cite{2020AJ....160..120J}; [11]\,\cite{2018A&A...613A..19D};  [12]\,\cite{2017AJ....153..174C}; [13]\,\cite{2018MNRAS.481.2779S}; [14]\,\cite{2011A&A...532A.124M}; [15]\,\cite{2014A&A...561A.137R}; [16]\,\cite{2024MNRAS.527.5293M}; [17]\,\cite{2014ApJ...793...77C}; [18]\,\cite{2007MNRAS.374..999J}

\item[a] The values of these parameters represent the median percentiles of the PDF, and the lower and upper limits represent the 16th and 84th percentiles, respectively. 
\item[b] The parallax was not well constrained, so we took the information from the literature \citep{Neumann2023A&A, 2015ApJ...806..265H}.
\item[c] The upper and lower limits of the eccentricities for these two sources are used in the comparison of the $P_{\rm orb}-e-v_{\rm pec}^{\rm z=0}$ distribution, as detailed in Section \ref{subsec:porb}. 

\end{tablenotes}
\end{threeparttable}
\end{center}
\end{table*}

\begin{figure*}[t]
\vspace{-0.5cm}
\gridline{\hspace{-1.0cm}\fig{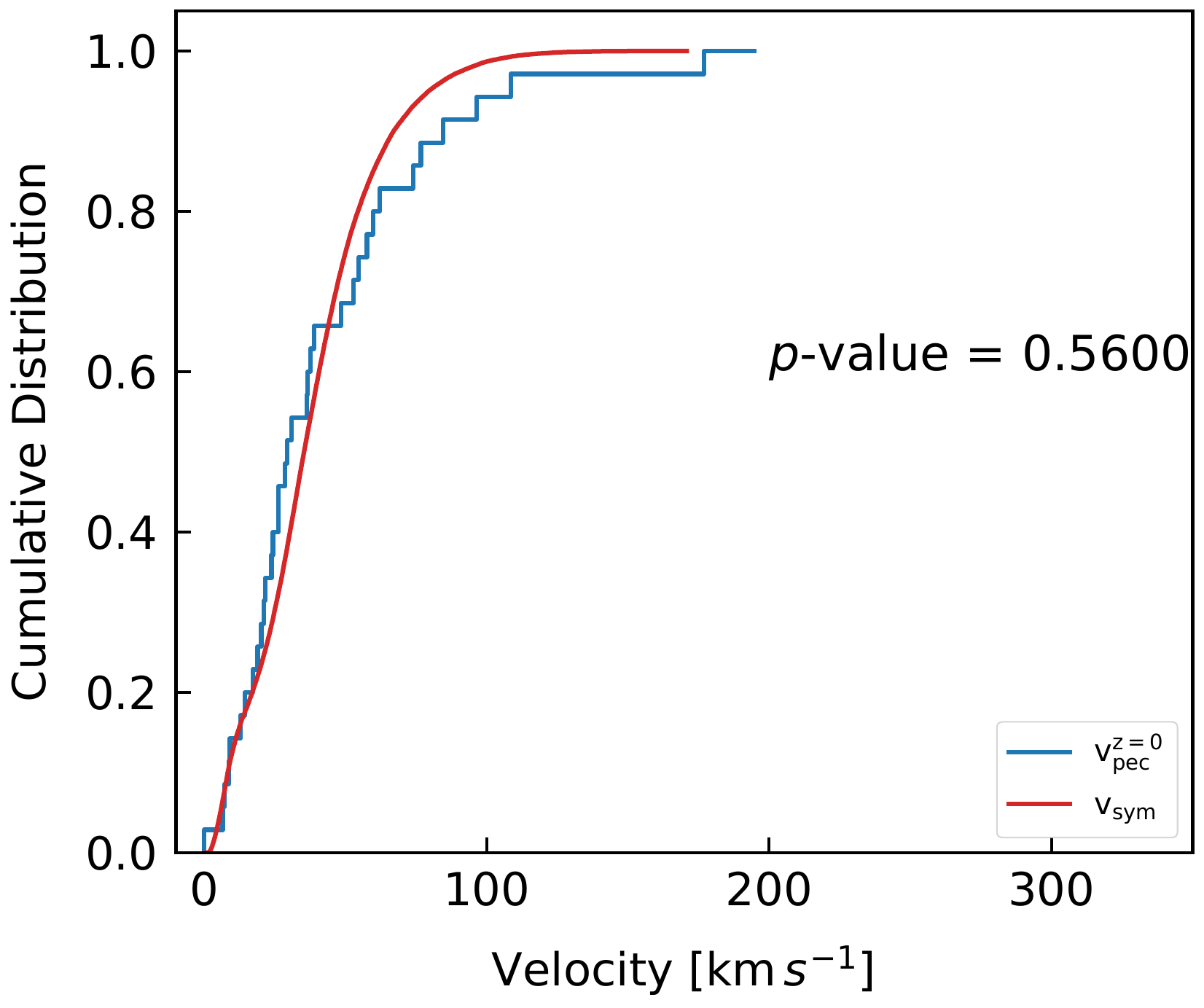}{0.33\textwidth}{$\sigma_1$ = 150 km\,s$^{-1}$, $\sigma_2$ = 30 km\,s$^{-1}$} \hspace{-0.4cm}\fig{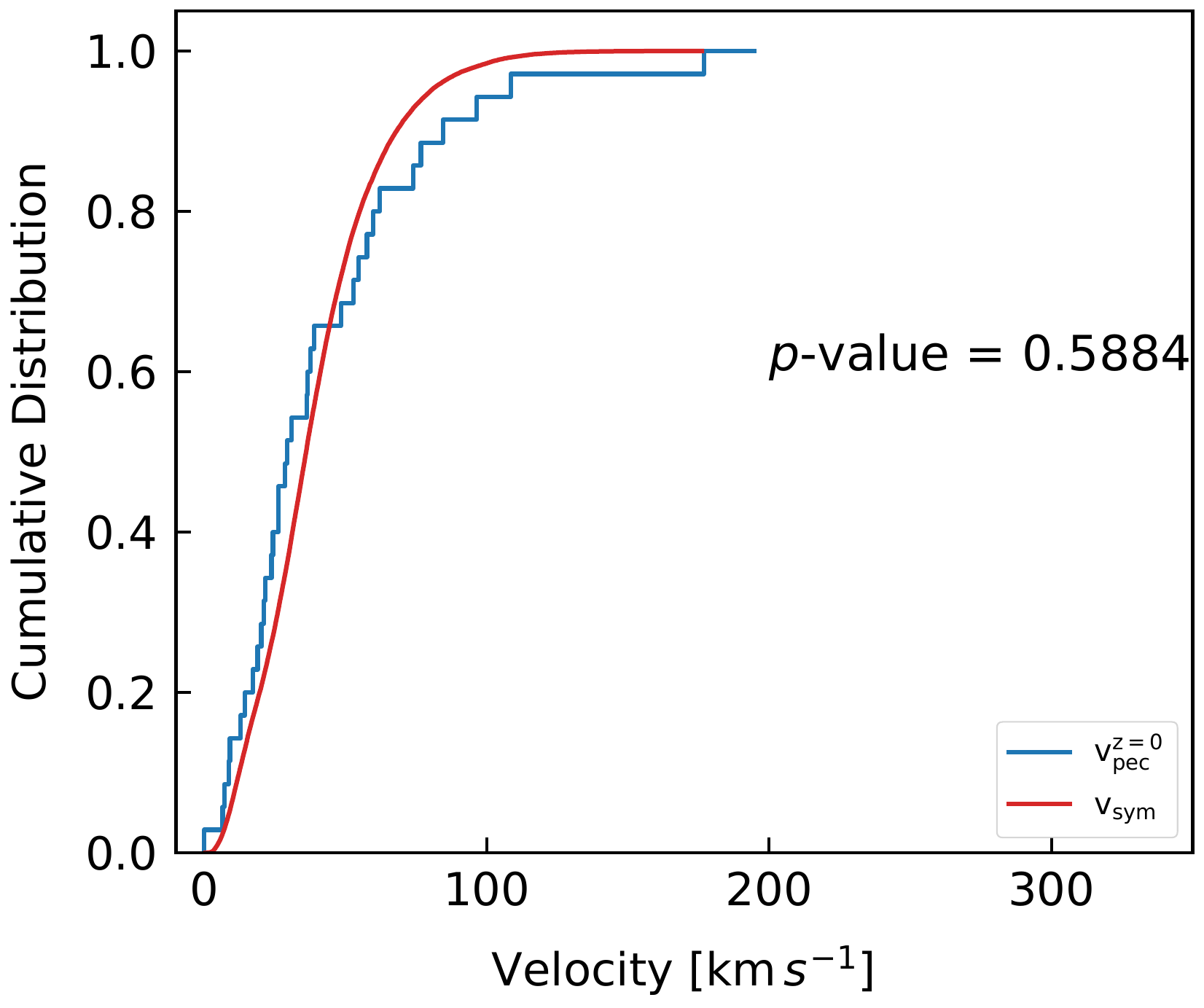}{0.33\textwidth}{$\sigma_1$ = 150 km\,s$^{-1}$, $\sigma_2$ = 50 km\,s$^{-1}$}
 \hspace{-0.4cm}\fig{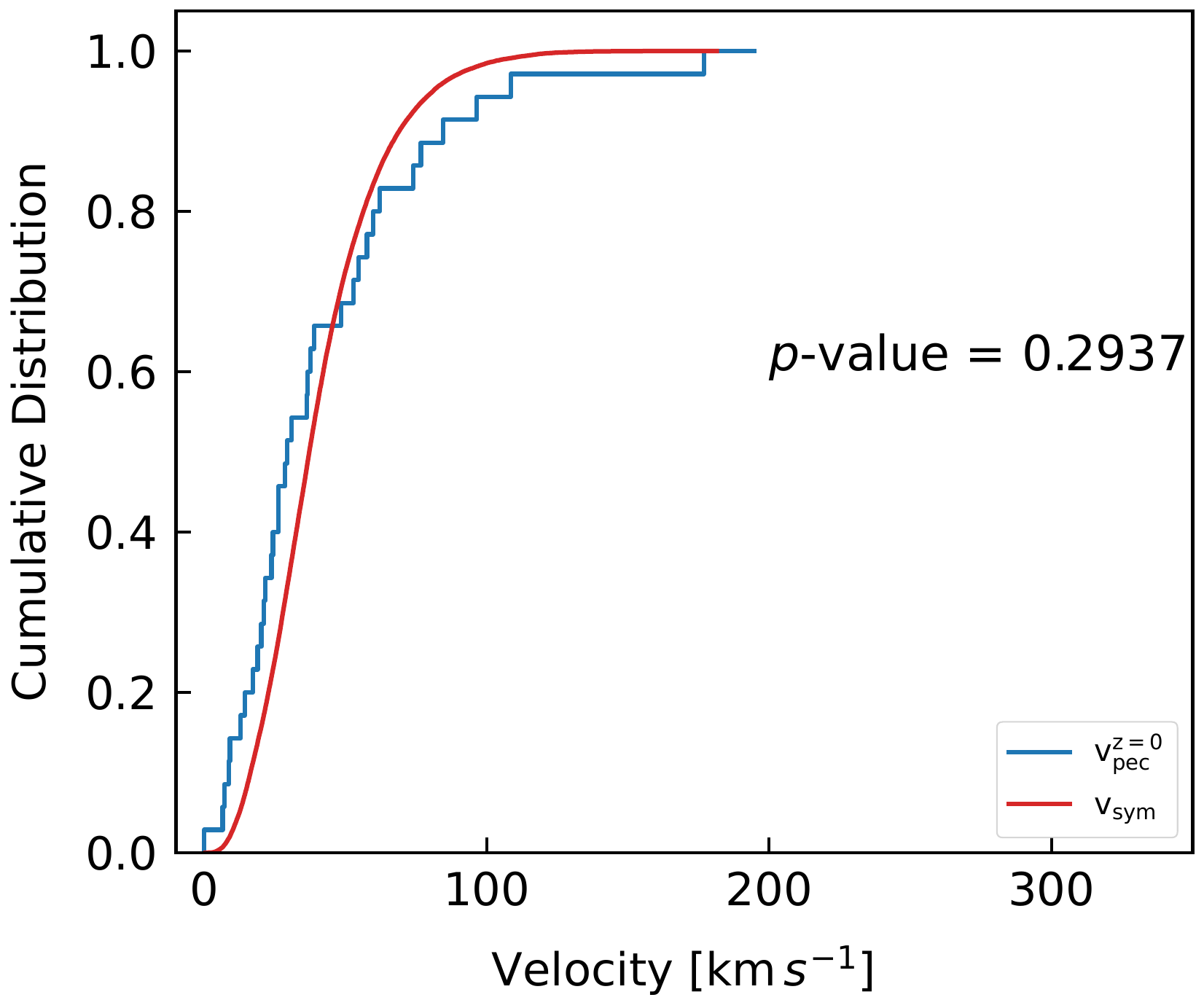}{0.33\textwidth}{$\sigma_1$ = 150 km\,s$^{-1}$, $\sigma_2$ = 80 km\,s$^{-1}$}
 }
\gridline{\hspace{-1.0cm}\fig{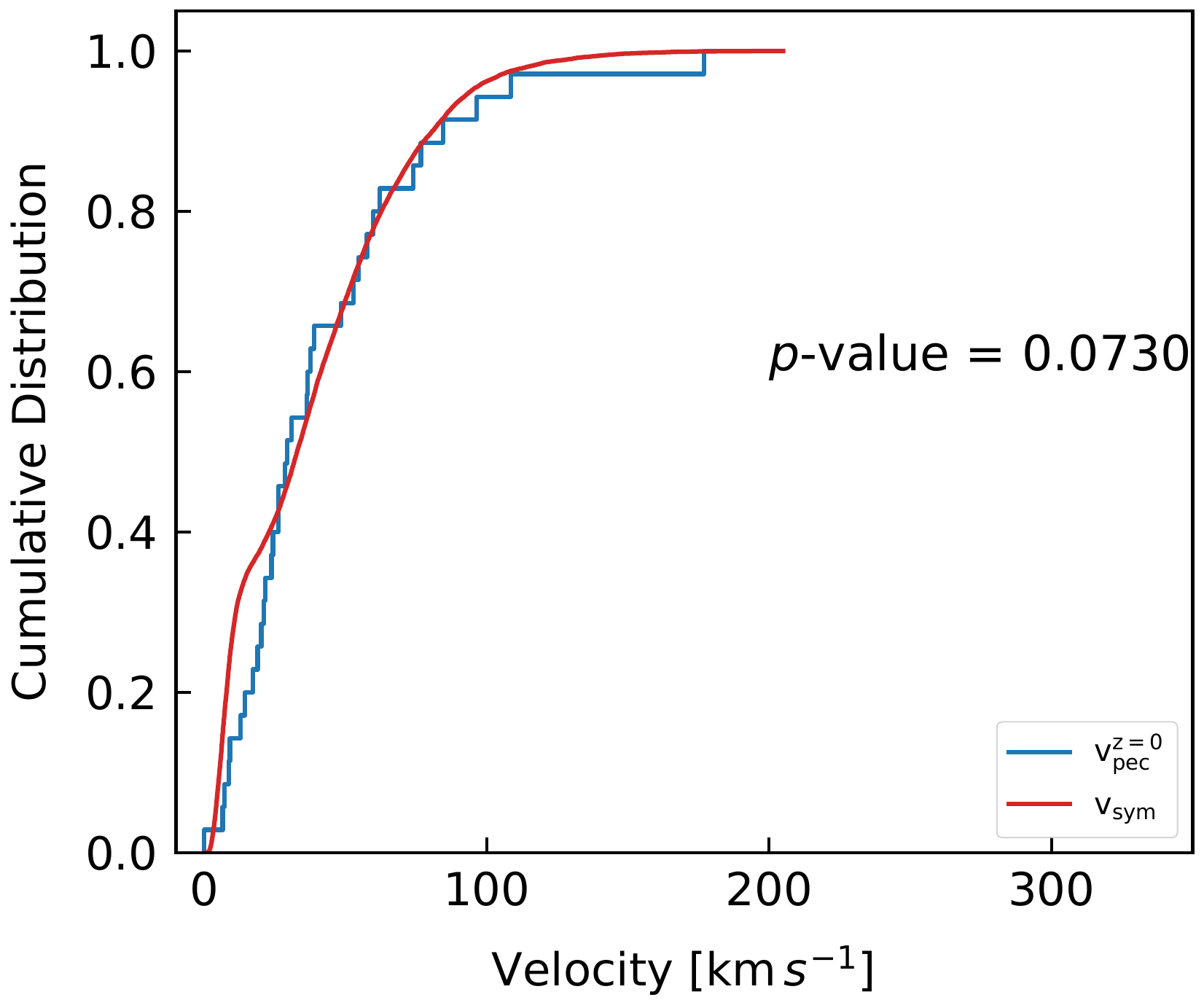}{0.33\textwidth}{$\sigma_1$ = 265 km\,s$^{-1}$, $\sigma_2$ = 30 km\,s$^{-1}$}
\hspace{-0.4cm}\fig{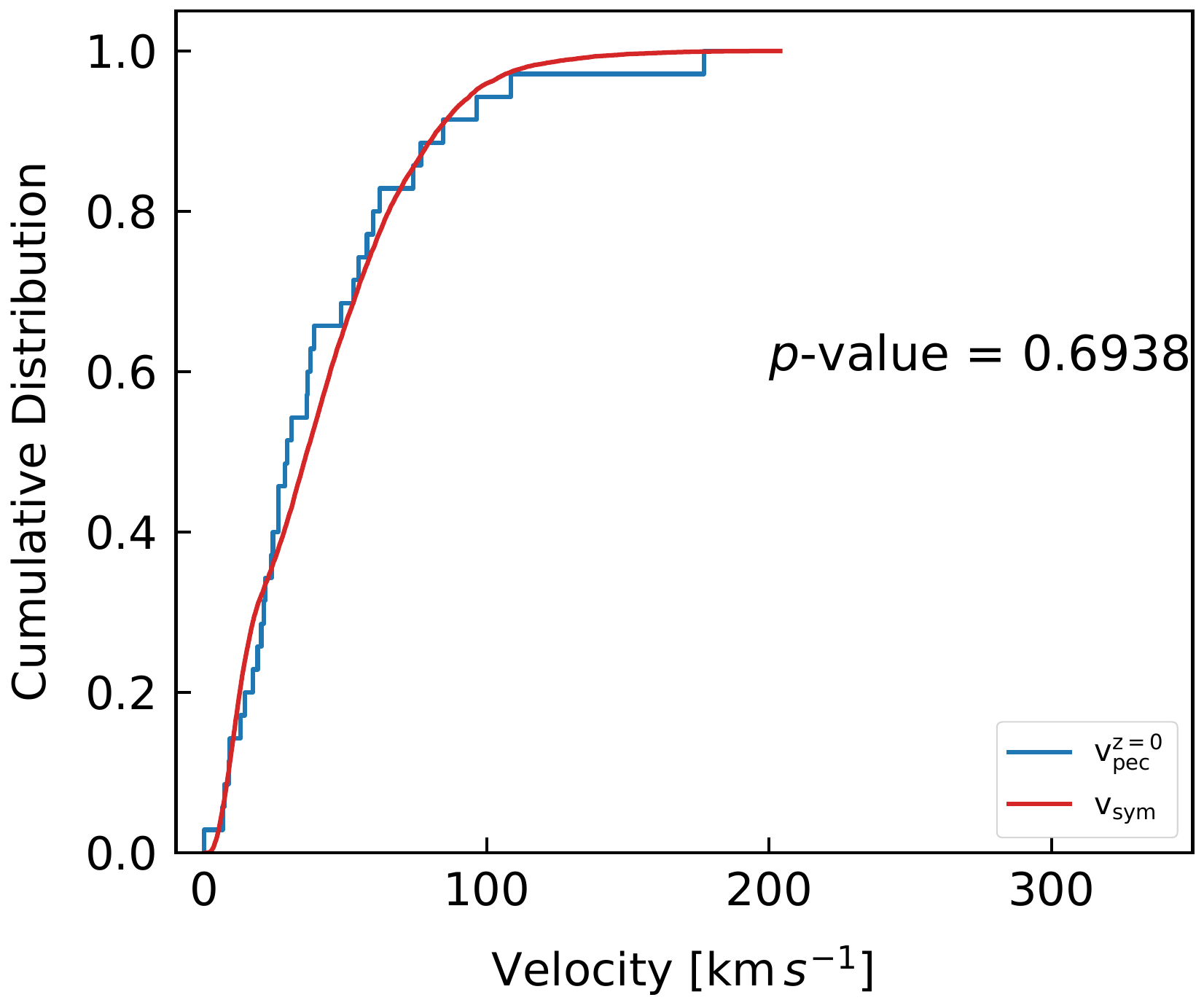}{0.33\textwidth}{$\sigma_1$ = 265 km\,s$^{-1}$, $\sigma_2$ = 50 km\,s$^{-1}$}
 \hspace{-0.4cm}\fig{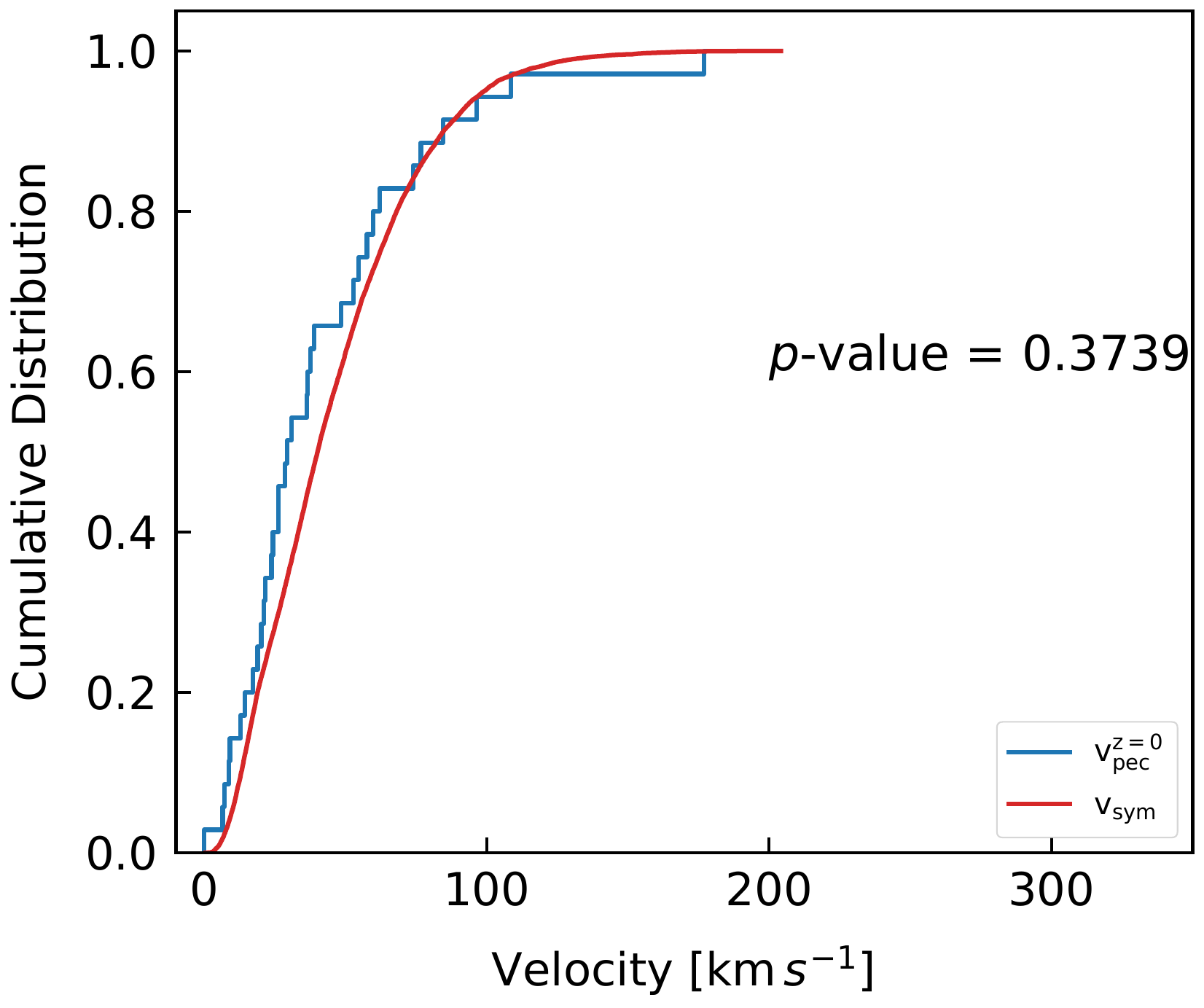}{0.33\textwidth}{$\sigma_1$ = 265 km\,s$^{-1}$, $\sigma_2$ = 80 km\,s$^{-1}$}
 } 
\gridline{\hspace{-1.0cm}\fig{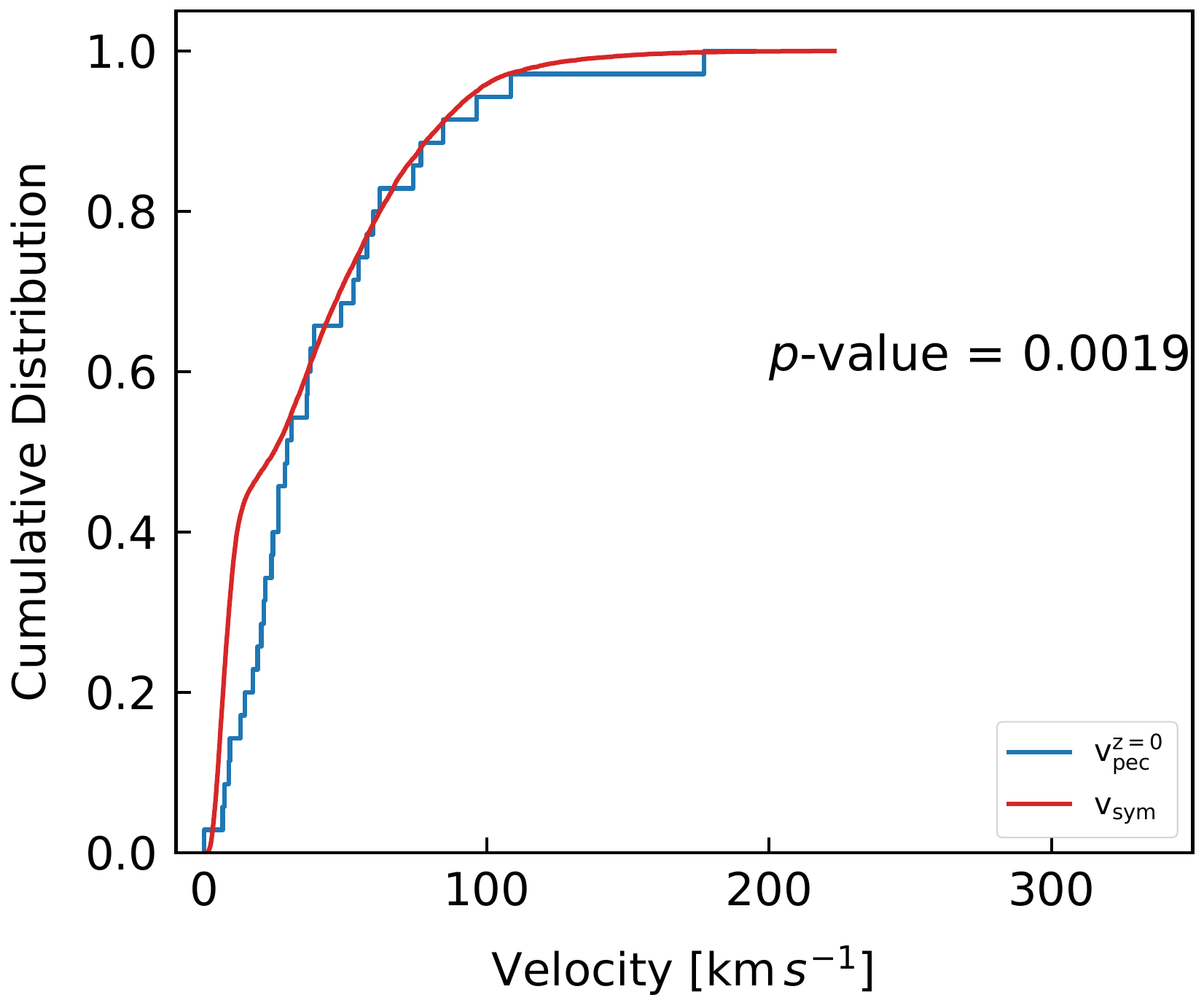}{0.33\textwidth}{$\sigma_1$ = 320 km\,s$^{-1}$, $\sigma_2$ = 30 km\,s$^{-1}$}
\hspace{-0.4cm}\fig{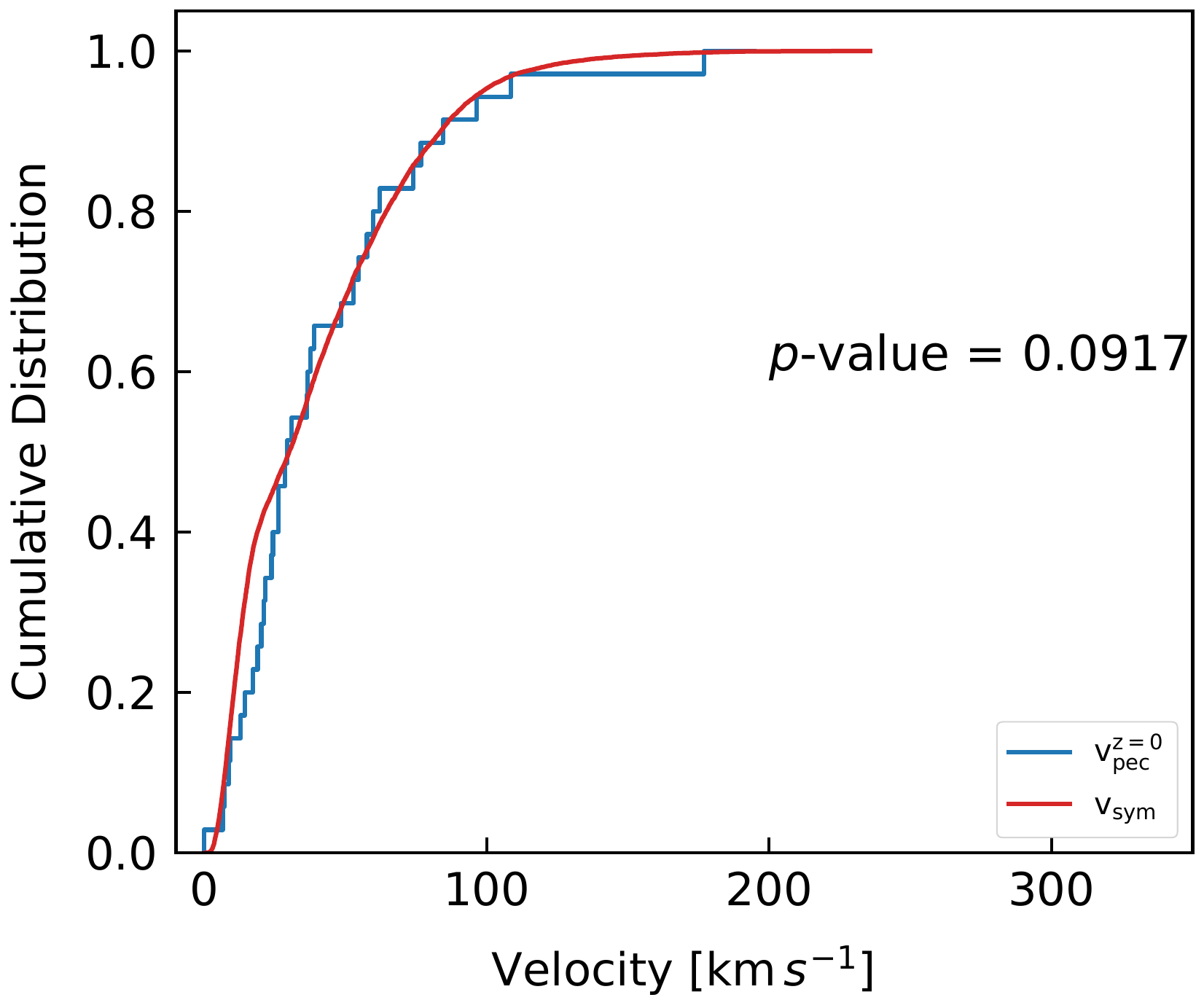}{0.33\textwidth}{$\sigma_1$ = 320 km\,s$^{-1}$, $\sigma_2$ = 50 km\,s$^{-1}$}
 \hspace{-0.4cm}\fig{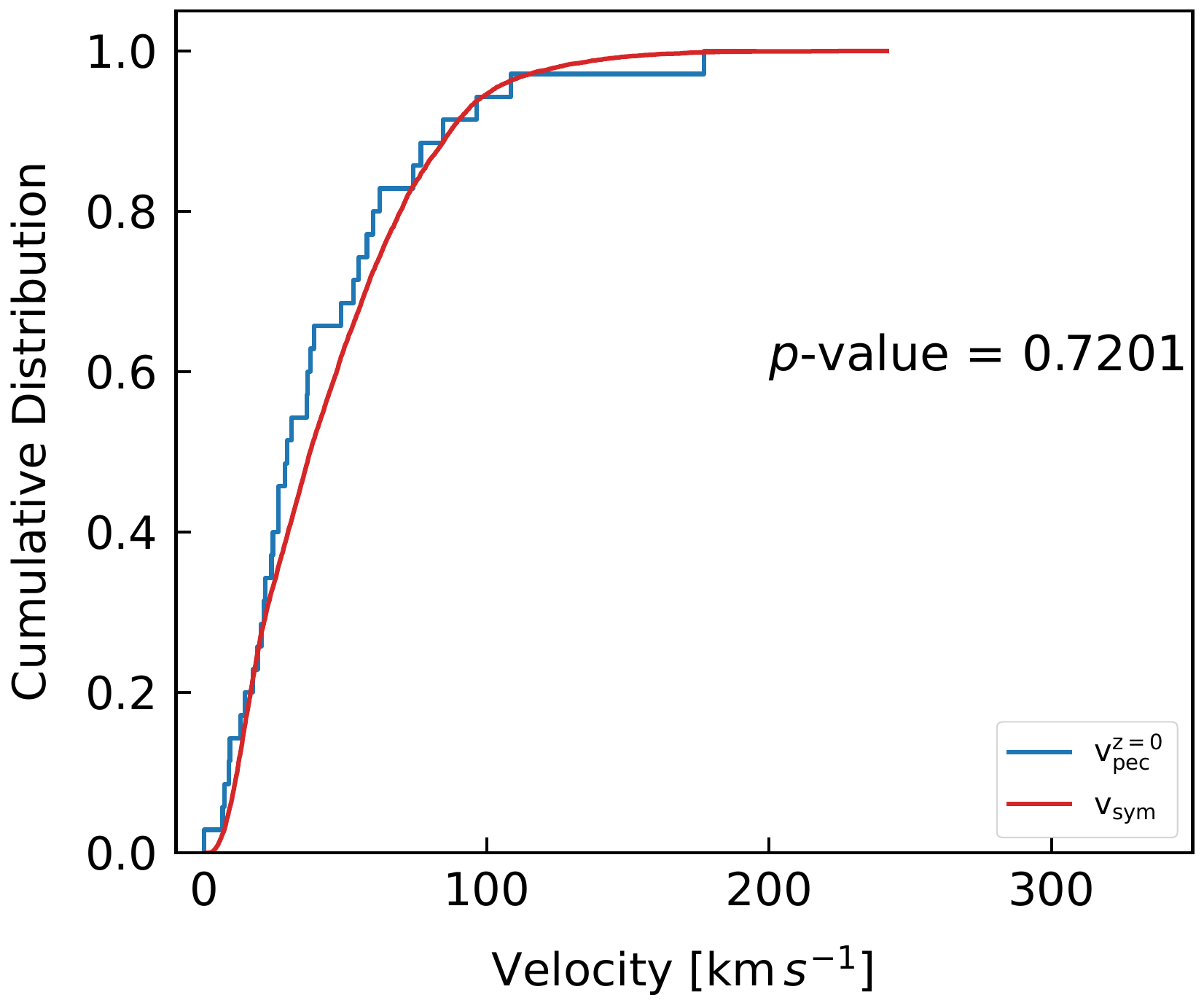}{0.33\textwidth}{$\sigma_1$ = 320 km\,s$^{-1}$, $\sigma_2$ = 80 km\,s$^{-1}$}
 }  
 \gridline{\hspace{-1.0cm}\fig{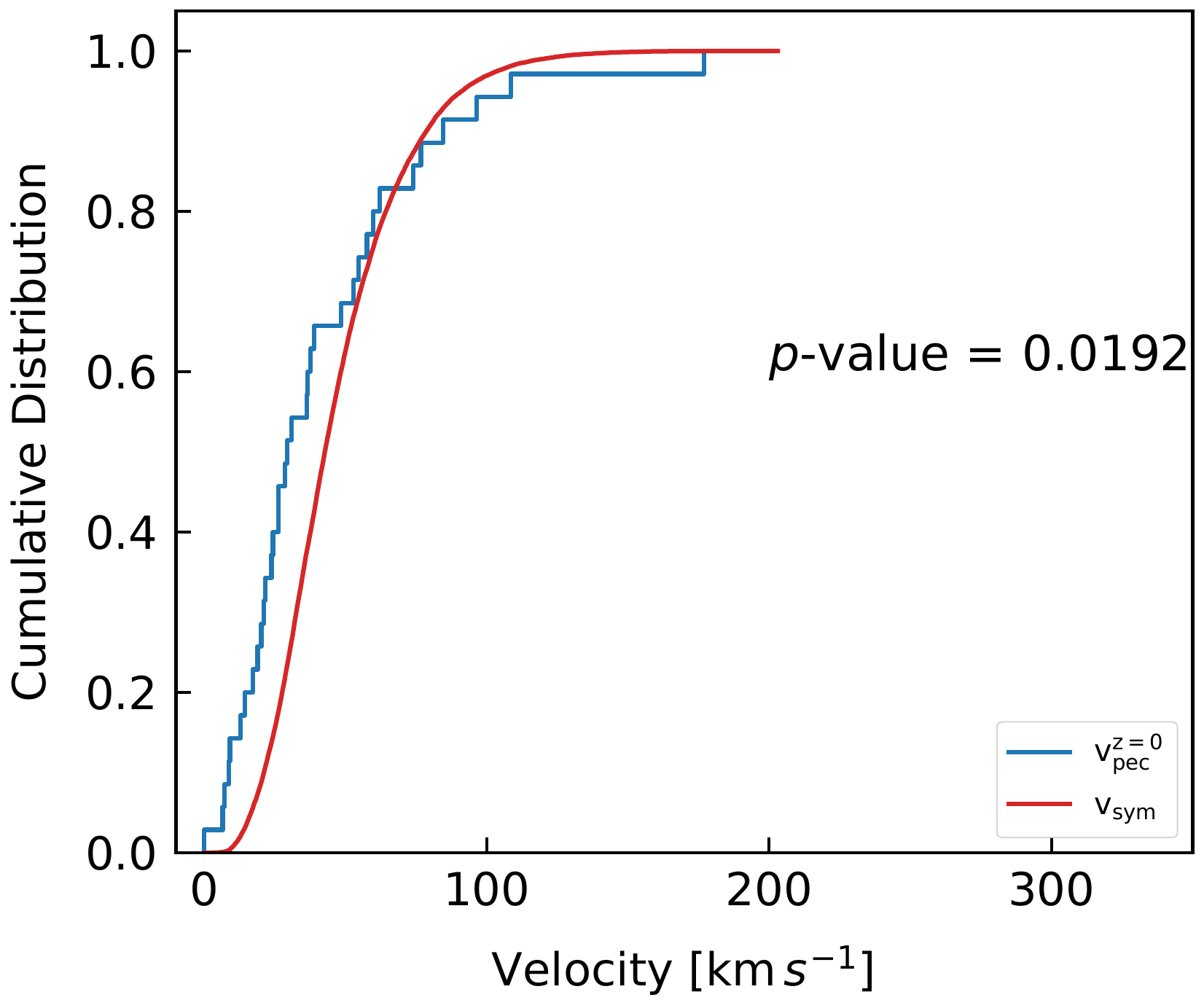}{0.33\textwidth}{$\sigma$ = 190 km\,s$^{-1}$}
\hspace{-0.4cm}\fig{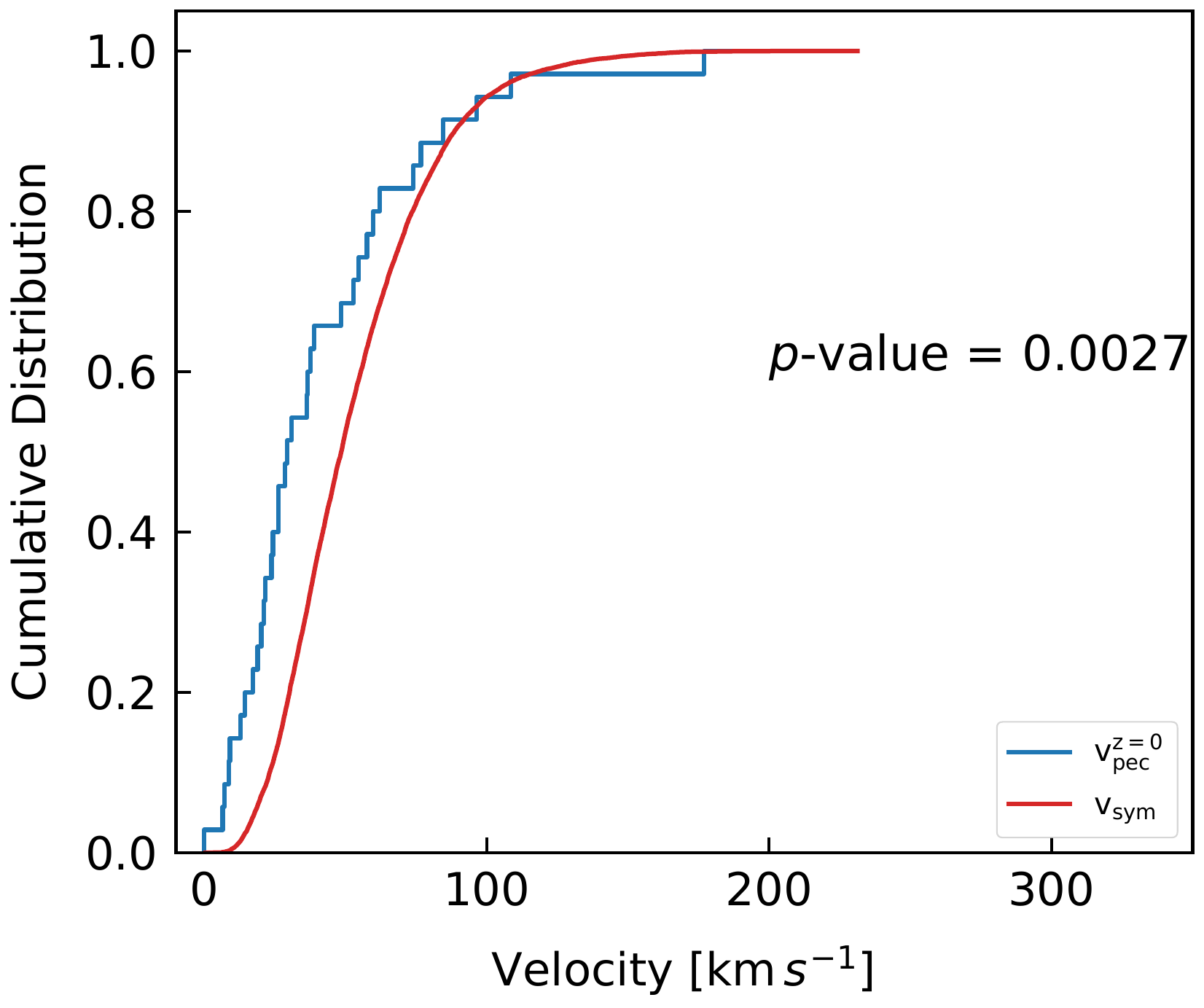}{0.33\textwidth}{$\sigma$ = 265 km\,s$^{-1}$}
 }
\caption{Comparison of the cumulative distributions for the modeled $v_{\rm sym}$ (red line) of NS HMXBs with $M_{\rm ecs}=(1.83-2.25)$ $M_{\odot}$ and derived $v_{\rm sym}$ (blue line) of 36 observed NS HMXBs. The $p$-value is also displayed for each model.}
\label{fig:vk_A}
\end{figure*}

\begin{figure*}[t]
\vspace{-0.5cm}
\gridline{\hspace{-1.0cm}\fig{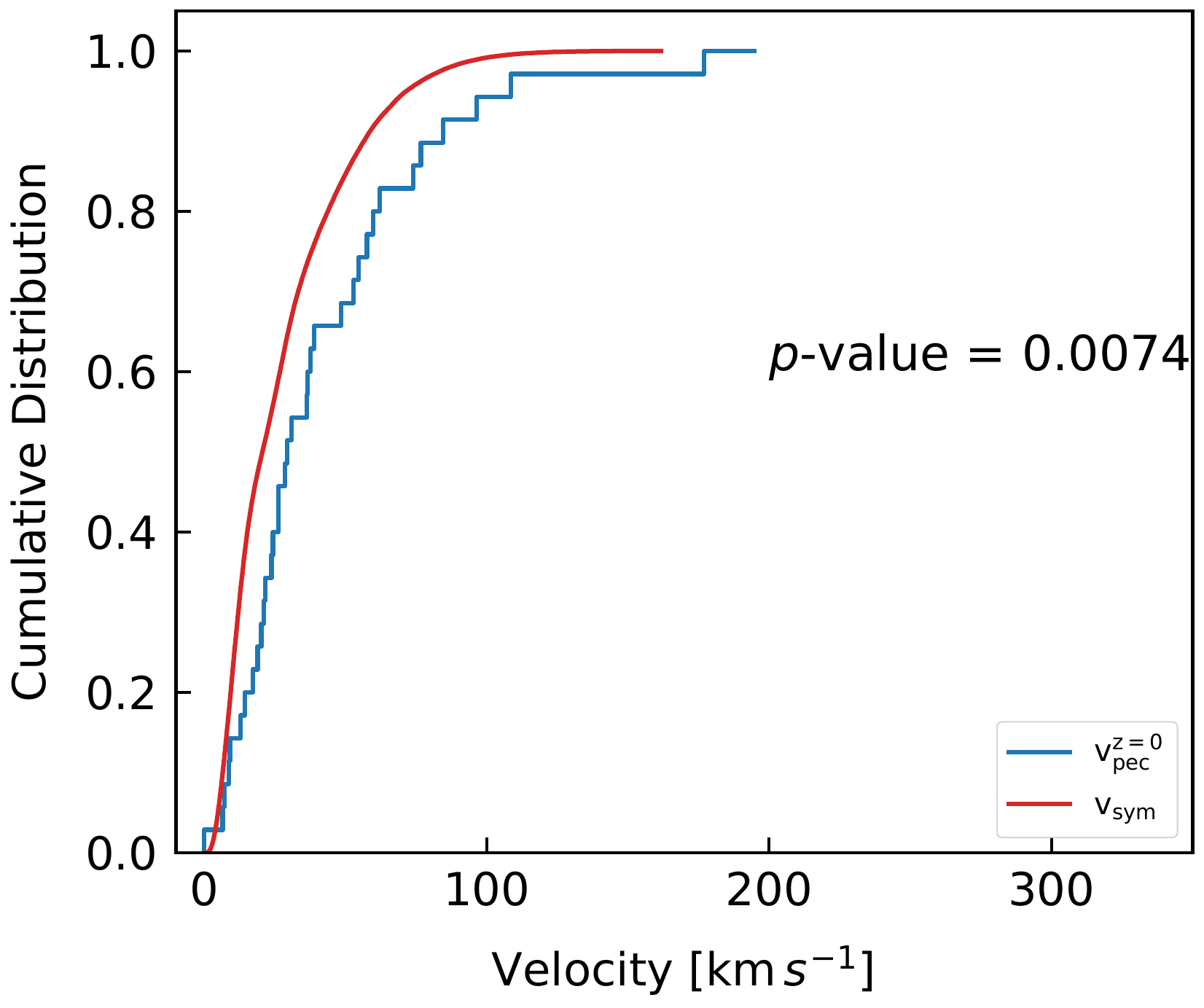}{0.33\textwidth}{$\sigma_1$ = 150 km\,s$^{-1}$, $\sigma_2$ = 30 km\,s$^{-1}$} \hspace{-0.4cm}\fig{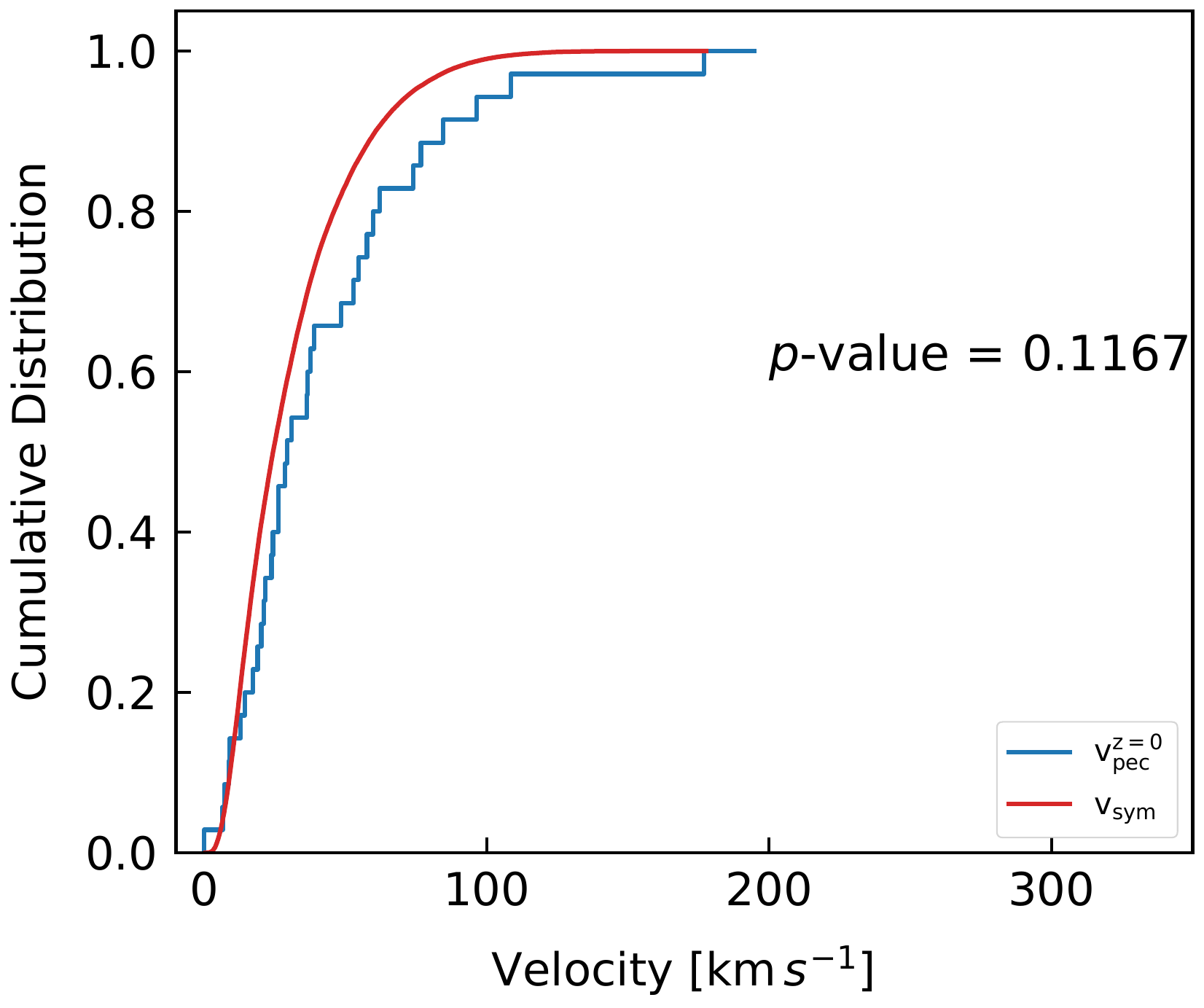}{0.33\textwidth}{$\sigma_1$ = 150 km\,s$^{-1}$, $\sigma_2$ = 50 km\,s$^{-1}$}
 \hspace{-0.4cm}\fig{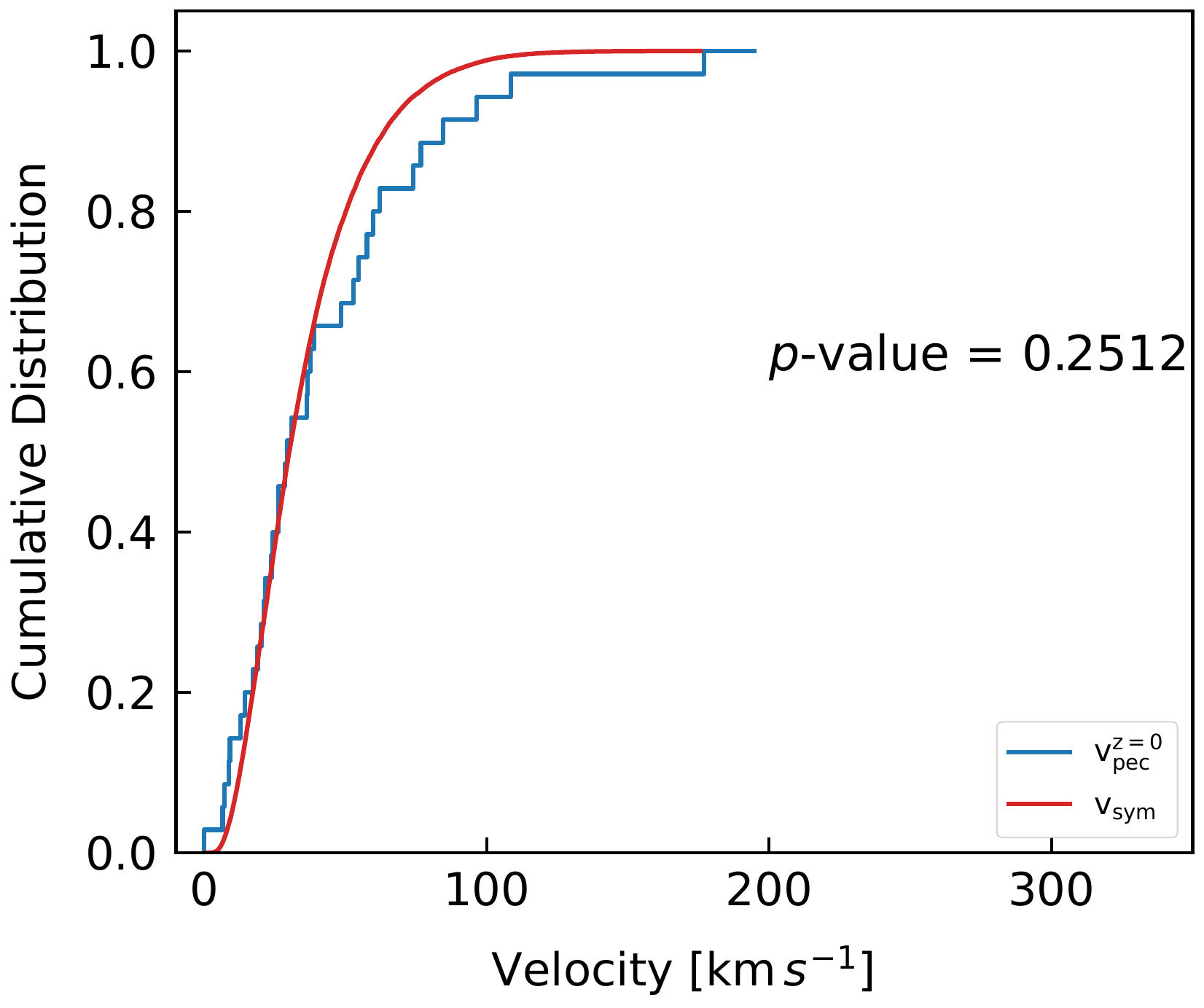}{0.33\textwidth}{$\sigma_1$ = 150 km\,s$^{-1}$, $\sigma_2$ = 80 km\,s$^{-1}$}
 }
\gridline{\hspace{-1.0cm}\fig{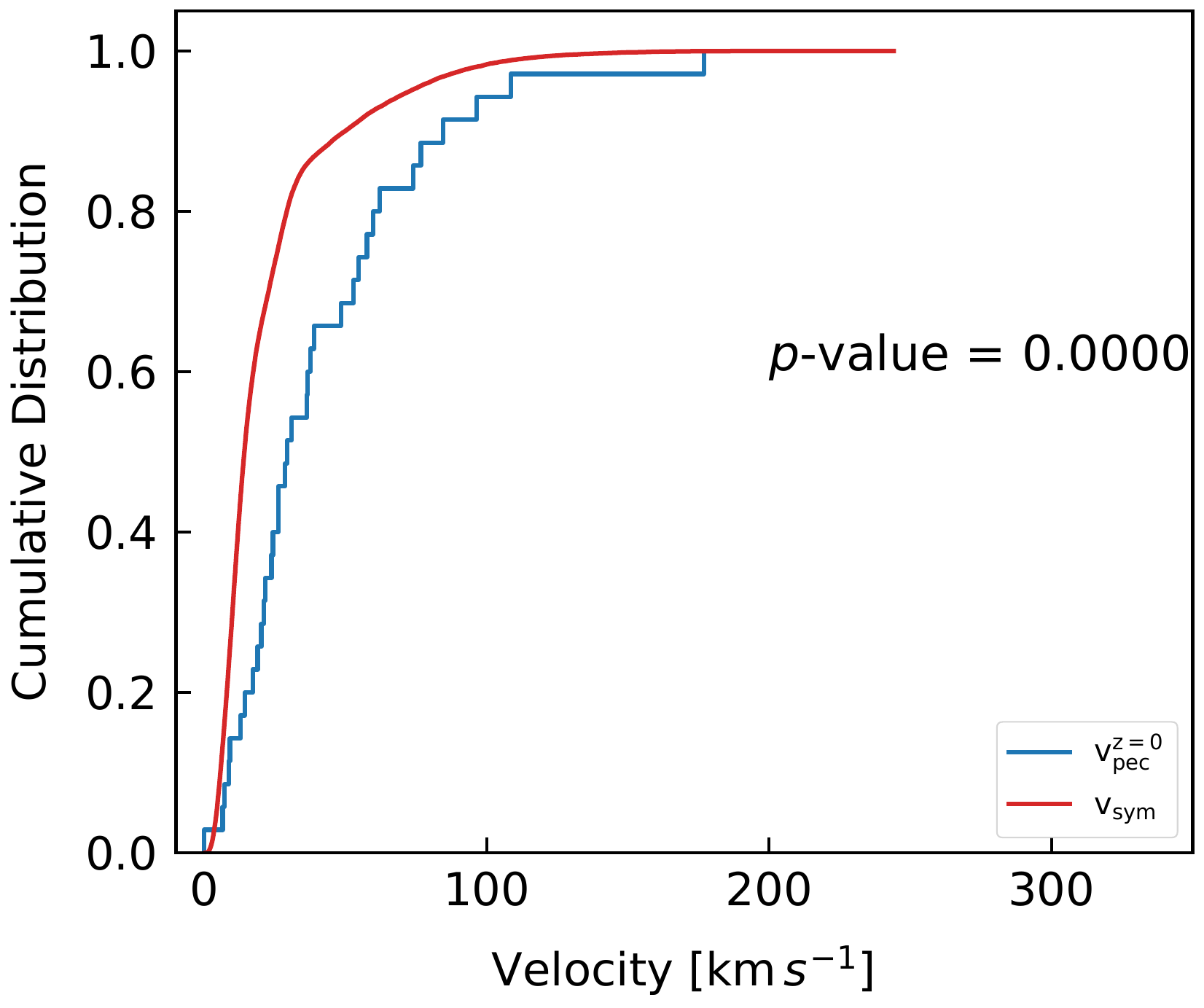}{0.33\textwidth}{$\sigma_1$ = 265 km\,s$^{-1}$, $\sigma_2$ = 30 km\,s$^{-1}$}
\hspace{-0.4cm}\fig{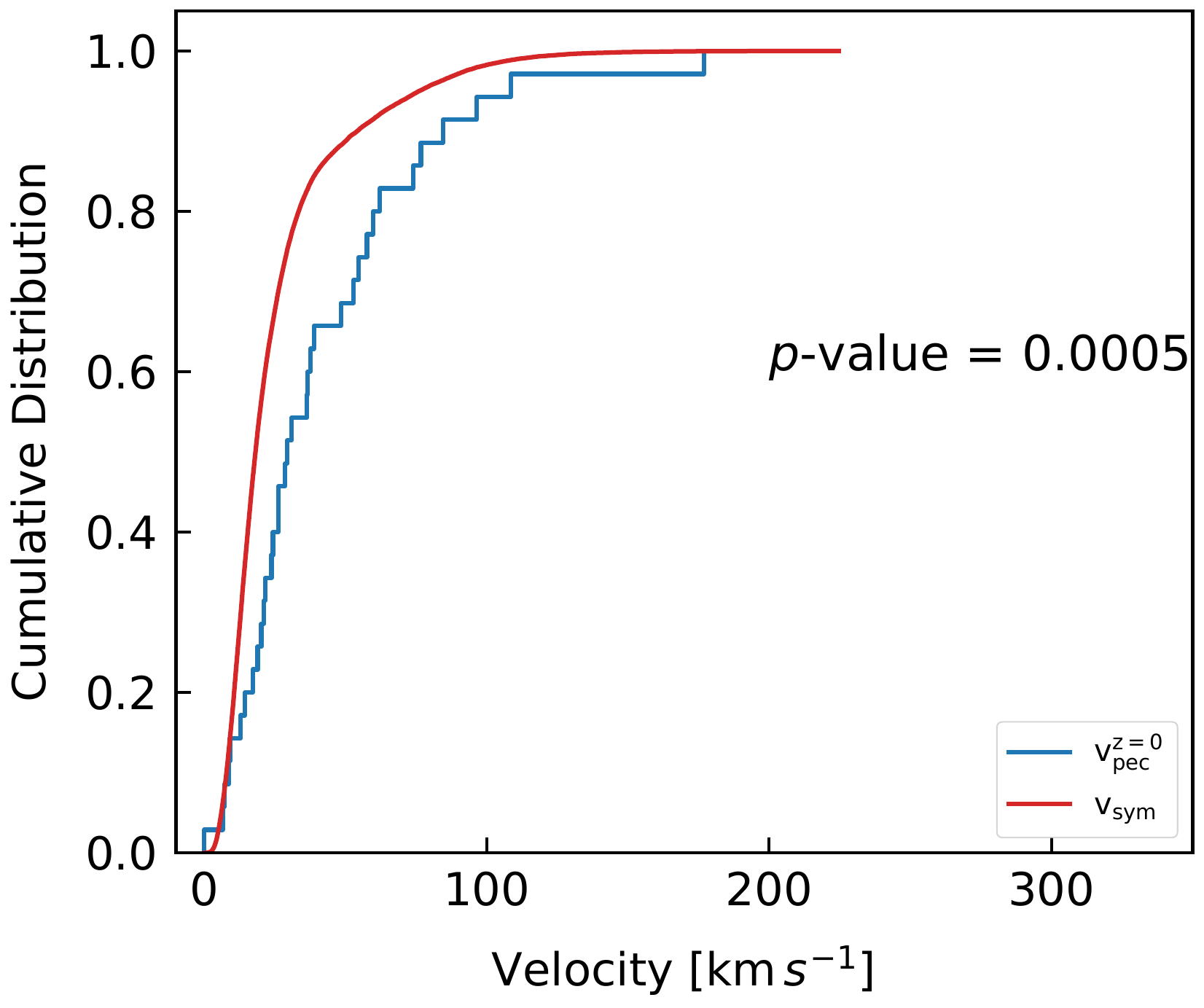}{0.33\textwidth}{$\sigma_1$ = 265 km\,s$^{-1}$, $\sigma_2$ = 50 km\,s$^{-1}$}
 \hspace{-0.4cm}\fig{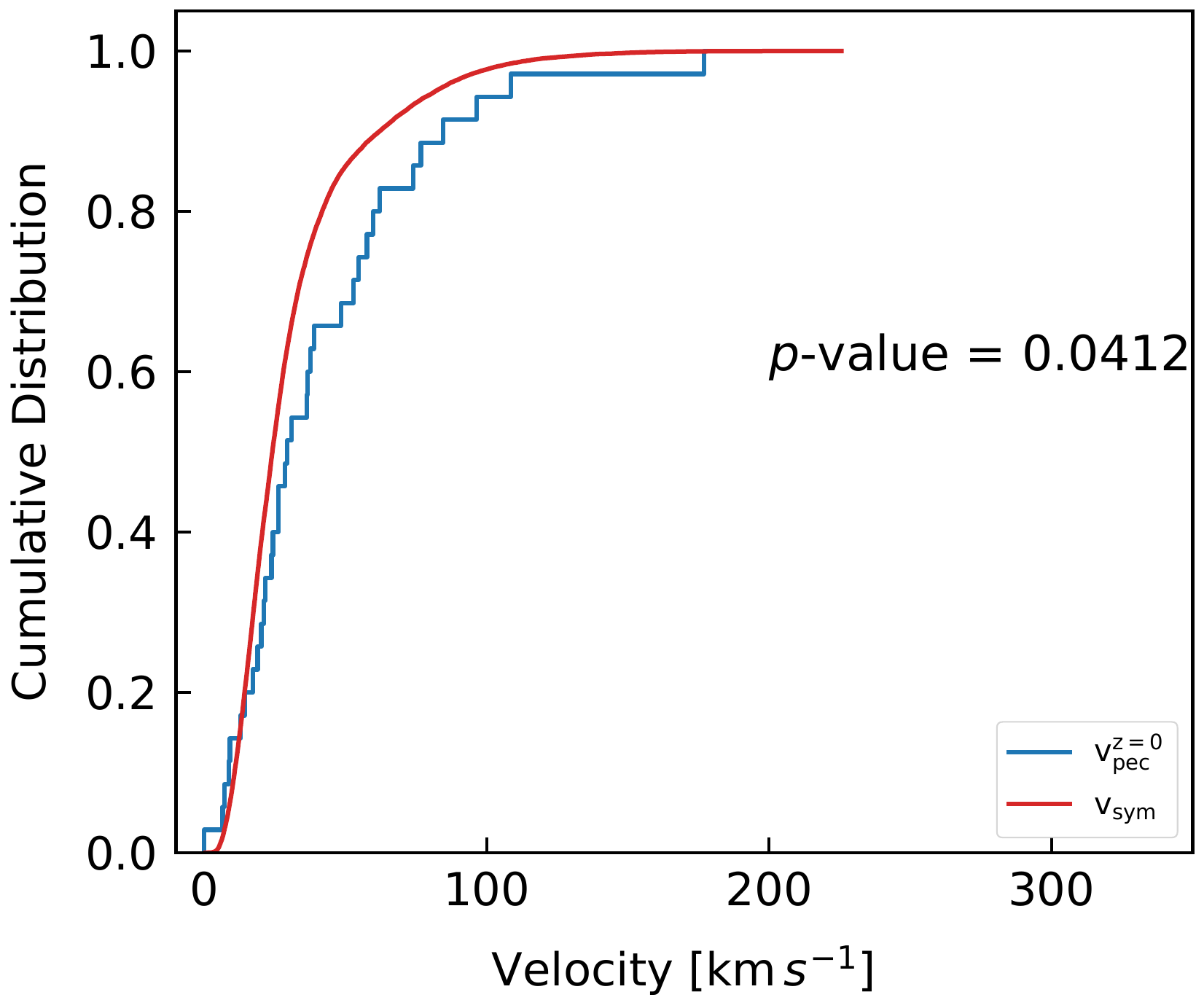}{0.33\textwidth}{$\sigma_1$ = 265 km\,s$^{-1}$, $\sigma_2$ = 80 km\,s$^{-1}$}
 } 
\gridline{\hspace{-1.0cm}\fig{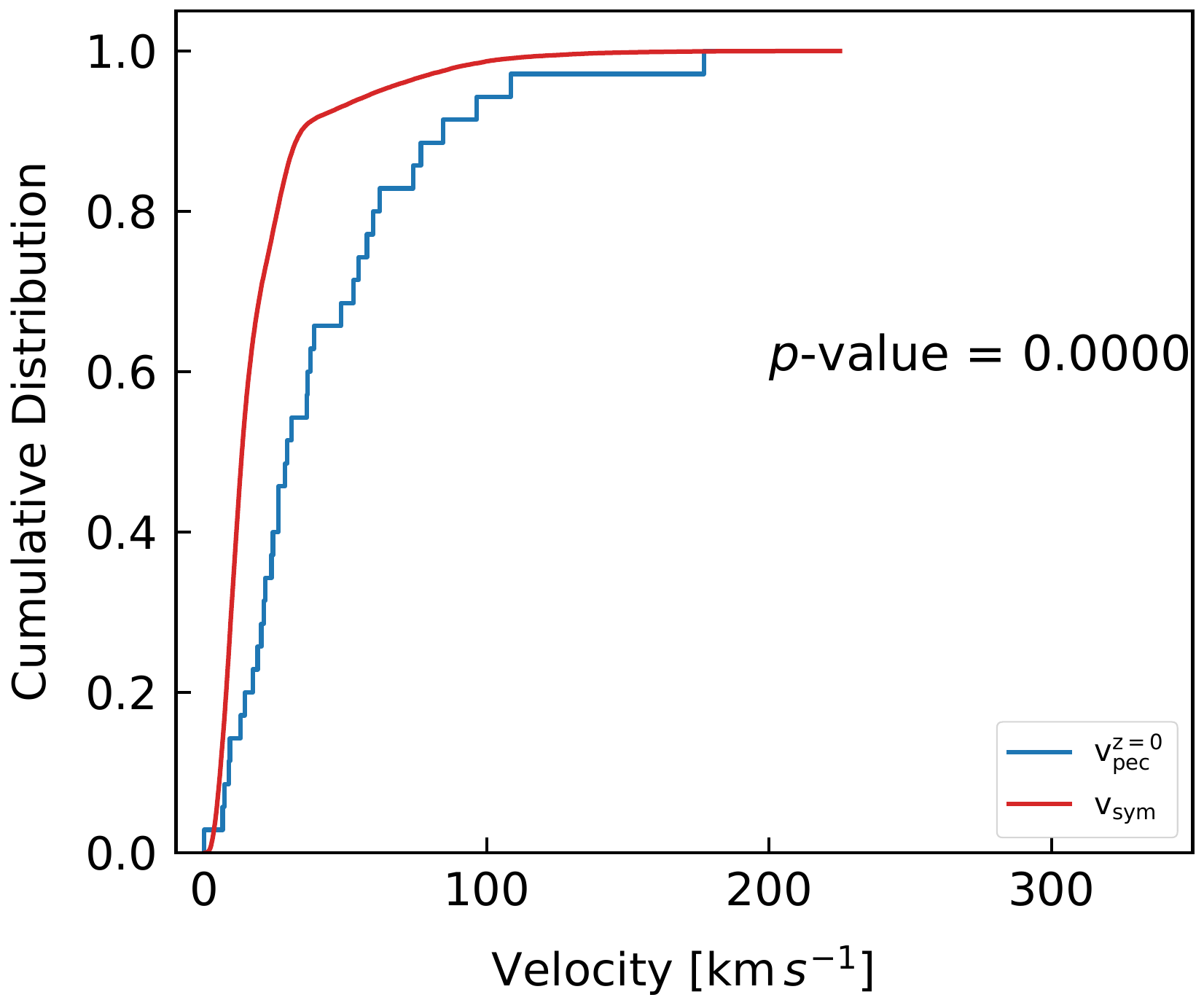}{0.33\textwidth}{$\sigma_1$ = 320 km\,s$^{-1}$, $\sigma_2$ = 30 km\,s$^{-1}$}
\hspace{-0.4cm}\fig{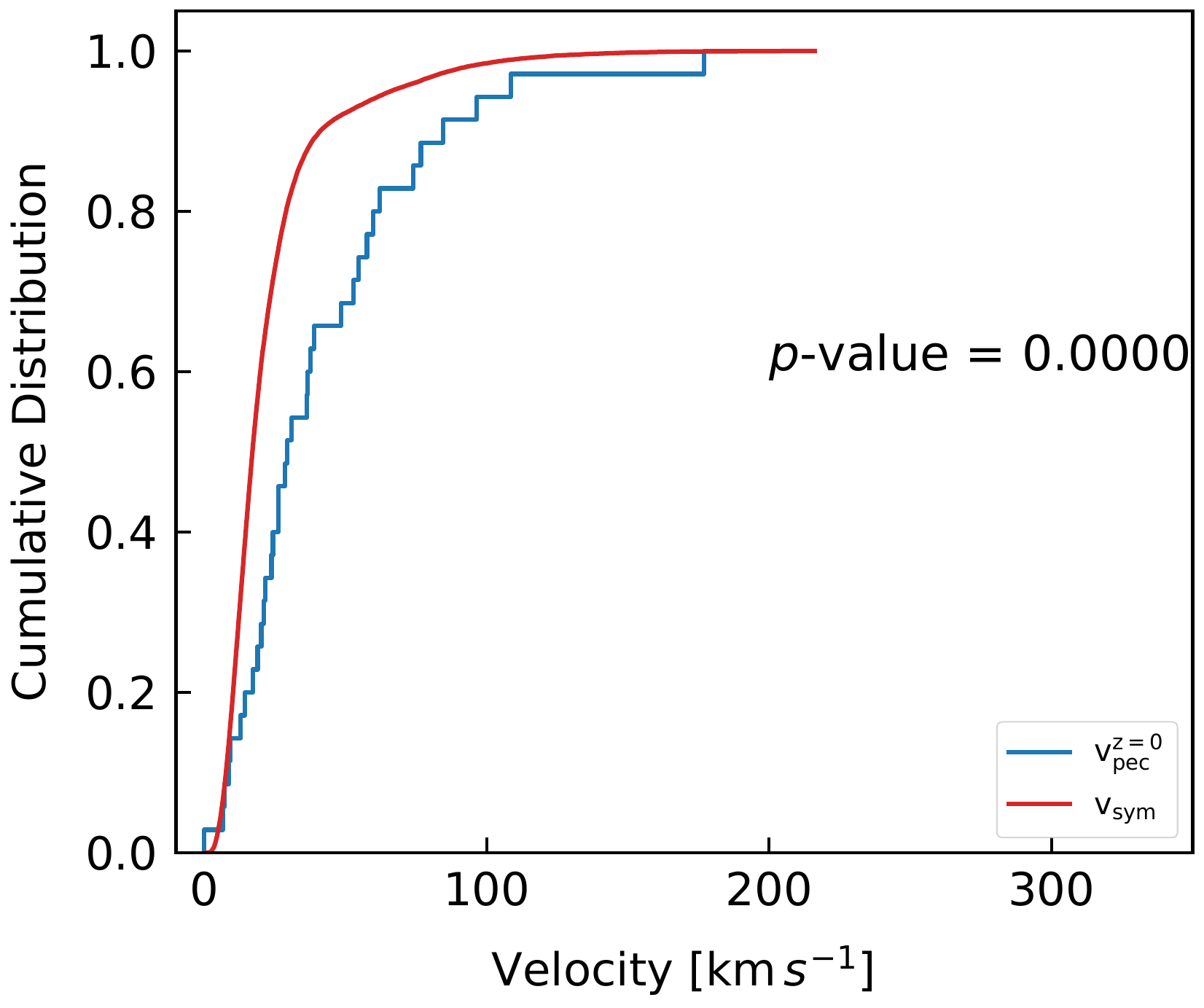}{0.33\textwidth}{$\sigma_1$ = 320 km\,s$^{-1}$, $\sigma_2$ = 50 km\,s$^{-1}$}
 \hspace{-0.4cm}\fig{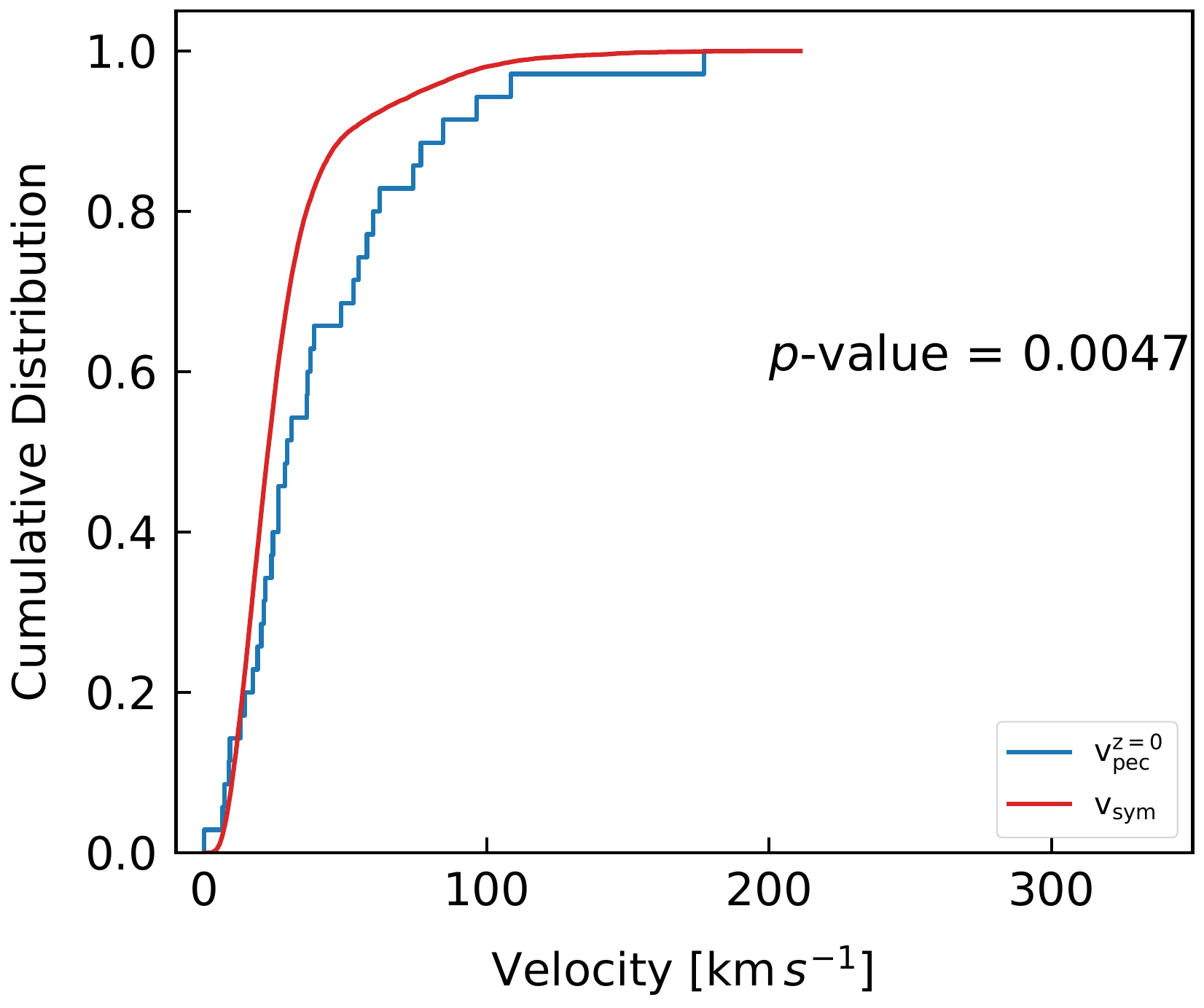}{0.33\textwidth}{$\sigma_1$ = 320 km\,s$^{-1}$, $\sigma_2$ = 80 km\,s$^{-1}$}
 }  
 \gridline{\hspace{-1.0cm}\fig{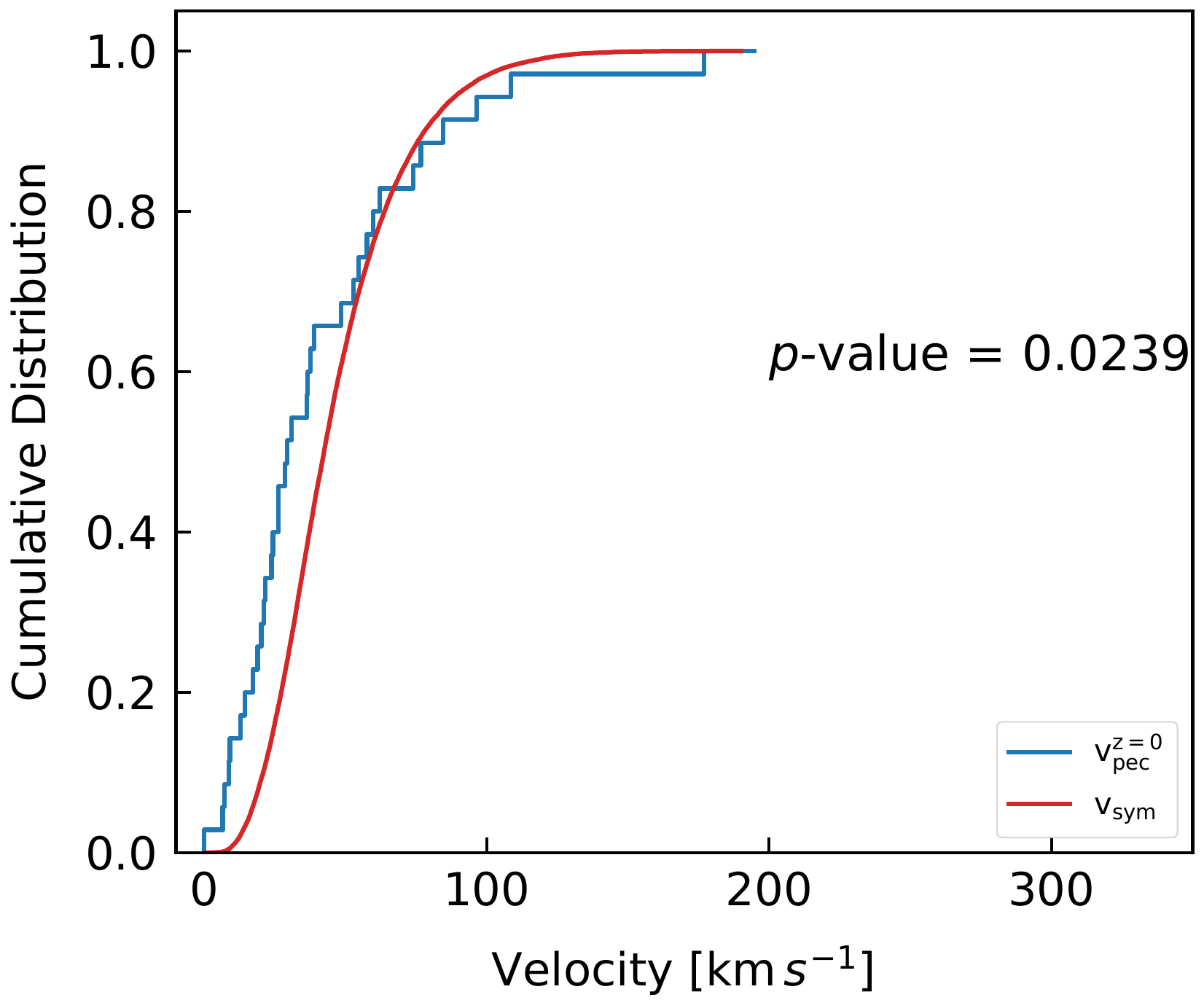}{0.33\textwidth}{$\sigma$ = 190 km\,s$^{-1}$}
\hspace{-0.4cm}\fig{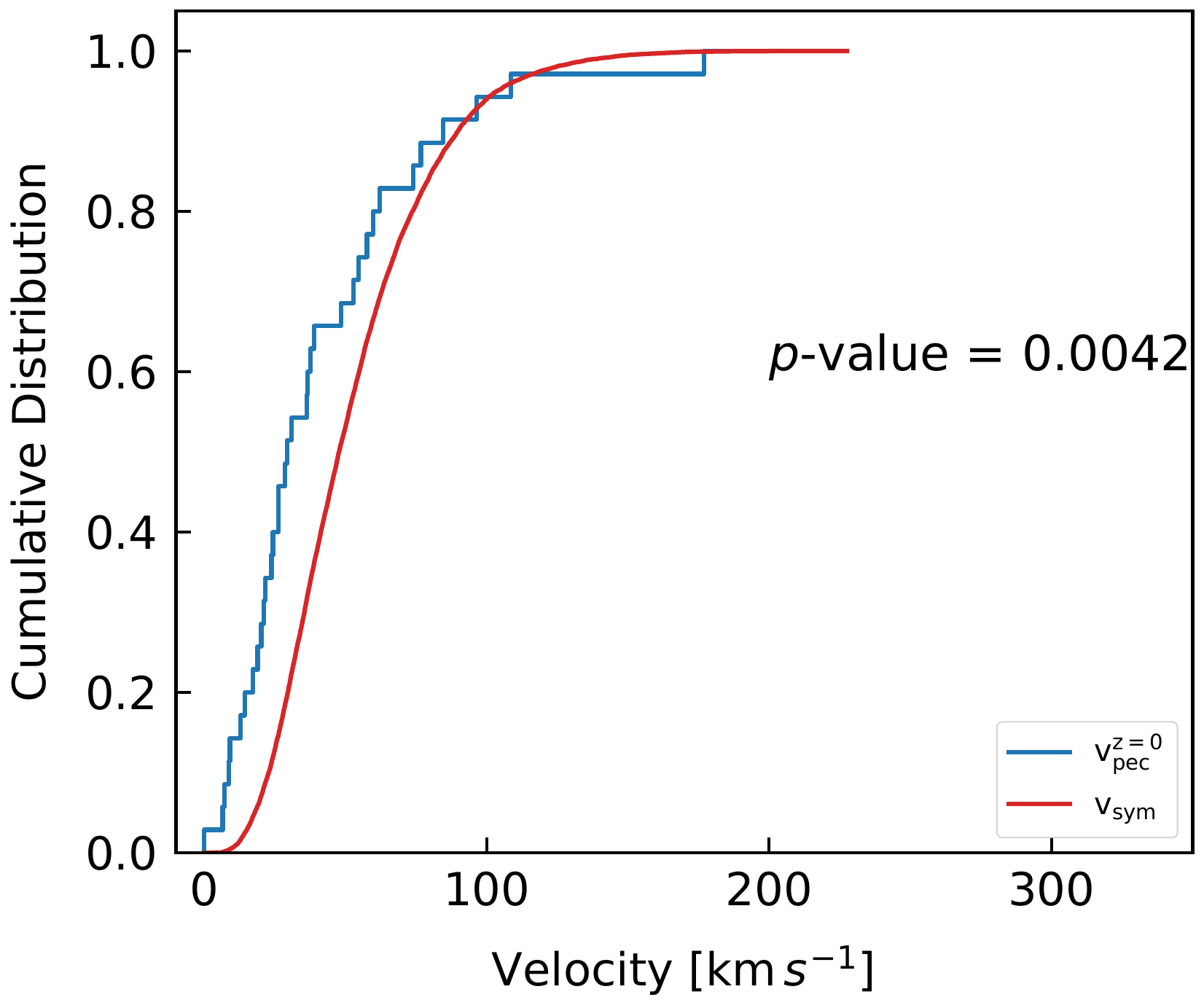}{0.33\textwidth}{$\sigma$ = 265 km\,s$^{-1}$}
 }
\caption{Same as Figure \ref{fig:vk_A}, but with $M_{\rm ecs}=(1.83-2.75)$ $M_{\odot}$.}
\label{fig:vk_B}
\end{figure*}

\section{NS HMXB Sample and Their Velocities} \label{sec:sample}
\subsection{The Sample} \label{subsec:sample}
We use the \textit{XRBcats} catalog\footnote{http://astro.uni-tuebingen.de/$\sim$xrbcat/} compiled by \cite{Neumann2023A&A} and  the \textit{HMXBwebcat} catalog\footnote{https://binary-revolution.github.io/HMXBwebcat/} compiled by \cite{Fortin2024A&A} to identify sources not recorded in \cite{Zhaoyue2023MNRAS}. The sources in the two catalogs have been cross-matched with \textit{Gaia} Data Release 3 \citep[\textit{Gaia} DR3;][]{GAIA2023A&A}\footnote{https://gea.esac.esa.int/archive/}, enabling the direct selection of sources with known proper motions ($\mu_\alpha$ in R.A. and $\mu_\delta$ in decl.) and parallaxes ($\varpi$). We then search for the available systemic radial velocity ($\gamma$) measurements for these sources, resulting in a total of 16 sources which are listed in Table \ref{tab:table1}. Only 11 of them are identified as NS HMXBs, and the others could be either NS HMXBs or BH HMXBs. The parameters include the eccentricities, orbital periods, and potential peculiar velocities at birth of the 11 confirmed NS HMXBs and 25 NS HMXBs from \cite{Zhaoyue2023MNRAS}. They are used for comparison with the simulated results, as described in Section \ref{sec:result}.

\subsection{Distances $(d)$}
\label{subsec:dis}
The distance ($d$) of the selected sources is determined according to  their measured \textit{Gaia} trigonometric parallaxes. For the three sources listed in Table \ref{tab:table1} with poorly constrained parallaxes, i.e., $|\sigma_{\varpi}/\varpi| > $ 0.5 (they actually have negative parallax values), we adopt the distances from the literature. For the sources with well-constrained $\varpi$, we calculate $d$ in two ways. If $\varpi \geq 2$ mas (only for gam Cas), $d$ is directly obtained by inverting $\varpi$ while its uncertainty is propagated from the uncertainty of $\varpi$, i.e., $d = 1/\varpi$, and $\sigma_{d} = \sigma_{\varpi}/\varpi^2$.  If $\varpi<2$ mas, a Bayesian method is used to infer $d$. An exponential prior of $d$ follows 
\begin{equation}
    \label{eq:prior}
    p(d) = \frac{1}{2L^3} d^2 \exp{\Big(-\frac{d}{L}\Big)},
\end{equation}
where $L$ is the scaling factor \citep{Astraatmadja2016ApJ, Luri2018A&A, Gandhi2019MNRAS}.  \cite{Zhaoyue2023MNRAS} evaluated $L$ to be $1.97_{-0.05}^{+0.05}$~kpc based on a sample of 125 binary systems, and we adopt this value in our work. 
Assuming that $\sigma_{\varpi}$ is introduced only by the random error, the observed $\varpi$ follows a Gaussian distribution with the mean value of the inverse of $d$, expressed as 
\begin{equation}
    \label{eq:likelihood}
    p(\varpi \vert d, \sigma_{\varpi}) = \frac{1}{\sqrt{2\pi} \sigma_{\varpi}} \exp{\bigg [- \frac{1}{2\sigma_{\varpi}^2}\Big(\varpi - \frac{1}{d}\Big)^2 \bigg]}.
\end{equation}
According to the Bayes' theorem, the posterior distribution of $d$ is
\begin{equation}
    \label{eq:posterior}
    p(d \vert \varpi, \sigma_{\varpi}) = \frac{p(\varpi \vert d, \sigma_{\varpi})p(d)}{p(\varpi)}
    \propto p(\varpi \vert d, \sigma_{\varpi})p(d).
\end{equation}
Using the observed $\varpi$ and its corresponding $\sigma_{\varpi}$, we can get the posterior probability density function (PDF) of $d$ for each source. The calculated median values of the PDF and the lower and upper errors corresponding to the 16th and 84th percentiles, respectively, are listed in the third column of Table \ref{tab:table2}.

\subsection{Peculiar Velocity $(v_{\rm pec})$ and Potential Peculiar Velocity at Birth $(v_{\rm pec}^{z=0})$} \label{subsec:peculiar_v}

Peculiar velocity ($v_{\rm pec}$) is the 3D velocity of a source relative to its local Galactic motion. Since we have the position, distance, proper motions, and radial velocity for each source, we can obtain its 3D location in the Galaxy and 3D motion relative to the Sun. The Galactic constants used in our work include the projected distance of the Sun from the Galactic center, $R_{0} = 8.34\pm0.16$ kpc, the Galactic rotation speed of at the radius of the Sun's orbit, $\mathit{\Theta}_{0} = 240\pm8$ km\,s$^{-1}$, the Sun's velocity toward the Galactic center, $U_{\odot} = 10.7\pm1.8$ km\,s$^{-1}$, the Sun's velocity in the direction of the Galactic rotation, $V_{\odot} = 15.6\pm6.8$ km\,s$^{-1}$, and the Sun's velocity toward the north Galactic pole, $W_{\odot} = 8.9\pm0.9$ km\,s$^{-1}$ \citep{Reid2014ApJ}. Following the method in \cite{Reid2009ApJ}, we assume Gaussian distributions for $d$, $\mu_\alpha\cos\delta$, $\mu_\delta$, $\gamma$, and the Galactic constants to get $10^6$ values of all these parameters. The Gaussian distribution of $d$ is sampled by taking the median value as the Gaussian mean and the average of the lower and upper errors corresponding to the 16th and 84th percentiles, respectively, as the standard deviation. These values of the Gaussian distributions are then used to obtain the distribution of $v_{\rm pec}$.  The Galactic coordinates (\textit{l}, \textit{b}) for all sources can be directly obtained from \textit{Gaia} DR3. Then, $\mu_\alpha\cos\delta$ and $\mu_\delta$ can be converted into proper motions in the Galactic coordinates ($\mu_{l}\cos{b}$, $\mu_{b}$). By multiplying $d$, the linear velocities in the Galactic longitude and altitude directions can be obtained by
\begin{equation}
    \label{eq:propermotion_to_speed}
    \begin{array}{ll}
    v_l = d \mu_{l} \cos{b}, \\
    v_b = d \mu_{b}.
    \end{array}
\end{equation}
These components can then be converted to the components in Cartesian Galactic coordinates $(U_1, V_1, W_1)$ through a rotation:
\begin{equation}
    \label{eq:spherical_to_Cartesian}
    \begin{array}{lll}
         U_1 = (\gamma \cos{b} - v_b \sin{b})\cos{l} - v_l \sin{l}, \\
         V_1 = (\gamma \cos{b} - v_b \sin{b})\sin{l} + v_l \cos{l}, \\
         W_1 = v_b \cos{b} + \gamma \sin{b},
    \end{array}
\end{equation}
where the values of $\gamma$ are listed in the second column of Table \ref{tab:table2}. It should be noted that $(U_1, V_1, W_1)$ are relative to the Sun. To convert the heliocentric velocity to the Galactocentric one, the motion of the Sun and the rotation velocity of the LSR should be added \citep{Reid2009ApJ}:
\begin{equation}
    \label{eq:U2}
    \begin{array}{lll}
         U_2 = U_1 + U_{\odot}, \\
         V_2 = V_1 + V_{\odot} + \Theta_{0}, \\
         W_2 = W_1 + W_{\odot}.
    \end{array}
\end{equation}
Finally, rotating the velocity in the Galactic plane and subtracting the Galactic rotation at the location of the source ($\Theta_{\rm S}$) by using the MWPotential2014 potential from the \texttt{GALPY} package \citep{Bovy2015ApJS}, we can obtain the three components of peculiar velocity:
\begin{equation}
    \label{eq:Us}
    \begin{array}{lll}
         U_{\rm S} = U_2 \cos{\beta} - V_2 \sin{\beta}, \\
         V_{\rm S} = V_2 \cos{\beta} + U_2 \sin{\beta} - \Theta_{\rm S}, \\
         W_{\rm S} = W_2.
    \end{array}
\end{equation}
Here, $\beta$ is the angle between the source and the Sun from the Galactic center which can be derived from the Galactic coordinates of the source:
\begin{equation}
    \label{eq:beta}
    \sin{\beta} = \frac{d\cos{b}\sin{l}}{\sqrt{R_{0}^2 + (d\cos{b})^2 - 2R_{0}d\cos{b}\cos{l}}},
\end{equation}
where $R_{0}$ is the distance to the Galactic center. Finally, $v_{\rm pec}$ can be obtained by $v_{\rm pec} = \sqrt{(U_{\rm S}^2 + V_{\rm S}^2+W_{\rm S}^2)}$. We list the calculated median value of the PDF and the lower and upper errors corresponding
to the 16th and 84th percentiles, respectively, in the fourth column of Table \ref{tab:table2}.

The calculated $v_{\rm pec}$ represents the source's present-day velocity, not peculiar velocity at the moment of the SN explosion. Due to the influence of the Galactic potential, the position and velocity of the source will change with time. The average velocity every time the binary across the Galactic plane is generally used to represent the birth peculiar velocity ($v_{\rm pec}^{z=0}$), as HMXBs formed in the Galactic disk. In order to estimate $v_{\rm pec}^{z=0}$, we draw $10^6$ random samples from Gaussian distributions for $d$, $\mu_\alpha\cos\delta$, $\mu_\delta$, and $\gamma$ to initialize the orbital calculations. Then, we use \texttt{GALPY} to integrate their Galactic orbits backward for 10 Gyr \footnote{The integration back for 10 Gyr exceeds the ages of NS HMXBs by far. However, it was pointed out that $v_{\rm pec}^{z=0}$ is not influenced by the integration time  \citep{Atri2019MNRAS}.} and record $v_{\rm pec}$ of the binary each time it passes through the Galactic plane. 
Finally,  we collect the mean of the disk-crossing $v_{\rm pec}$ for each set of random parameters drawn from the Gaussian distributions to form the PDF of $v_{\rm pec}^{z=0}$. The calculated median values of the PDF and the lower and upper errors corresponding to the 16th and 84th percentiles, respectively, are listed in the fifth column of Table \ref{tab:table2}.

\subsection{Kick Velocity $(v_{\rm k})$} \label{subsec:kick_v}
The kick imparted to the newborn NS during an SN changes the velocity of the binary system. The relationship between the systemic velocity ($v_{\rm sym}$) and the kick velocity ($v_{\rm k}$) is given by \citep{Hurley2002MNRAS}
\begin{equation}
    \label{eq:vs}
    \vec{v}_{\rm sym} = \frac{M_1^\prime}{M_{\rm b}^\prime}\vec{v}_{\rm k} - \frac{\Delta M_1M_2}{M_{\rm b}^\prime M_{\rm b}}\vec{v}.
\end{equation}
Here $M_1$ and $M_1^\prime$ are the masses of the primary star before and after the SN explosion, respectively; $M_2$ is the mass of the secondary star; $M_{\rm b}$ and $M_{\rm b}^\prime$ are the total masses of the binary before and after the SN explosion, respectively; and $\Delta M_1=M_1-M_{1}^{\prime}$ is the ejecta mass from the primary star. As mentioned in Section \ref{subsec:SN_models}, $M_{1}^{\prime}$ is derived from the rapid SN model of \cite{Fryer2012ApJ} for CCSNe and set to $1.3$ $M_{\odot}$ for ECSNe. Finally, $\vec{v}$ is the relative orbital velocity at the moment of the SN explosion.

Since the angle between $v_{\rm k}$ and the orbital plane is unknown, it is hard to directly get $v_{\rm k}$ from observation. Instead, we can generate a large number of binary systems with randomly distributed kick velocities to model the distribution of  $v_{\rm sym}$.  By comparing the distributions of $v_{\rm sym}$ and $v_{\rm pec}^{z=0}$, we can examine which natal kick distribution can best match the observations. The methods and models are described in Section \ref{sec:method}.

\section{Methods}
\label{sec:method}
\subsection{Binary Population Synthesis Model}

We adopt the Monte Carlo population synthesis code {\tt BSE} initially developed by \cite{Hurley2002MNRAS} and modified by \cite{Kiel2006MNRAS} and \cite{Shao2014ApJ} to follow the formation of NS HMXBs. We evolve 10$^7$ primordial binaries with the star formation rate  $5\,M_{\odot}~$yr$^{-1}$  over the past 12 Gyr \citep{Smith1978A&A}. The initial parameters of the binaries include the primary mass ($M_1$), the mass ratio ($q$), and the binary separation ($a$). They are randomly drawn from the independent probability distribution functions. The initial $M_1$ is assumed to follow the \cite{Kroupa1993MNRAS} initial mass function in the range of $(3-100)$ $M_{\odot}$. $q$ is the ratio of the secondary mass and the primary mass ($M_2/M_1$), and its initial value is uniformly distributed in the range of $(0-1)$. The initial $a$ is set in the range of $(3-10^4)\,R_{\odot}$, following a logarithmically uniform distribution. The metallicity ($Z$) is set to 0.02.\footnote{ Since Galactic HMXBs are relatively young objects, it is commonly assumed that the HMXB populations are little affected by metallicity. However, the evolution of the massive progenitor stars depends on metallicity in a number of ways, especially through their winds. To examine these effects, \citet{Dray2006MNRAS} performed population simulations with different metallicities and showed that Galactic HMXBs can be well matched with the solar metallicity model.}  All binaries are assumed to be in circular orbits since initial eccentricities have minor influence on the following evolution.

In our simulation, we adopt the critical mass ratio ($q_{\rm cr}$) from \cite{Shao2014ApJ} to determine whether the mass transfer from the primary star to the secondary star through Roche-lobe overflow is dynamically stable. For unstable mass transfer, the binary enters a CE phase. We use the energy conservation equation from \cite{Webbink1984ApJ} to treat the CE evolution. The end product depends on the CE efficiency parameter ($\alpha_{\rm CE}$), which describes the fraction of the orbital energy that is used to expel the CE, and the envelope binding energy parameter ($\lambda$), which depends on the stellar density distribution as well as the evolutionary status. 
We set a constant $\alpha_{\rm CE} = 1.0 $ \footnote{In principle, the $\alpha_{\rm CE}$ value should vary with the evolutionary states of both the donor and accretor, as well as the initial orbit, but it is still poorly known and usually taken with a fixed value $\leq 1$ \citep{2013A&ARv..21...59I}. For a large fraction of HMXBs, the mass transfer between the NS/BH progenitor and the companion star is stable because the mass ratio is not extremely small, so our results are not sensitive to the value of $\alpha_{\rm CE}$. For example, \cite{Linden2010ApJ} found that the predicted number of bright HMXBs remains relatively constant despite the changes in $\alpha_{\rm CE}$.} and employ $\lambda$ values calculated by \cite{WangJia&Li2016RAA}. 

\subsection{SN models and Natal kicks}
\label{subsec:SN_models}
NSs can form from either CCSNe or ECSNe. In the {\tt BSE} code, the mass of the helium (He) core at the asymptotic giant branch base is generally used to discriminate the SN mechanisms. ECSNe are thought to arise from super-asymptotic giant branch stars with initial mass of $\sim (6.5-12)$ $M_{\odot}$, though the precise initial mass range for the ECSN progenitors is highly uncertain \citep{2017PASA...34...56D}. Moreover, ECSNe could be enhanced in binaries because envelope stripping by Roche-lobe overflow suppresses the second dredge-up \citep[][and references therein]{2017ApJ...850..197P,2021ApJ...920L..37W}. 
So we consider two possible He core mass ($M_{\rm ecs}$) ranges $(1.83-2.25)$ $M_{\odot}$ and $(1.83-2.75)$ $M_{\odot}$ for ECSNe, considering the uncertainties in the triggering conditions \citep{Shao2018ApJ}. The mass of the remnant NSs formed from CCSNe is obtained from the rapid SN model of \citet{Fryer2012ApJ} and the gravitational mass of NSs from ECSNe is set to $1.3$ $M_{\odot}$ considering that some baryonic mass of the 1.38 $M_{\odot}$ ONe core will be lost to form an NS. 

A total of 11 $v_{\rm k}$ distributions are constructed referring to previous investigations shown in Table \ref{tab:value}. We set up two single Maxwellian distributions, with dispersion the same as in \citet{Hansen1997} and \citet{Hobbs2005MNRAS}.  For the remaining nine velocity models, we adopt bimodal distributions as suggested by many subsequent studies on radio pulsars and XRBs \citep[e.g.,][]{Pfahl2002ApJ,Verbunt2017A&A, Atri2019MNRAS,2021MNRAS.508.3345I,Zhaoyue2023MNRAS}. They demonstrated the existence of both high-velocity (a few hundred km\,s$^{-1}$) and low-velocity ($\lesssim 100$ km\,s$^{-1}$) populations. Theoretically, this may reflect relatively large and small kicks in CCSNe and ECSNe, respectively \citep{Janka2017ApJ, Gessner2018ApJ, 2020LRCA....6....3M}. We accordingly sample random $v_{\rm k}$ from Maxwellian distributions
characterized by dispersion of $\sigma_1$ and $\sigma_2$ for CCSNe and ECSNe, respectively. We also assume that the directions of $v_{\rm k}$ follow an isotropic distribution in the polar coordinates. Both the high- and low-velocity components in our bimodal distributions are characterized by three typical dispersion values, respectively. We select the post-SN systems with the companion mass greater than $8$ $M_{\odot}$ for the observed HMXBs, as such massive stars can produce strong stellar winds and outflows during their lifetimes to power X-ray emission of the accreting NSs \citep{Chaty2022abn}. This criterion is consistently applied to the HMXB samples collected from the catalogs of \citet{Neumann2023A&A} and \citet{Fortin2024A&A}.

\section{Results} \label{sec:result}
\subsection{Comparison of the Velocity Distributions} \label{subsec:velocity}

Figures~\ref{fig:vk_A} and \ref{fig:vk_B} compare the cumulative distribution functions of the simulated $v_{\rm sym}$ (red lines) and the derived medium $v_{\rm pec}^{z=0}$ values of 36 observed NS HMXBs (blue lines) with $M_{\rm ecs} = (1.83-2.25)$ $M_{\odot}$ and $(1.83-2.75)$ $M_{\odot}$, respectively. We apply the Kolmogorov-Smirnov (K-S) test to evaluate the consistency of these two distributions. The K-S test is a statistical method used to test whether an empirical distribution conforms to a theoretical distribution or to compare two empirical distributions for significant differences. The $p$-value in the K-S test is generally used to determine whether two distributions can be considered to originate from the same underlying distribution, with a significance level ($\alpha$) of 0.05. We use the function \texttt{ks\underline{ }2samp} in \texttt{scipy} \citep{Virtanen2020NatMe} to calculate the $p$-values for the 11 $v_{\rm k}$ distributions. The results are displayed in Figures \ref{fig:vk_A} and \ref{fig:vk_B} and in Table \ref{tab:value}. We also present the weights of NSs produced by ECSNe, which represent their relative contributions to the combined distribution. The results show that the weights of NSs from ECSNe vary with $v_{\rm k}$. Several models have $p$-values greater than 0.05, indicating that there are not enough evidence in the data to reject the null hypothesis that the observed distribution is consistent with the model predictions. Notably, many fits appear statistically acceptable, particularly those with moderate  $p-$values. For instance, two representative cases demonstrate high compatibility: (1) $\sigma_1 = 320$ km\,s$^{-1}$, $\sigma_2 = 80$ km\,s$^{-1}$, $M_{\rm ecs}=(1.83-2.25)$ $M_{\odot}$ (with $p-$value $= 0.8662$), and (2) $\sigma_1 = 265$ km\,s$^{-1}$, $\sigma_2 = 50$ km\,s$^{-1}$, $M_{\rm ecs}=(1.83-2.25)$ $M_{\odot}$ (with $p-$value $= 0.6295$). While the $p$-value serves as an effective preliminary filter, eliminating most incompatible models, the selection of the optimal model requires consideration of additional criteria. From a purely statistical perspective, models with $M_{\rm ecs}=(1.83-2.25)$ $M_{\odot}$ seem to be more consistent with the observed velocity distribution compared to those with $M_{\rm ecs}=(1.83-2.75)$ $M_{\odot}$. However, it is hard to directly select the best model only from the velocity distribution. Combining other parameters might help us exclude less plausible models to get the best one, as discussed in Section \ref{subsec:porb}.

\subsection{Comparison of the $P_{\rm orb}$-$e$-$v_{\rm pec}^{z=0}$ distributions} \label{subsec:porb}
\begin{figure*}[t]
\vspace{-0.5cm}
\gridline{\hspace{-1.0cm}\fig{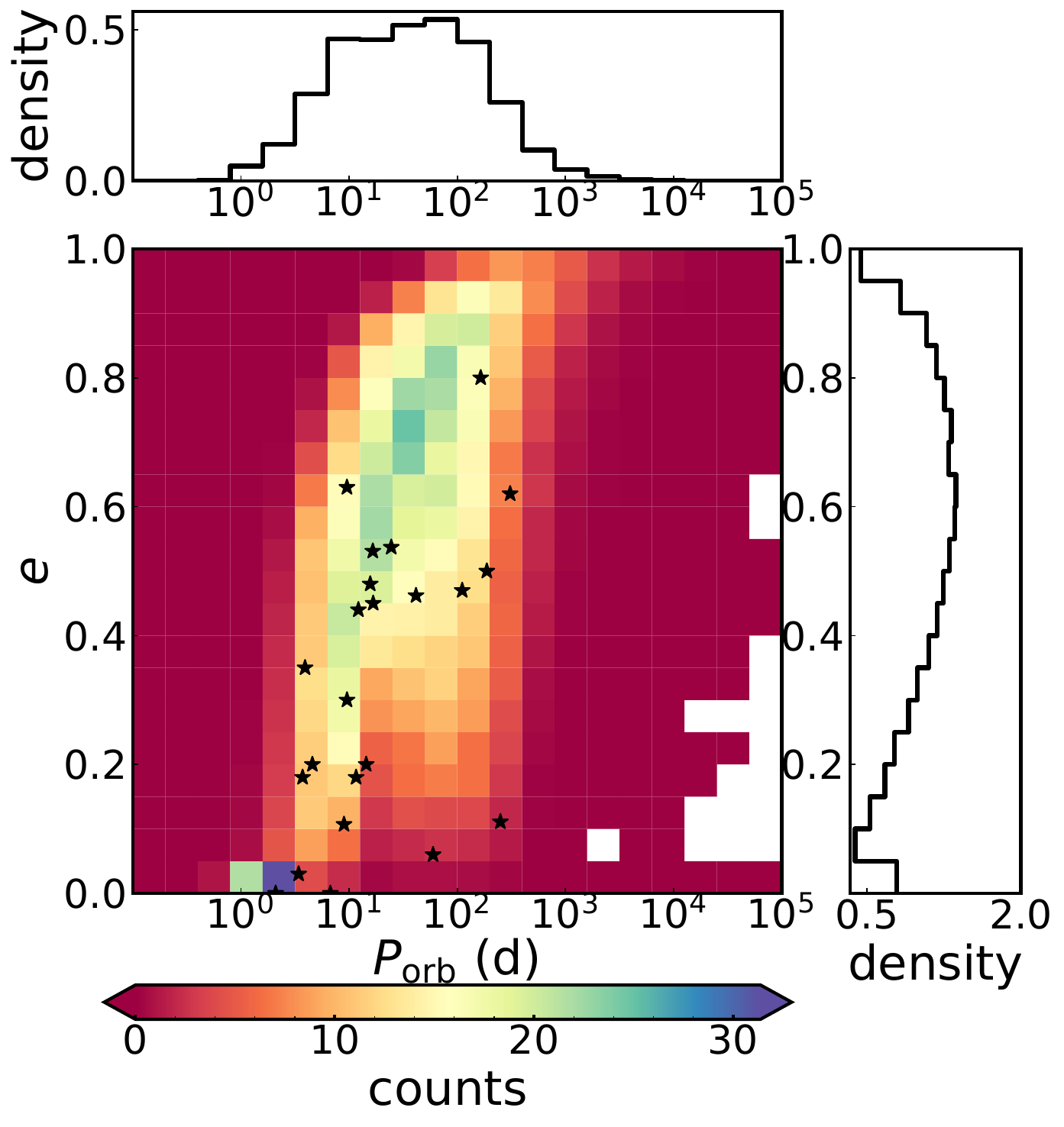}{0.27\textwidth}{$\sigma_1$ = 150 km\,s$^{-1}$, $\sigma_2$ = 30 km\,s$^{-1}$} \hspace{-0.4cm}\fig{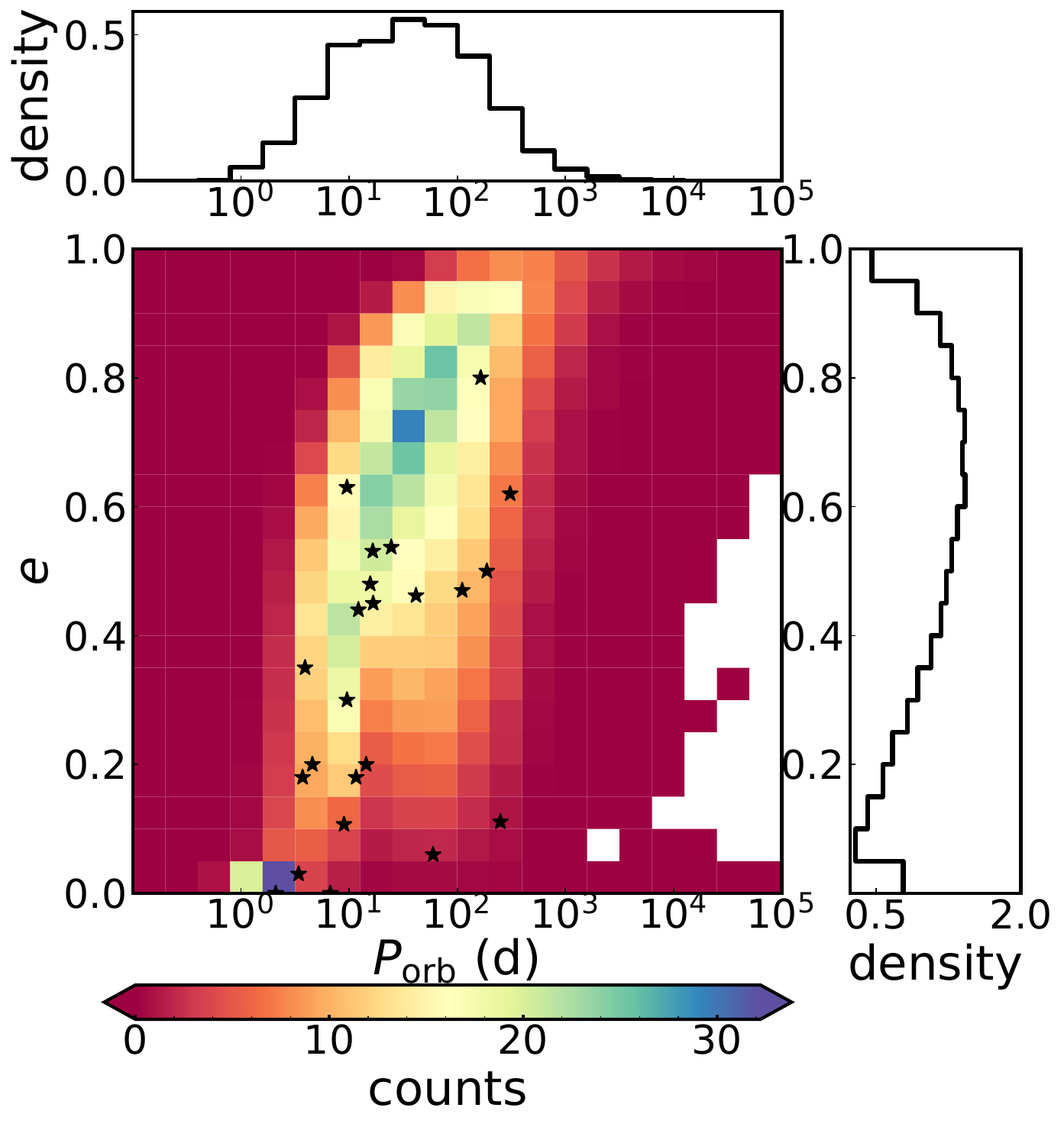}{0.27\textwidth}{$\sigma_1$ = 150 km\,s$^{-1}$, $\sigma_2$ = 50 km\,s$^{-1}$}
 \hspace{-0.4cm}\fig{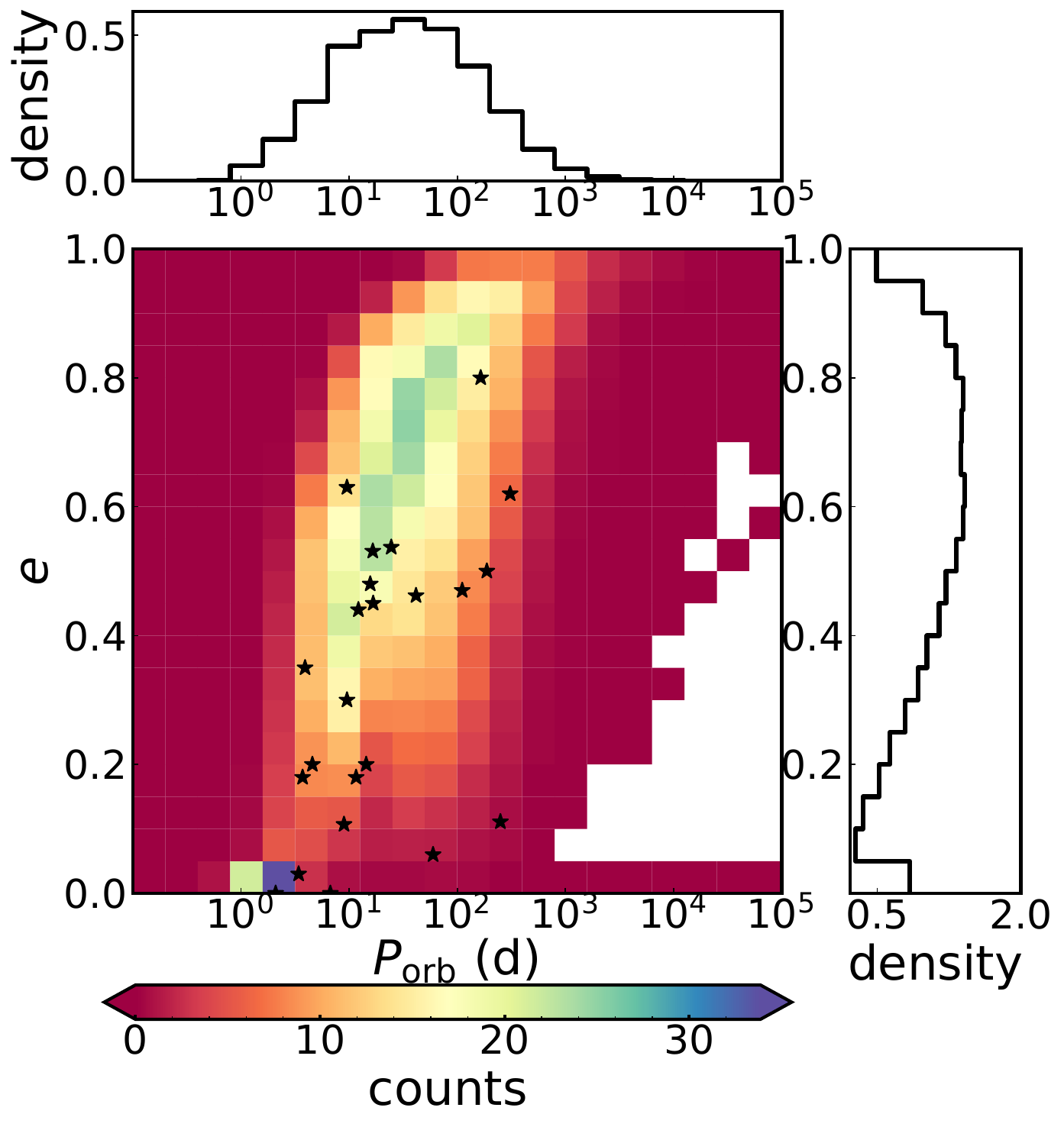}{0.27\textwidth}{$\sigma_1$ = 150 km\,s$^{-1}$, $\sigma_2$ = 80 km\,s$^{-1}$}
 }
\gridline{\hspace{-1.0cm}\fig{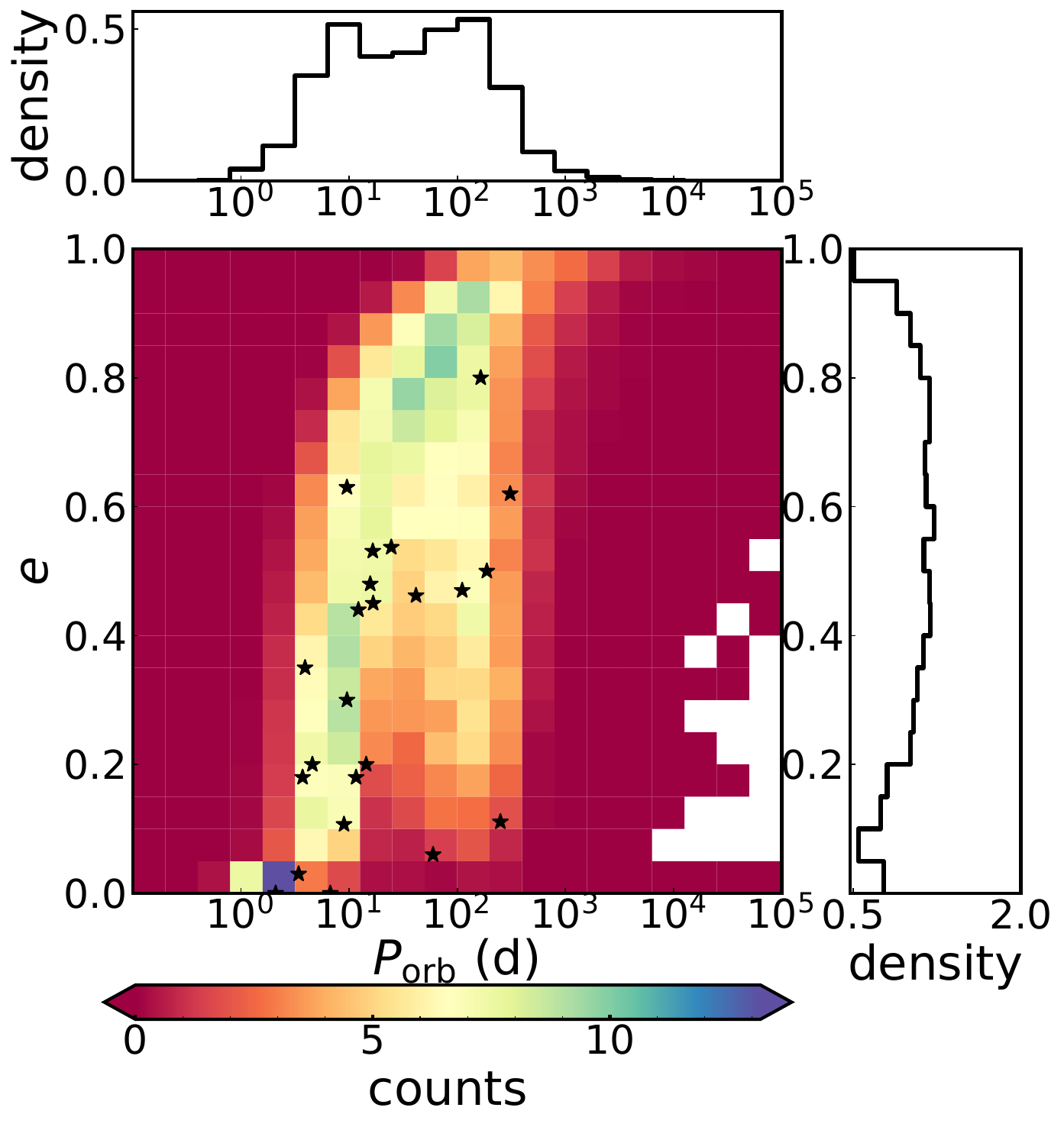}{0.27\textwidth}{$\sigma_1$ = 265 km\,s$^{-1}$, $\sigma_2$ = 30 km\,s$^{-1}$}
\hspace{-0.4cm}\fig{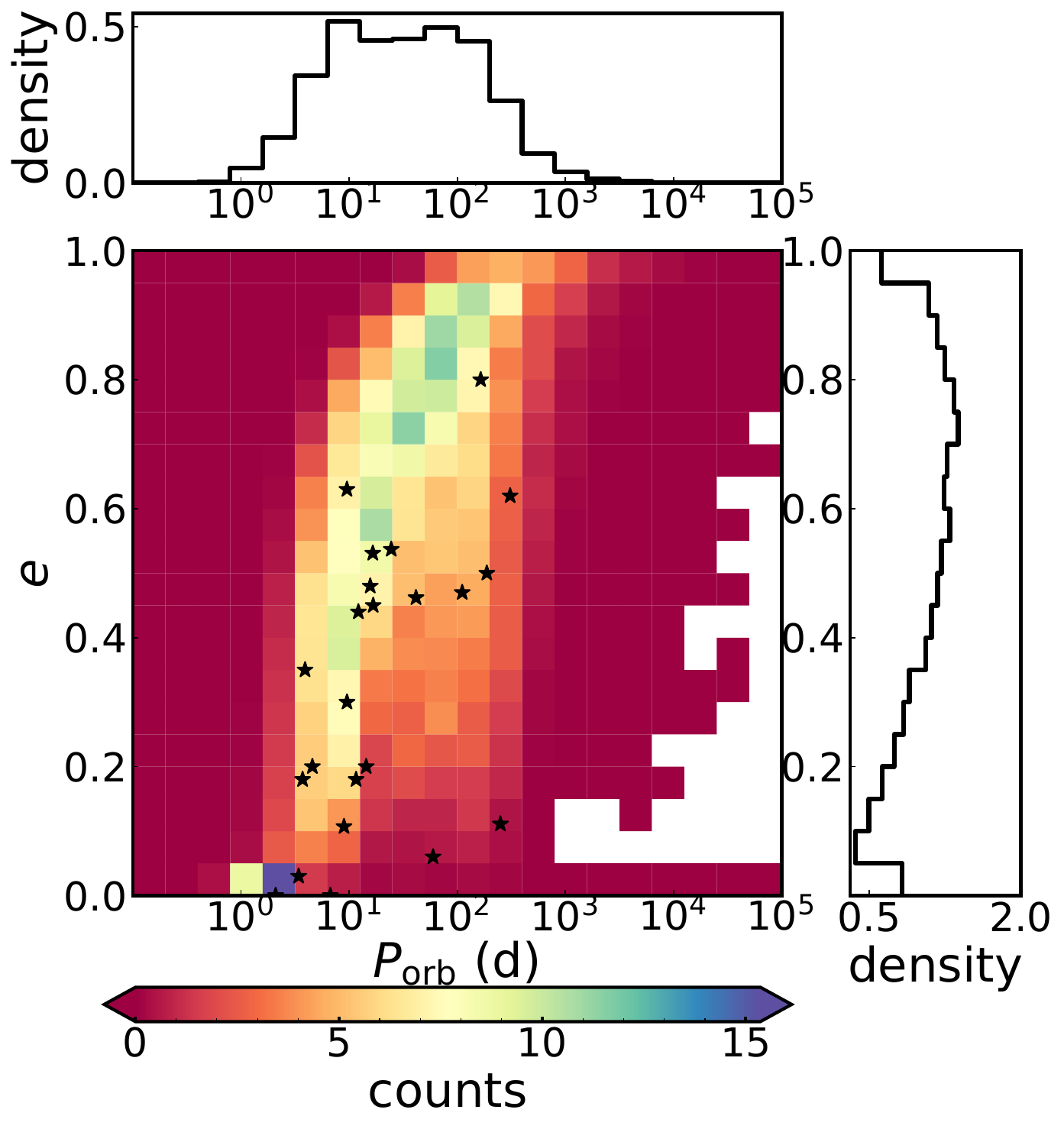}{0.27\textwidth}{$\sigma_1$ = 265 km\,s$^{-1}$, $\sigma_2$ = 50 km\,s$^{-1}$}
 \hspace{-0.4cm}\fig{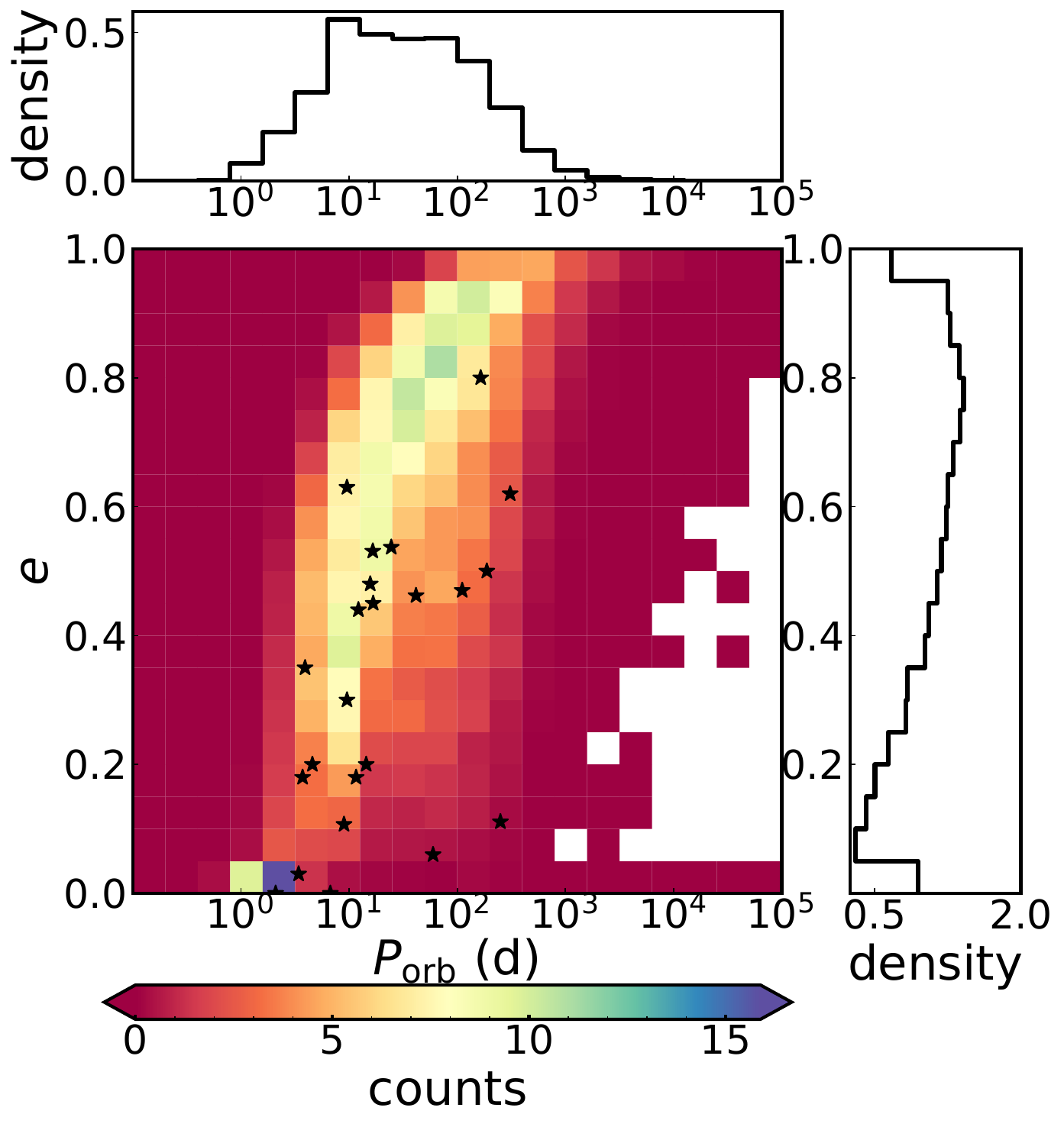}{0.27\textwidth}{$\sigma_1$ = 265 km\,s$^{-1}$, $\sigma_2$ = 80 km\,s$^{-1}$}
 } 
\gridline{\hspace{-1.0cm}\fig{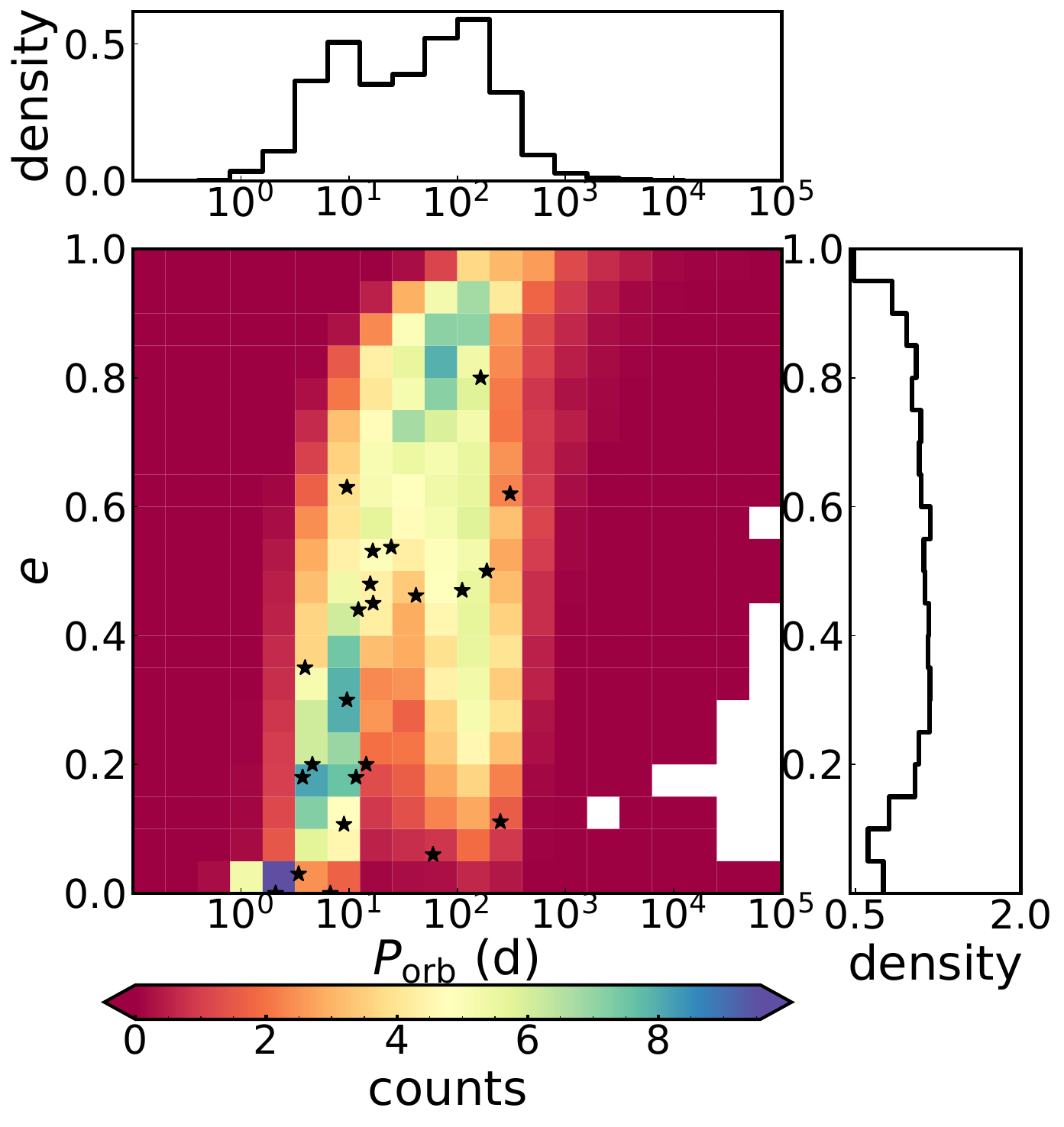}{0.27\textwidth}{$\sigma_1$ = 320 km\,s$^{-1}$, $\sigma_2$ = 30 km\,s$^{-1}$}
\hspace{-0.4cm}\fig{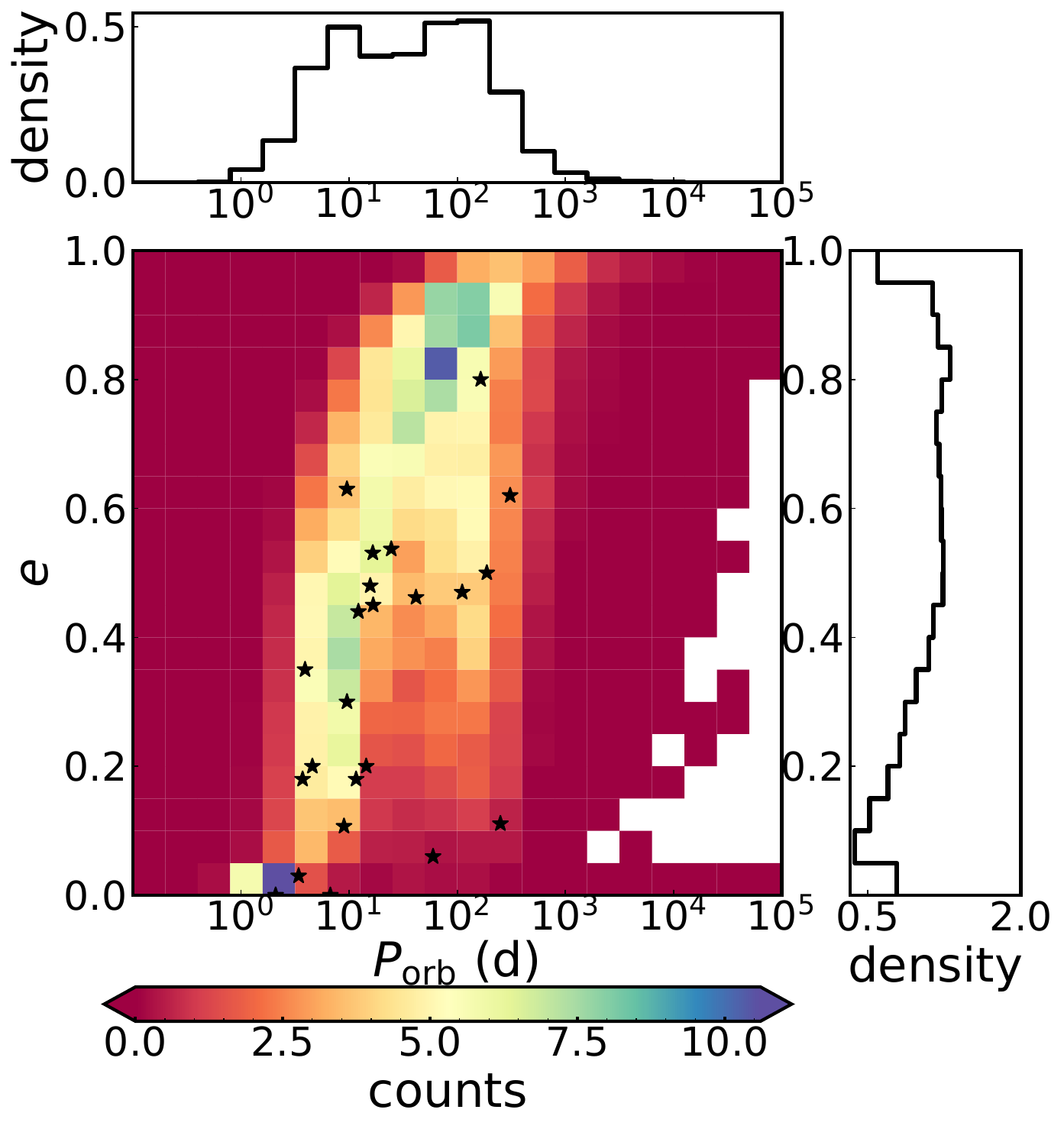}{0.27\textwidth}{$\sigma_1$ = 320 km\,s$^{-1}$, $\sigma_2$ = 50 km\,s$^{-1}$}
 \hspace{-0.4cm}\fig{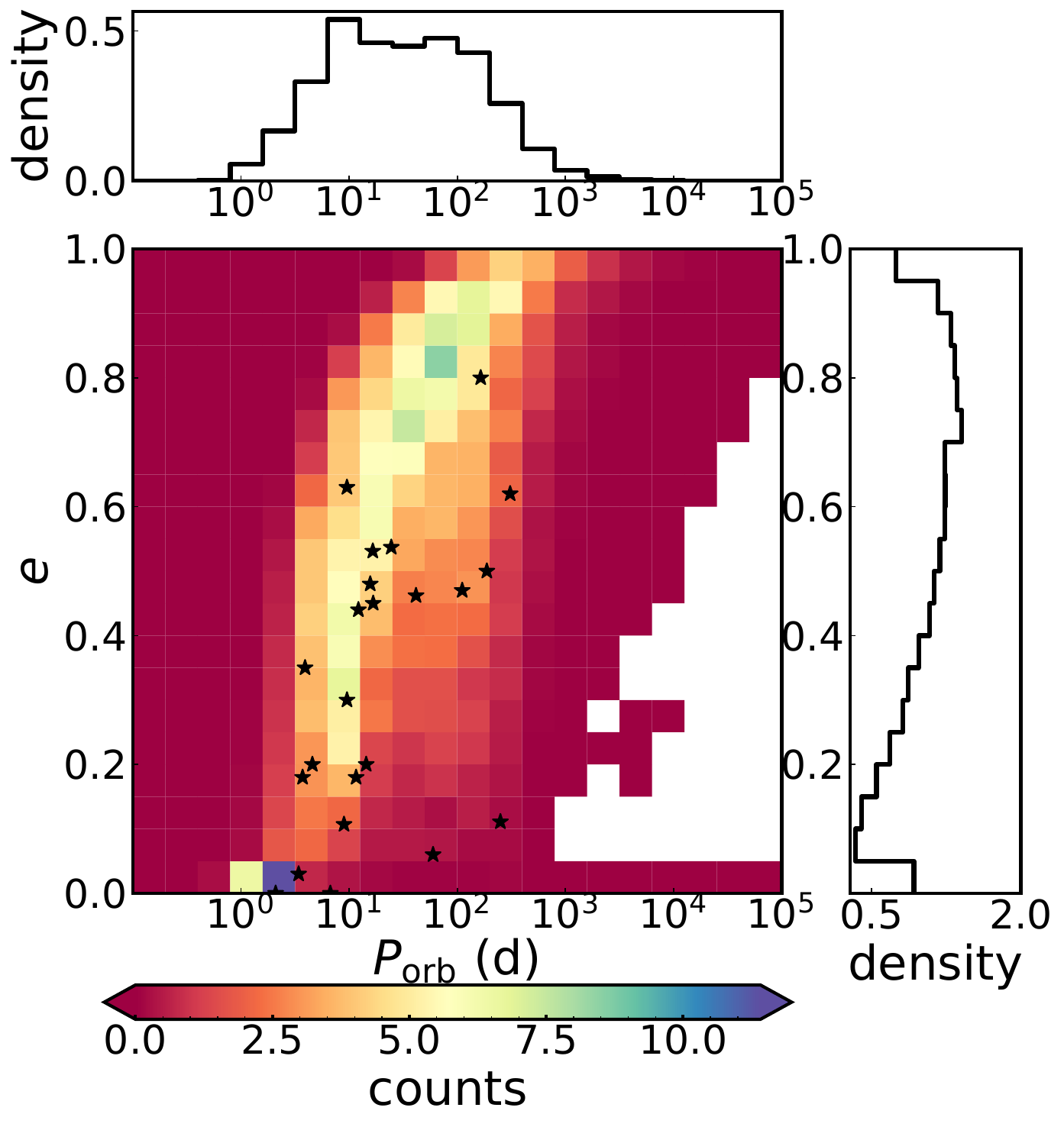}{0.27\textwidth}{$\sigma_1$ = 320 km\,s$^{-1}$, $\sigma_2$ = 80 km\,s$^{-1}$}
 }  
 \gridline{\hspace{-1.0cm}\fig{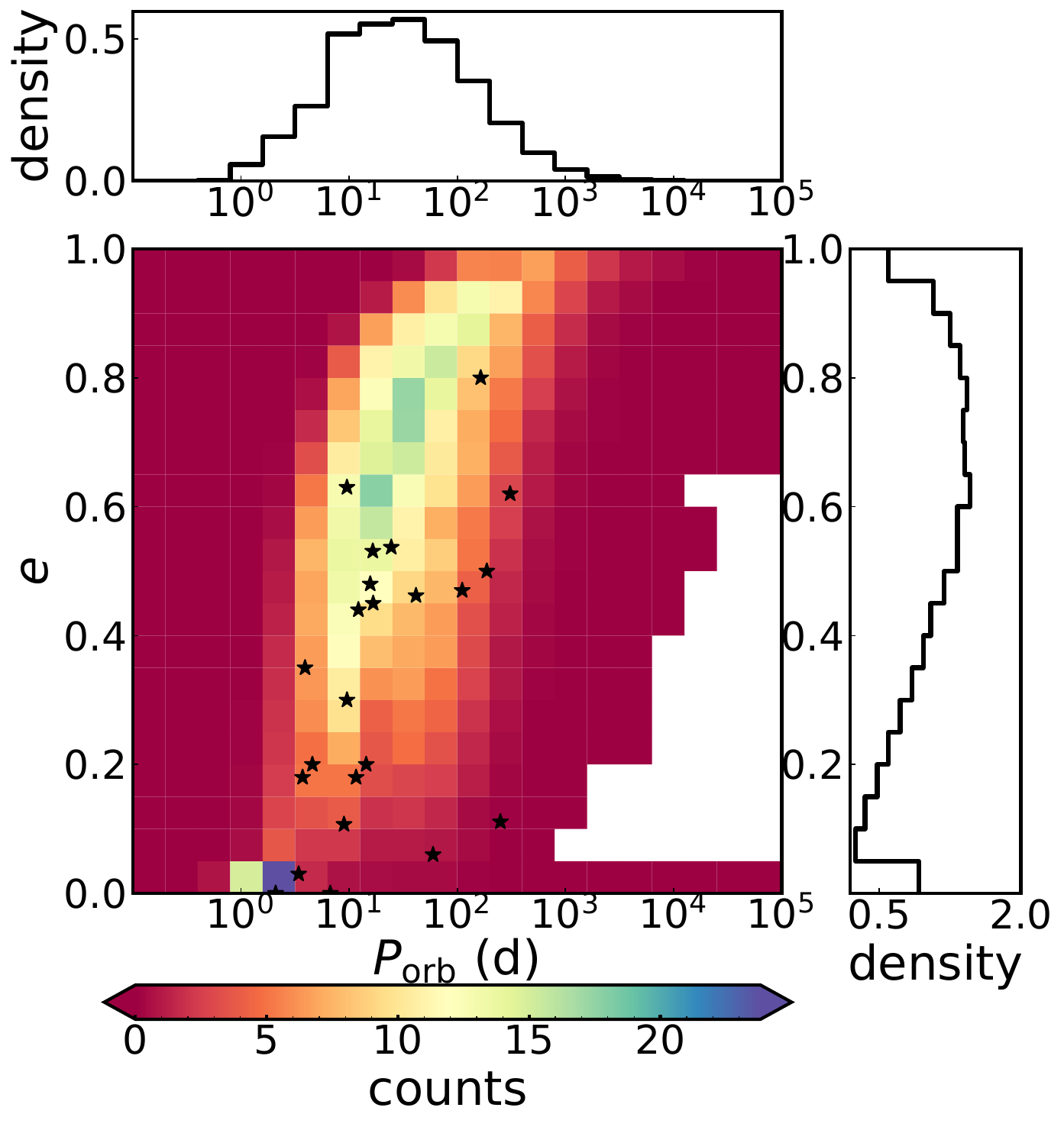}{0.27\textwidth}{$\sigma$ = 190 km\,s$^{-1}$}
\hspace{-0.4cm}\fig{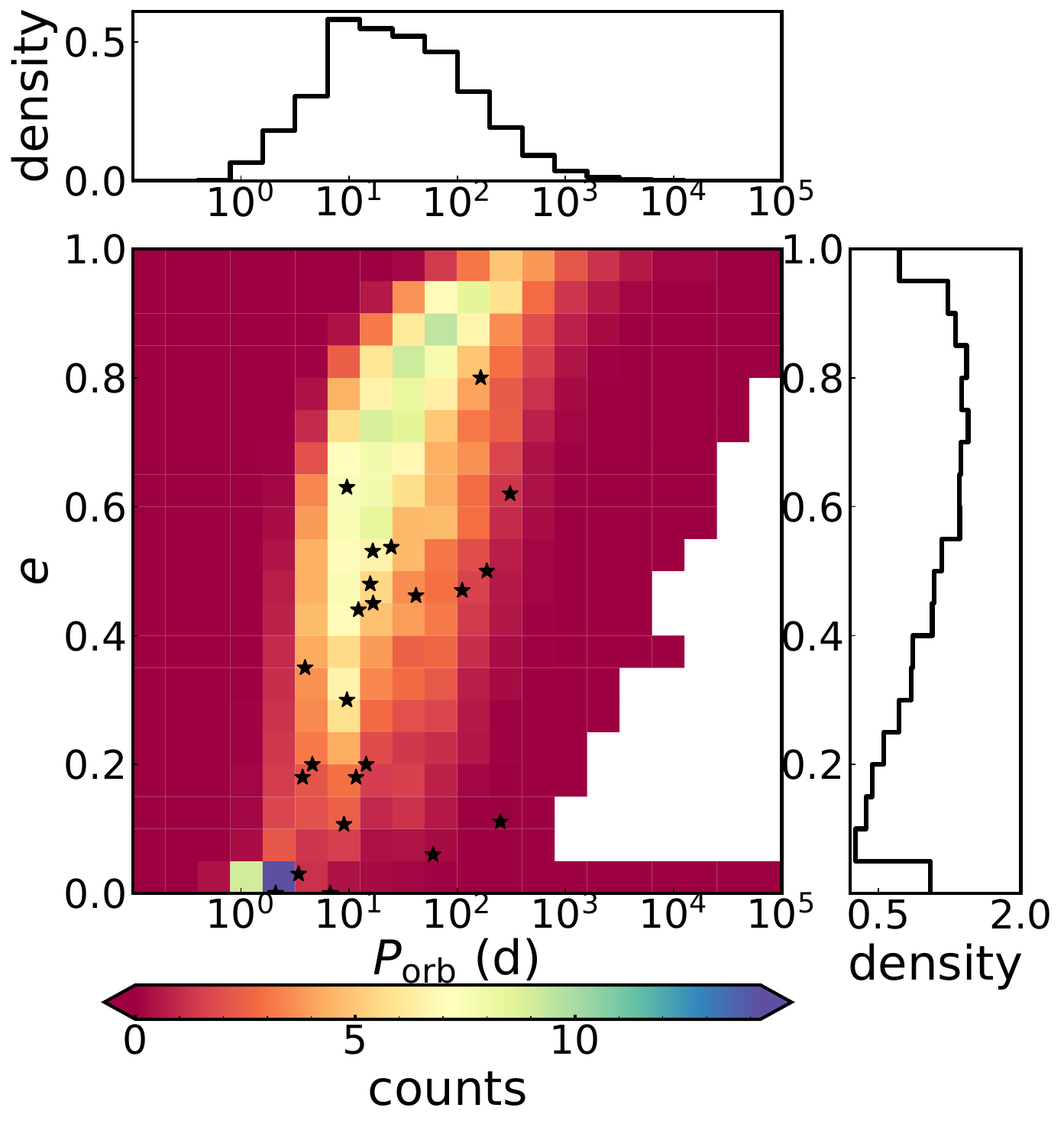}{0.27\textwidth}{$\sigma$ = 265 km\,s$^{-1}$}
 }
 \caption{Distributions of the modeled $P_{\rm orb}$ and $e$ of NS HMXBs with  $M_{\rm ecs}=(1.83-2.25)$ $M_{\odot}$. The black stars represent the observed NS HMXBs.}
\label{fig:Pe_A}
\end{figure*}

\begin{figure*}[t]
\vspace{-0.5cm}
\gridline{\hspace{-1.0cm}\fig{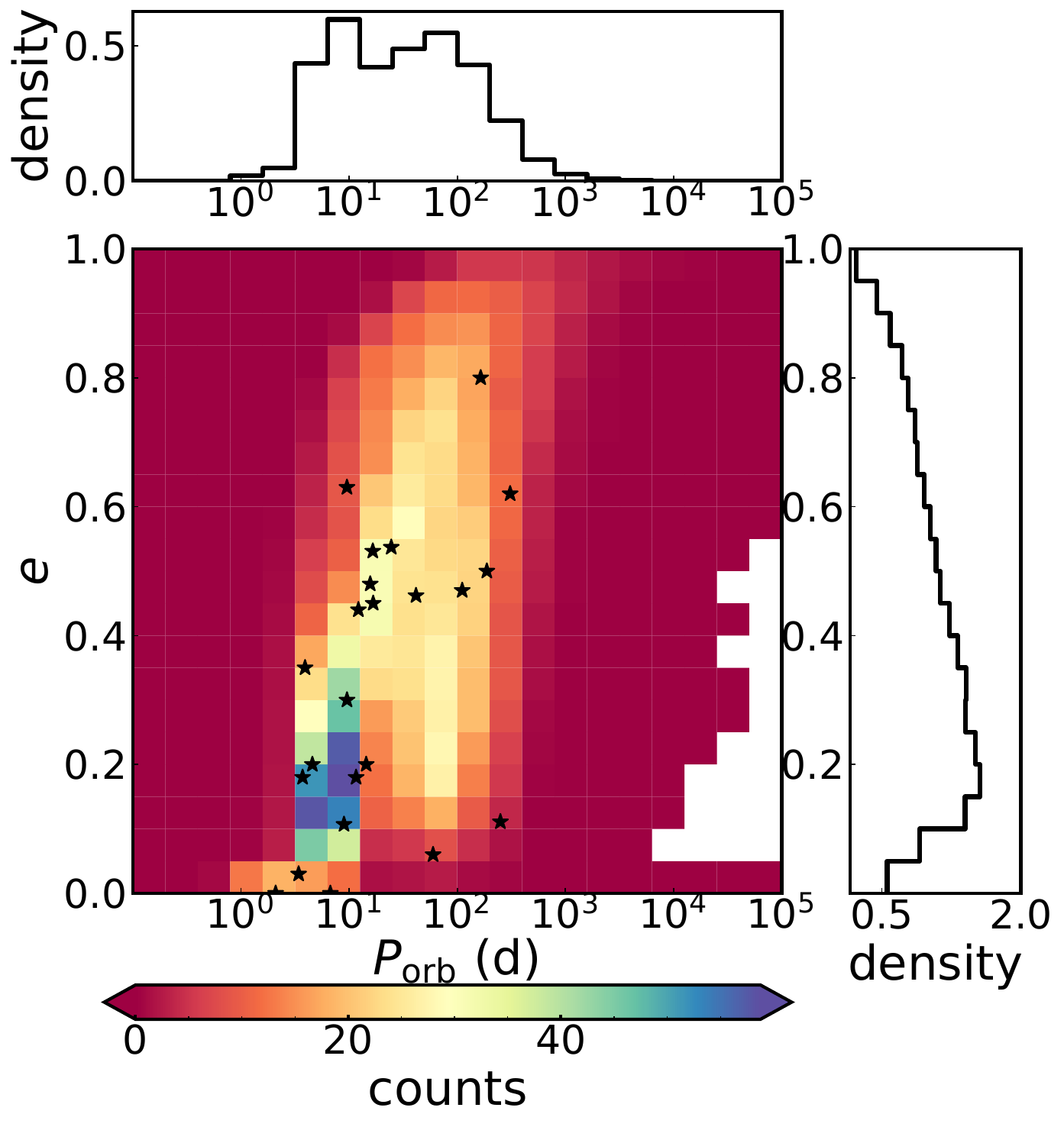}{0.27\textwidth}{$\sigma_1$ = 150 km\,s$^{-1}$, $\sigma_2$ = 30 km\,s$^{-1}$} \hspace{-0.4cm}\fig{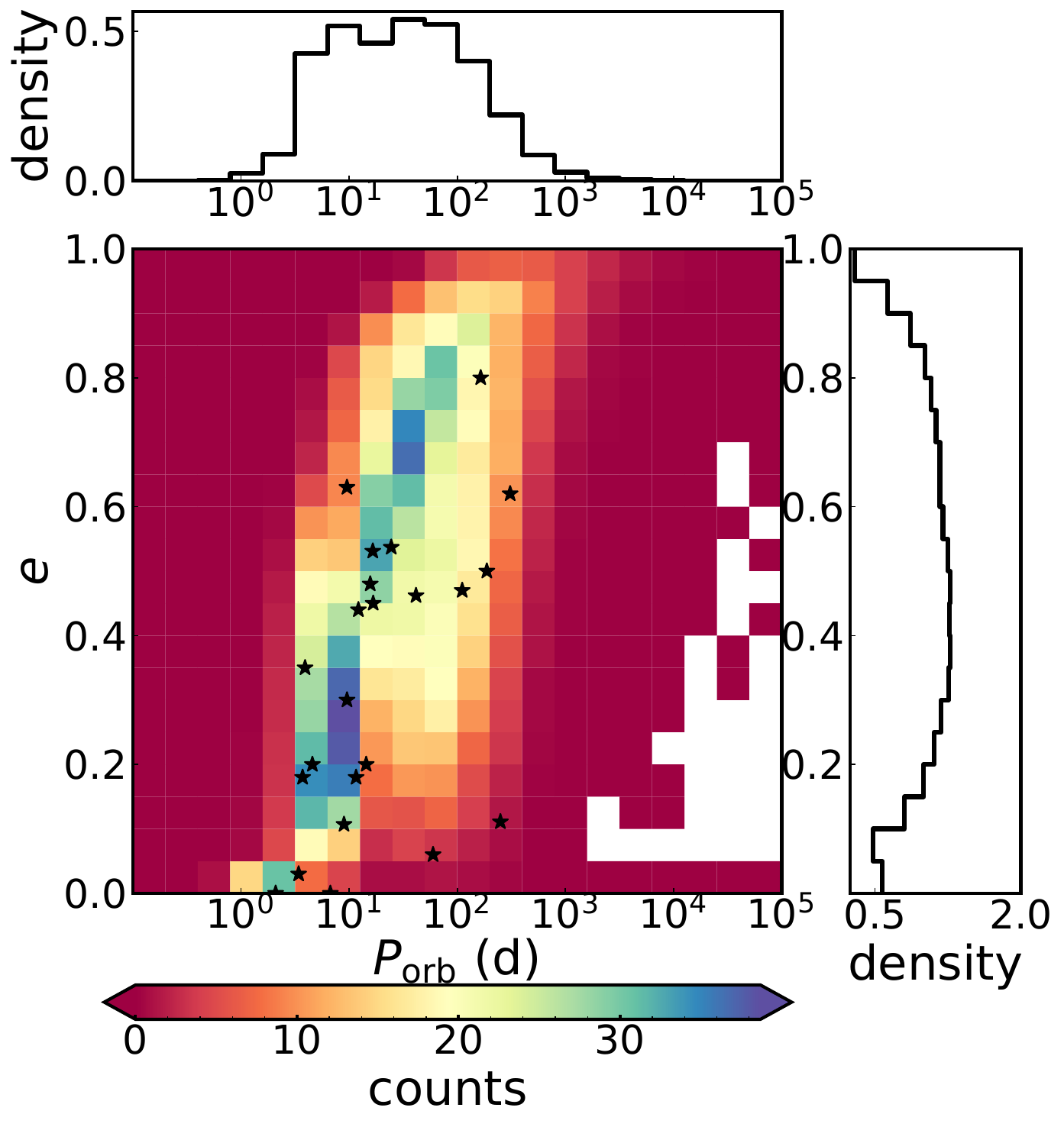}{0.27\textwidth}{$\sigma_1$ = 150 km\,s$^{-1}$, $\sigma_2$ = 50 km\,s$^{-1}$}
 \hspace{-0.4cm}\fig{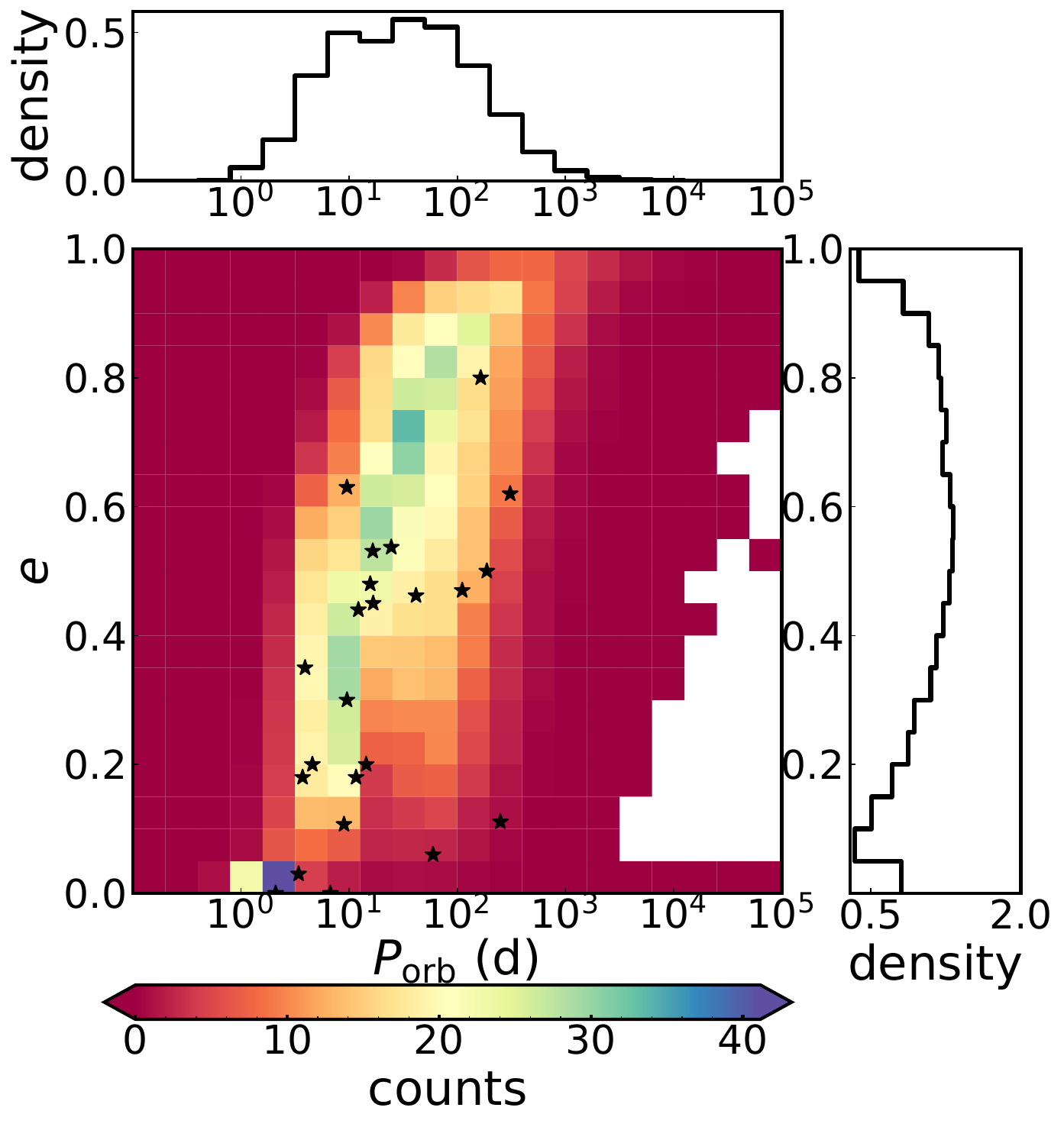}{0.27\textwidth}{$\sigma_1$ = 150 km\,s$^{-1}$, $\sigma_2$ = 80 km\,s$^{-1}$}
 }
\gridline{\hspace{-1.0cm}\fig{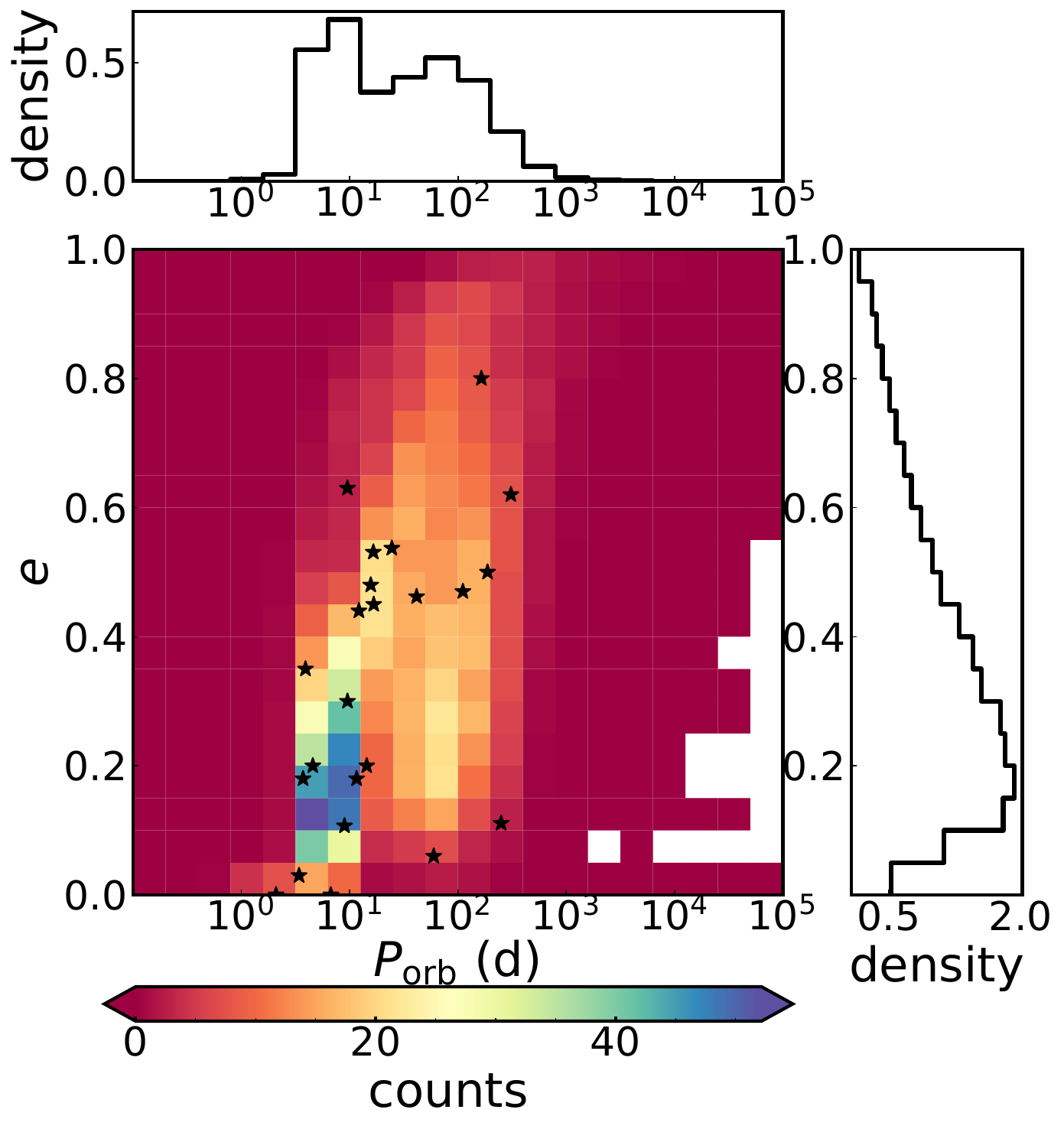}{0.27\textwidth}{$\sigma_1$ = 265 km\,s$^{-1}$, $\sigma_2$ = 30 km\,s$^{-1}$}
\hspace{-0.4cm}\fig{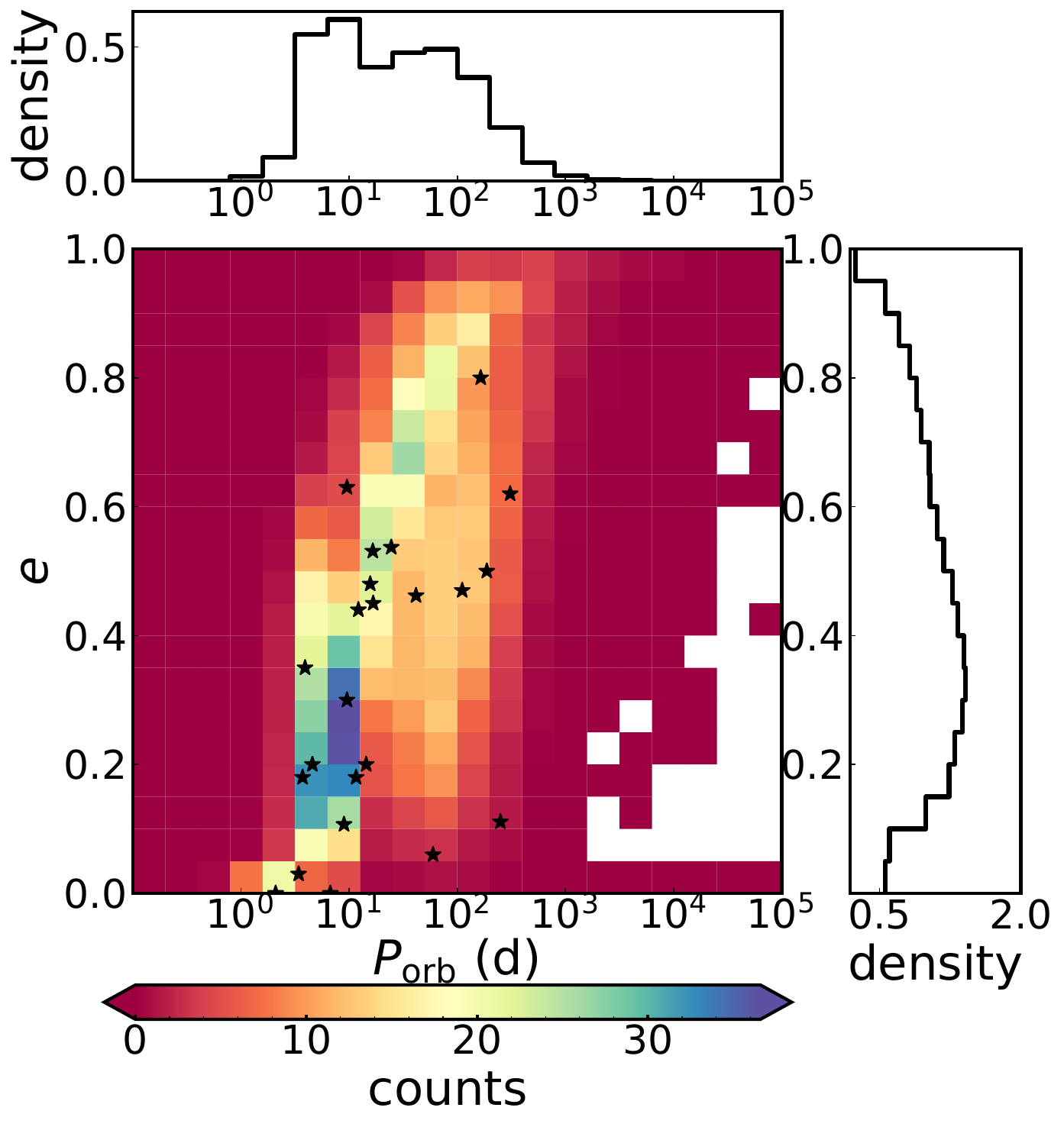}{0.27\textwidth}{$\sigma_1$ = 265 km\,s$^{-1}$, $\sigma_2$ = 50 km\,s$^{-1}$}
 \hspace{-0.4cm}\fig{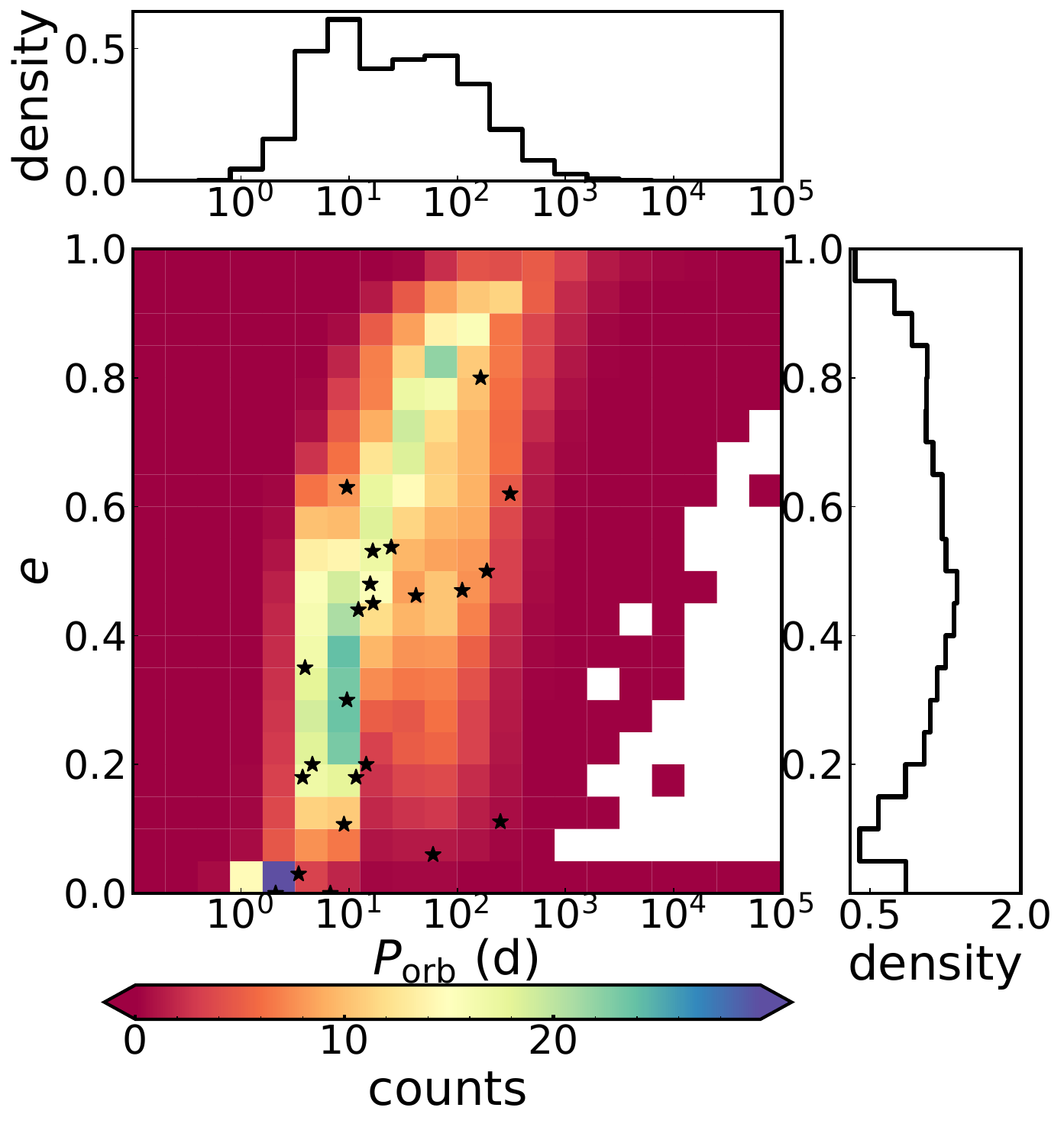}{0.27\textwidth}{$\sigma_1$ = 265 km\,s$^{-1}$, $\sigma_2$ = 80 km\,s$^{-1}$}
 } 
\gridline{\hspace{-1.0cm}\fig{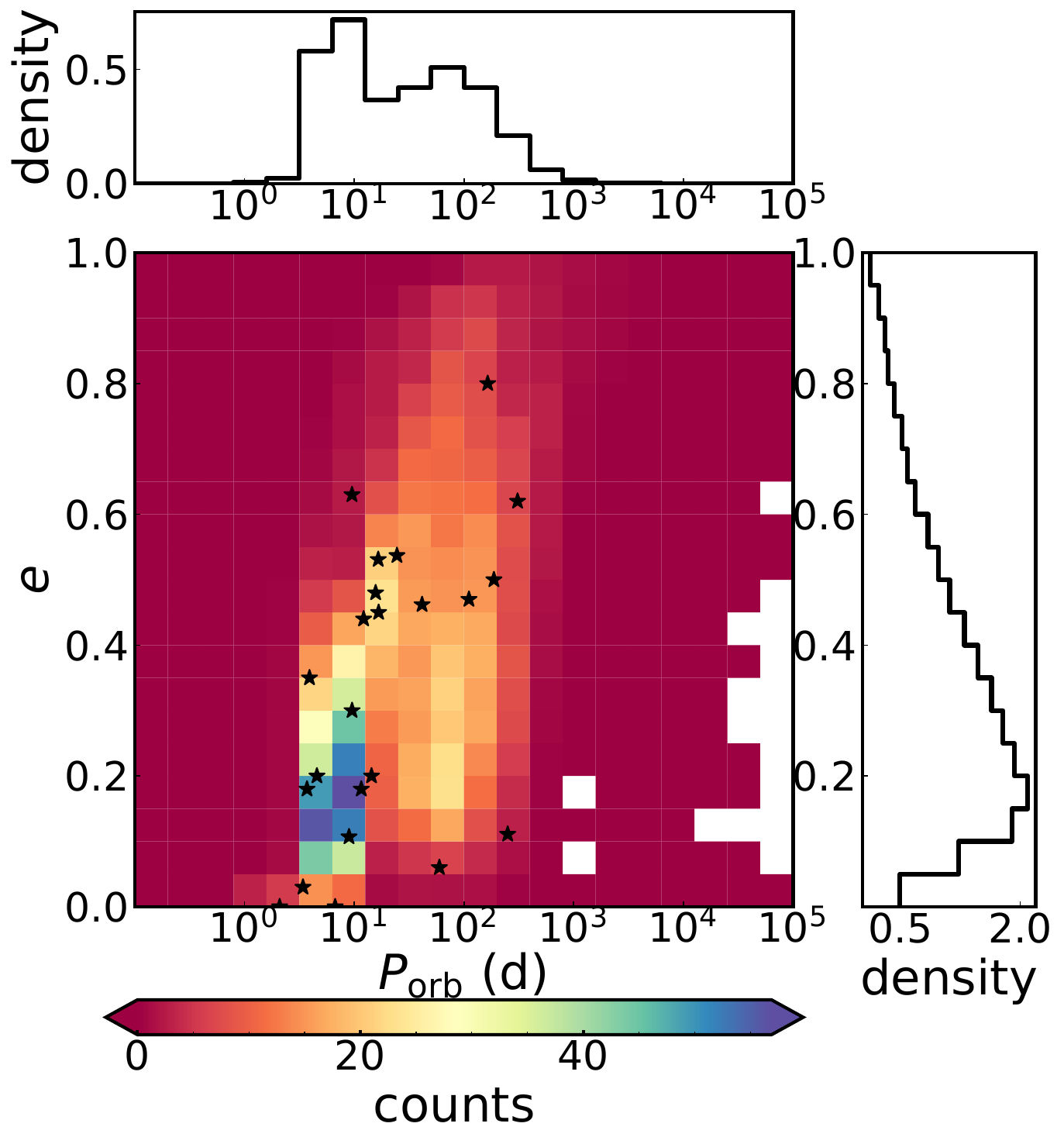}{0.27\textwidth}{$\sigma_1$ = 320 km\,s$^{-1}$, $\sigma_2$ = 30 km\,s$^{-1}$}
\hspace{-0.4cm}\fig{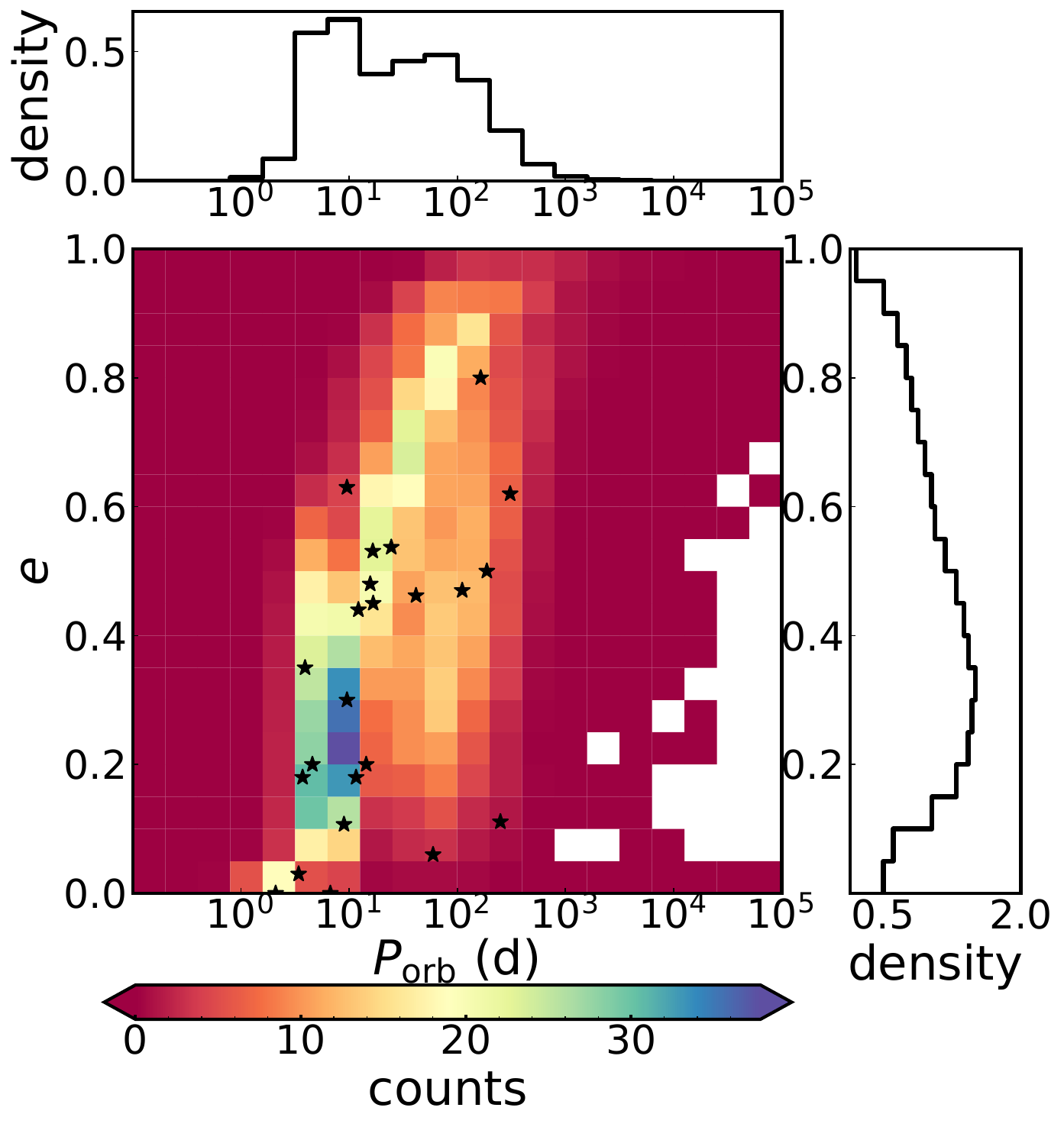}{0.27\textwidth}{$\sigma_1$ = 320 km\,s$^{-1}$, $\sigma_2$ = 50 km\,s$^{-1}$}
 \hspace{-0.4cm}\fig{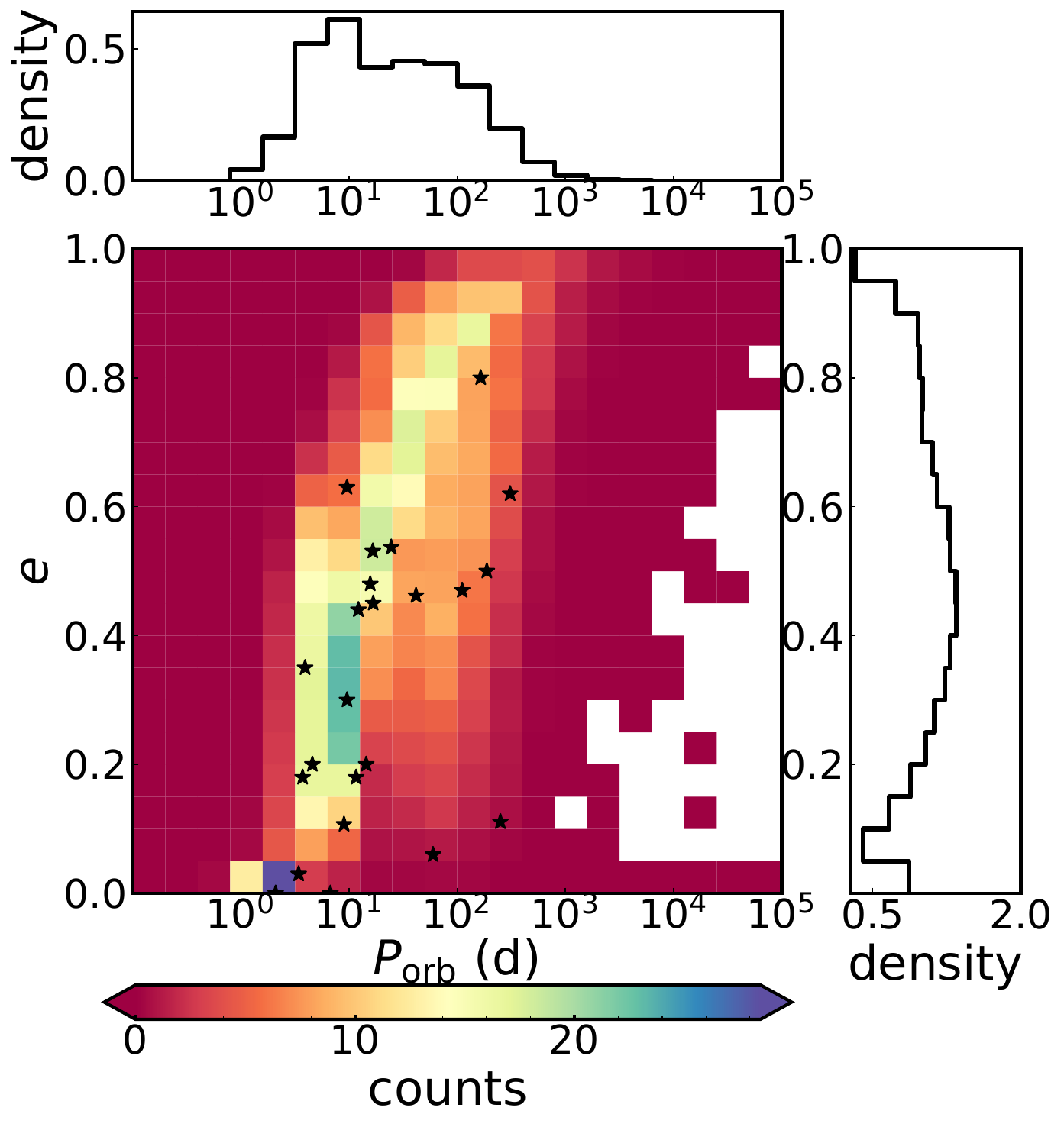}{0.27\textwidth}{$\sigma_1$ = 320 km\,s$^{-1}$, $\sigma_2$ = 80 km\,s$^{-1}$}
 }  
 \gridline{\hspace{-1.0cm}\fig{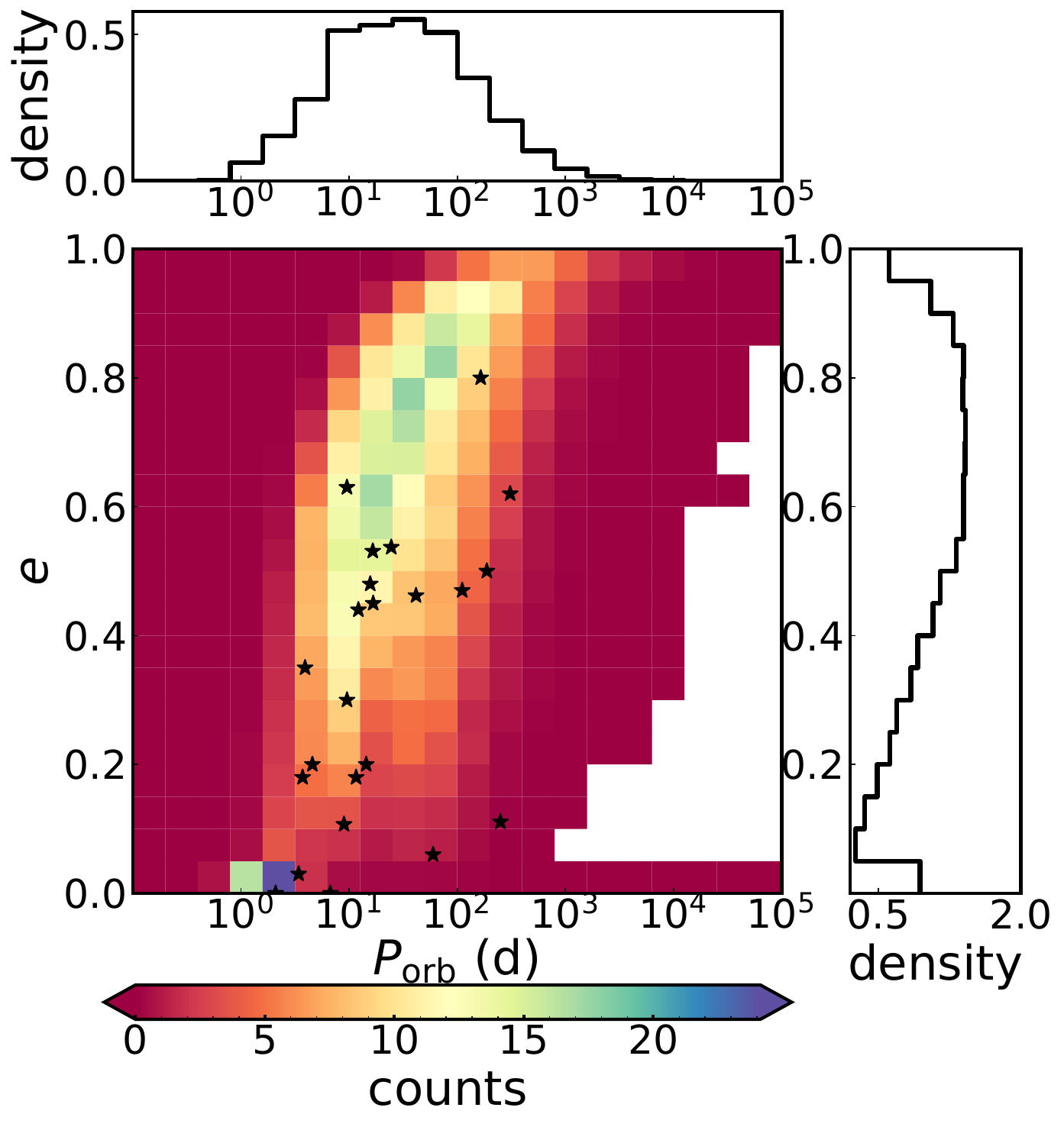}{0.27\textwidth}{$\sigma$ = 190 km\,s$^{-1}$}
\hspace{-0.4cm}\fig{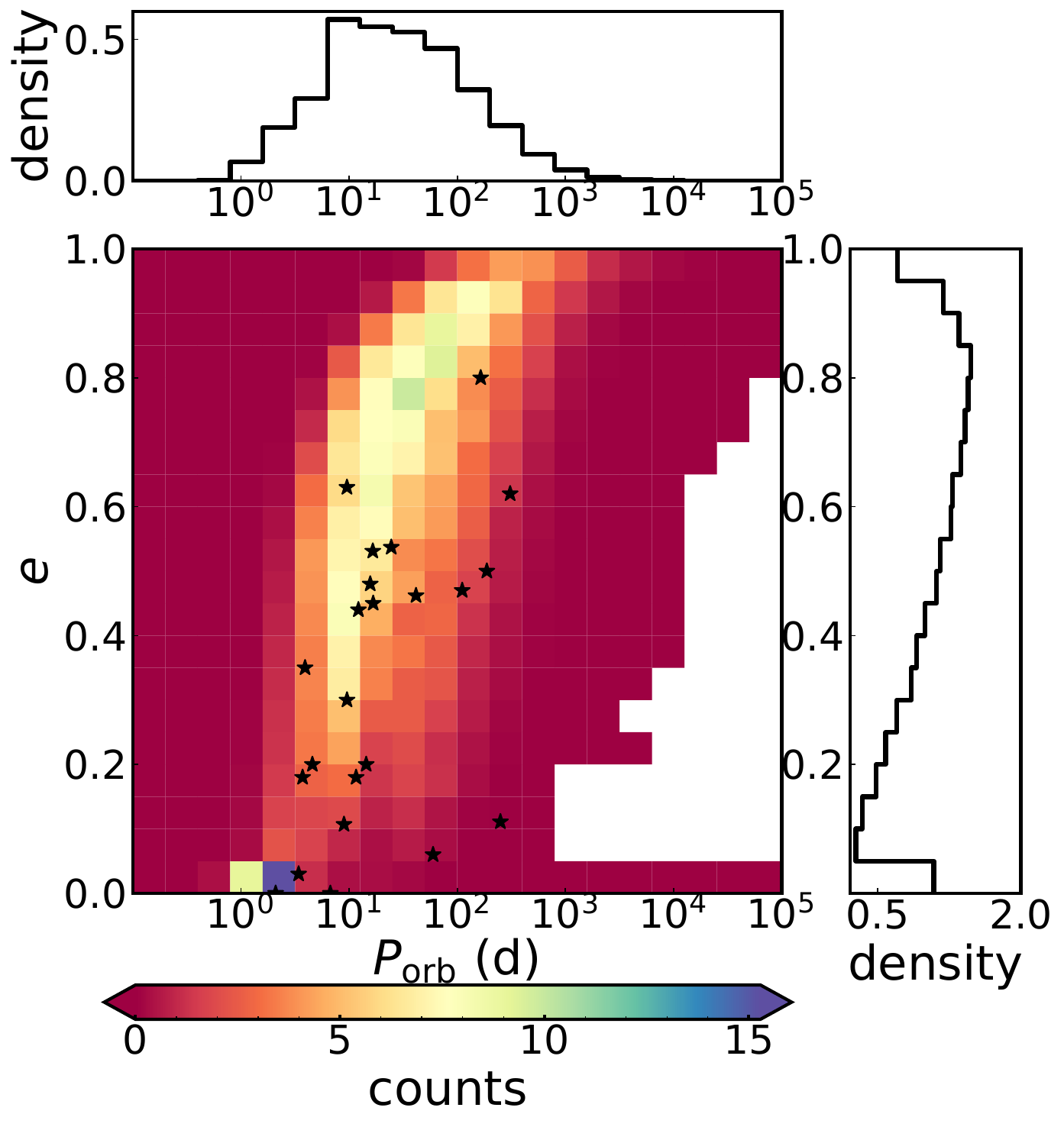}{0.27\textwidth}{$\sigma$ = 265 km\,s$^{-1}$}
 }
\caption{Same as Figure \ref{fig:Pe_A}, but with $M_{\rm ecs}=(1.83-2.75)$ $M_{\odot}$.}
\label{fig:Pe_B}
\end{figure*}

We compare the simulated and observed orbital period-eccentricity ($P_{\rm orb}$-$e$) distributions for NS HMXBs in Figures~\ref{fig:Pe_A} and \ref{fig:Pe_B}, with the input parameters same as in Figures~\ref{fig:vk_A} and \ref{fig:vk_B}, respectively. We search $P_{\rm orb}$ and $e$ values for the 36 NS HMXBs utilized in our velocity distribution analysis. Among them, only 24 sources have valid $P_{\rm orb}$ and $e$ measurement. Besides the available $P_{\rm orb}$ values in \cite{Zhaoyue2023MNRAS}'s catalog, we also search \citet{Fortin2024A&A}'s catalog  for the unrecorded $e$ values. We list the the $P_{\rm orb}$ and/or $e$ values for 11 HMXBs in the last two columns of Table \ref{tab:table2}.
We employ the Bayesian analysis method \citep{Andrews2015ApJ} to find out the highest posterior probability in the $P_{\rm orb}$-$e$-$v_{\rm pec}^{z=0}$ parameter space. The posterior probability of model $M$
given the observed data $D$ is described by
\begin{equation}
    \label{eq:probability}
    P(M|D) = \frac{P(D|M)P(M)}{P(D)},
\end{equation}
where $D$ represents the data of $P_{\rm orb}$, $e$, and $v_{\rm pec}^{z=0}$, $M$ represents the models we set, $P(D)$ is the normalizing constant independent of the models, $P(M)$ is the prior probability of a specific model, and $P(D|M)$ is the likelihood of the observed data given that the model is true, denoted as $\Lambda(D)$. Note that $P(M)/P(D)$ can be normalized as a constant since we assume that all models have the same prior probability, so $\Lambda(D)$ can directly reflect the value of $P(M|D)$. Since all the observed systems are independent, $\Lambda(D)$ is expressed as
\begin{equation}
    \label{eq:lambda_D}
    \Lambda(D) = \Lambda({\rm log} \ P_{{\rm orb}}, e, v_{\rm pec}^{\rm z=0}) = \prod \limits_{i} P(\log P_{{\rm orb},i}, e_{i}, v_{{\rm pec},i}^{\rm z=0}|M),
\end{equation}
where $P(\log P_{{\rm orb},i}, e_{i}, v_{\rm pec}^{\rm z=0}|M)$ is the probability density for every specific observed point in the $P_{\rm orb}$-$e$-$v_{\rm pec}^{z=0}$ parameter space. 

We apply kernel density estimation with a Gaussian kernel and follow \citet{2015mdet.book.....S}’s rule with \texttt{Scipy} to obtain the PDF. Given the values of $P_{\rm orb}$, $e$, and $v_{\rm pec}^{z=0}$ in the PDF, we calculate the values of $\Lambda(D)$ for the 11 $v_{\rm k}$ distribution models and list them in Table \ref{tab:value}. We find that when both dispersions of the $v_{\rm k}$ distributions are below 200\,km\,s$^{-1}$, the $\Lambda(D)$ values are extremely small, with some approaching zero. Regarding the $P_{\rm orb}$-$e$-$v_{\rm pec}^{z=0}$ distribution, the best model corresponds to the one with the highest likelihood $\Lambda(D)$, i.e., $M_{\rm ecs}=(1.83-2.25)$ $M_{\odot}$ and bimodal kick velocity distribution with $\sigma_1 = 320$ km\,s$^{-1}$ and $\sigma_2 = 80$ km\,s$^{-1}$. This is in agreement with the result in Section~\ref{subsec:velocity}.

In addition to comparing $\Lambda(D)$ in the 3D ($P_{\rm orb}$-$e$-$v_{\rm pec}^{z=0}$) parameter space, we also check its performance in the 2D ($P_{\rm orb}$-$e$) parameter space with the same procedure. The best velocity distribution is a bimodal Maxwellian distribution with $\sigma_1$ = 265 km\,s$^{-1}$ and $\sigma_2$ = 30 km\,s$^{-1}$, and $M_{\rm ecs}$ in the range of $(1.83-2.75)$ $M_{\odot}$, as  shown in Table~\ref{tab:value}. However, the $p$-value for the $v_{\rm pec}$ distributions is 0 in this case, indicating disagreement between the velocity distribution and the $P_{\rm orb}$-$e$ distribution. When all the key parameters (i.e., $v_{\rm pec}^{\rm z=0}$, $P_{\rm orb}$, and $e$) are involved, the current data seem to better match the model of bimodal Maxwellian distribution characterized by $\sigma_1$ = 320 km\,s$^{-1}$, $\sigma_2$ = 80 km\,s$^{-1}$, and $M_{\rm ecs}$ = $(1.83-2.25)$ $M_{\odot}$ compared to other models.

\begin{table*}
\centering
\caption{Statistical test of the $v_{\rm k}$ models for NS HMXBs.}
\renewcommand{\arraystretch}{1.5}
\begin{tabular}{c| |c|c|c|c||c|c|c|c}
    \hline
    \hline
     $v_{\rm k}$ distribution & \multicolumn{4}{c||}{$M_{\rm ecs}$ = (1.83-2.25)~$M_{\odot}$}& \multicolumn{4}{c}{$M_{\rm ecs}$ = (1.83-2.75)~$M_{\odot}$}  \\
     \cline{2-9}
     (km\,s$^{-1}$) &$p$-value & Weight $^{\rm a}$ & $\log\Lambda(P_{\rm orb}, e, v_{\rm pec}^{\rm z=0})$ &$\log\Lambda(P_{\rm orb}, e)$  & $p$-value & Weight$^{\rm a}$ & $\log\Lambda(P_{\rm orb}, e, v_{\rm pec}^{\rm z=0})$ &$\log\Lambda(P_{\rm orb}, e)$\\
    \hline

    $\sigma_{\rm 1}$ = 150, $\sigma_{\rm 2}$ = 30 & 0.3463 & 15.46$\%$ & $-$inf &$-$9.21 & 0.0050 & 60.52$\%$ &-inf &$-$6.22\\
     $\sigma_{\rm 1}$ = 150, $\sigma_{\rm 2}$ = 50 &  0.3690 & 13.00$\%$ & $-$inf &$-$9.99 & 0.0617 &56.32$\%$ & $-$inf &$-$7.59\\    
    $\sigma_{\rm 1}$ = 150, $\sigma_{\rm 2}$ = 80& 0.3659 &8.52$\%$ & $-$inf &$-$10.84 & 0.1471 & 47.80$\%$ & $-$inf & $-$8.88\\
    $\sigma_{\rm 1}$ = 265, $\sigma_{\rm 2}$ = 30& 0.0602 & 34.68$\%$ & $-$inf & $-$8.08 & 0.0000 &82.63$\%$ & $-$inf & $-$5.81\\
    $\sigma_{\rm 1}$ = 265, $\sigma_{\rm 2}$ = 50& 0.6295 &28.73$\%$ & $-$inf & $-$10.10 & 0.0003 & 80.04$\%$ & $-$inf & $-$6.68\\
    $\sigma_{\rm 1}$ = 265, $\sigma_{\rm 2}$ = 80& 0.4808 & 21.00$\%$ & $-$inf & $-$10.65 & 0.0223 &74.00$\%$ & $-$inf & $-$8.54\\
    $\sigma_{\rm 1}$ = 320, $\sigma_{\rm 2}$ = 30&  0.0013 & 45.24$\%$ & $-$274.54 & $-$8.27 &  0.0000 & 88.94$\%$ & $-$inf & $-$5.86\\
    $\sigma_{\rm 1}$ = 320, $\sigma_{\rm 2}$ = 50&  0.0727 & 39.78$\%$ & $-$266.04 & $-$9.41 & 0.0000 & 86.65$\%$ & $-$inf & $-$6.46\\
    $\sigma_{\rm 1}$ = 320, $\sigma_{\rm 2}$ = 80& 0.8662 & 28.83$\%$ & $-$188.08 & $-$10.34 & 0.0021 & 81.96$\%$ & $-$inf & $-$8.15\\
    $\sigma$ = 190& 0.0262 & 3.32$\%$ & $-$inf & $-$11.59 & 0.0340 & 34.15$\%$ & $-$inf & $-$11.56\\
    $\sigma$ = 265& 0.0042 & 3.63$\%$ & $-$inf &$-$14.85 & 0.0066 & 38.42$\%$ & $-$inf & $-$12.91\\
    \hline
    \hline
\end{tabular}
\begin{tablenotes}
\item[] $^{\rm a}$ For NSs produced by ECSNe.
\end{tablenotes}
\label{tab:value}
\end{table*}

\section{Conclusions and Discussion} \label{sec:conclusion}
In this work, we perform the BPS study to investigate the motion of NS HMXBs. We construct 11 possible $v_{\rm k}$ distributions for newborn NSs and compare the simulated systemic velocities with the derived peculiar velocities of 36 NS HMXBs. We employ the Bayesian analysis method to calculate the posterior probability $\Lambda(D)$ for different $v_{\rm k}$ distributions in the $P_{\rm orb}$-$e$-$v_{\rm pec}^{z=0}$ parameter space. 

Our simulation results are based on the assumption that ECSNe produce systematically weaker kicks than CCSNe. In this framework, the best model is characterized by $M_{\rm ecs}$ in the range of $(1.83-2.25)$ $M_{\odot}$ and a bimodal Maxwellian $v_{\rm k}$ distribution with $\sigma_1 = 320$ km\,s$^{-1}$ (CCSNe) and $\sigma_2 = 80$ km\,s$^{-1}$ (ECSNe). The corresponding proportions of the two components are $71\%$ and $29\%$, respectively. These numbers align with the statistical results obtained by \citet{Verbunt2017A&A} for young pulsars, and are also compatible with the theoretical studies of the kicks in CCSNe \citep[$\lesssim$~1000  km\,s$^{-1}$,][]{Janka2017ApJ,2020LRCA....6....3M,Tauris2023book}  and ECSNe \citep[$\lesssim$~100 km\,s$^{-1}$;][]{Gessner2018ApJ}. 

We obtain the $v_{\rm k}$ distribution different from previous studies such as \cite{Zhaoyue2023MNRAS} because we have identified several high-velocity systems (e.g., Swift J0243.6$+$6124) in our expanded sample of HMXBs. In comparison, $v_{\rm pec}^{\rm z=0}$ of the HMXBs population in \cite{Zhaoyue2023MNRAS} are all below 110 km\,s$^{-1}$.
Notably, the $\sim 312$ km\,s$^{-1}$ $v_{\rm pec}^{\rm z=0}$ of Swift J0243.6+6124, an ultra-luminous XRB in the Galaxy \citep{Tsygankov2018MNRAS, Reig2020A&A}, makes it stand out within the sample. Assuming that the masses of the exploding helium star, the resultant NS, and the companion star are 3.6 $M_{\odot}$,  1.4 $M_{\odot}$, and 16 $M_{\odot}$, respectively, we can estimate the ratio of $v_{\rm k}$ to the orbital velocity ($v_{\rm orb,i}$) of the pre-SN binary  to be less than 2.3 with the method of \cite{Hills1983ApJ}. If taking ${v_{\rm k}}$/${v_{\rm orb,i}} = 2.3$, we obtain the orbital separation of the pre-SN binary to be $\sim 3.0\,R_\sun$ and $v_{\rm k}\sim 2500$ km\,s$^{-1}$ for a random kick \citep{Tauris1996A&A}. Since no numerical simulations have observed kicks more than 2000 km\,s$^{-1}$ \citep{Scheck2006A&A, Janka2017ApJ}, this implies that Swift J0243.6$+$6124 might have been formed in an extraordinary way. On the other hand, we cannot rule out the possibility that its abnormal velocity reported by the Apache Point Observatory Galactic Evolution Experiment \citep{2020AJ....160..120J} could be inaccurate. However, even excluding Swift J0243.6+6124, we find that the best $v_{\rm k}$ model is the same as before. 

Our study is subject to several uncertainties including the initial parameter distributions of the primordial binaries, the mass transfer efficiency, and most importantly, the SN mechanisms. Detailed supernova simulations revealed that core-collapse outcomes exhibit complex behaviors beyond a simple dependence on initial stellar mass \citep[e.g.,][]{Sukhbold2016,Ertl2020,Schneider2023,Burrows2024}. Current recipes like those in \citet{Fryer2012ApJ} might not reflect the relation between the progenitor stars and the compact remnants, and natal kicks should depend on the progenitor properties at the time of the explosion rather than being drawn from some universal distributions \citep{Mandel2024}. From the observational perspective, enlarging the sample of NS HMXBs can help carry out more effective test of the theoretical models. The known HMXBs are relatively bright X-ray sources because the NSs are rapidly accreting from their companions, which means that the HMXBs stage occupies only a small portion of the binary evolution. Detection of faint or quiescent HMXBs population will significantly contribute to the NS XRB population and help understand the evolutionary history of massive binaries.

The lower end of the NS kicks during CCSNe is typically associated with the collapse of low-mass Fe cores, which is often linked to ultra-stripped supernovae (USSNe).  Their progenitors are stripped by degenerate companions, leaving the envelope of $\lesssim$ 0.2 $M_{\odot}$ \citep[e.g.,][]{Tauris2015MNRAS}. In our study, about $20\%$ NS HMXBs progenitors have outer envelope less than $0.2$ $M_{\odot}$. We have not separately considered  natal kicks during USSNe because it is still unclear whether NSs formed from USSNe are imparted a significantly smaller kick compared with those from CCSNe. For example, numerical simulation by
\citet{2021A&A...645A...5S} suggested that stripped stars on average
give rise to lower-mass NSs, higher explosion energies, and higher kick velocities (with $\sigma=315\pm 24$ km~s$^{-1}$).

Besides different SN mechanisms, there could be alternative explanations on the high- and low-kick velocity components. \cite{Pfahl2002ApJ} investigated the possible origin of HMXBs with low $e$ and long $P_{\rm orb}$, and proposed a phenomenological picture that the magnitude of $v_{\rm k}$ to an NS born in a binary system depends on the rotation rate of its immediate progenitor following mass transfer. Rapidly rotating pre-collapse cores produce NSs with relatively small kicks, and vice versa for slowly rotating cores. This model can be tested with detailed binary evolution calculation incorporating the spin evolution of the component stars, provided that the coupling between cores and envelopes in massive stars is well understood.
More recently, \cite{Hirai2024ApJ} showed that the motion of NSs could be influenced by multiple physical processes. These include: (1) immediate recoil due to the mass loss \citep[Blaauw kick;][]{Blaauw1961BAN}, (2) instantaneous kick associated with core collapse, and (3) the extended rocket mechanism that continues to accelerate NSs over long timescales after the initial kick. \cite{Hirai2024ApJ} argued in a qualitatively way that these kicks + rocket combinations may help form wide low-eccentricity NS binaries, depending on the mechanisms of the rocket effect \citep[i.e., electromagnetic rocket or neutrino rocket;][]{Harrison1975ApJ, Lai2001ApJ, Peng2004IAUS, Kojima2011ApJ, Li2022ApJ}. It is obviously worth doing detailed population studies taking into account the rocket effect.

We conclude that future work with more precise measurements on $P_{\rm orb}$ and $e$, larger NS HMXB sample, and more physical parameters  such as the companion mass, spin, and the X-ray luminosity will enhance our ability to refine the natal kick distribution of NSs.

\section*{acknowledgments}
We are grateful to an anonymous referee for insightful comments that helped improve the manuscript. We thank Yong Shao, Shi-Jie Gao, and Jian-Guo He for helpful discussions. This work was supported by the National Key Research and Development Programs of China (2021YFA0718500), and the Natural Science Foundation of China under grant No. 12041301 and 12121003.


\clearpage

\end{CJK*}
\end{document}